\pdfoutput=1

\documentclass[final]{elsarticle}

\usepackage{hyperref}

\journal{Journal of \LaTeX\ Templates}

\usepackage{amsmath}
\usepackage{amssymb}
\usepackage{subcaption}

\usepackage{grffile}

\usepackage[usenames,dvipsnames,svgnames,table]{xcolor}
\usepackage[subfigure]{tocloft}
\usepackage{changes}
\usepackage{xifthen}
\usepackage{wasysym}
\usepackage{algorithm}
\usepackage{algpseudocode}
\usepackage{amsopn}
\usepackage{bm}
\usepackage{multicol}

\makeatletter

\algnewcommand\NewParameterLine{%
	\newline \hspace*{\algorithmicindent} \hspace*{\algorithmicindent}
}

\algnewcommand\LeftComment[1]{%
	\Statex \vspace{0.5\baselineskip}\hspace{\ALG@thistlm}$\triangleright$  #1\hfill %
}

\algnewcommand\FirstLeftComment[1]{%
	\Statex \vspace{0.5\baselineskip}\hspace{\algorithmicindent}\hspace{\ALG@thistlm}$\triangleright$  #1\hfill %
}

\algnewcommand\FirstLeftCommentCont[1]{%
	\Statex \hspace{\algorithmicindent}\hspace{\ALG@thistlm}\phantom{$\triangleright$}  #1\hfill %
}
\algnewcommand\MultiLineState{%
	\Statex \hspace{\algorithmicindent}\hspace{\ALG@thistlm}%
}
\makeatother

\newcommand{\AND}
{\textbf{and \ }}

\newcounter{exampleCounter}
\refstepcounter{exampleCounter}

\renewcommand{\vec}[1]{\mathbf{#1}}

\newcommand{\mat}[1]{\mathbf{#1}}

\providecommand{\url}[1]{\texttt{#1}}

\newcommand{\norm}[2][]
{
	\ensuremath{\left\| #2\right\|
		\ifthenelse{\isempty{#1}}
		{}
		{_{{#1}}}}\xspace
}

\newcommand{\gradient}[1]
{\ensuremath{\nabla#1}\xspace
}

\newcommand{\hessian}[1]
{\ensuremath{\mat{H}#1}\xspace
}

\newcounter{todoListCounter}

\newcommand{\theTodoListCounter}
{
  \arabic{todoListCounter}
}

\newcommand{\notDone}
{\textcolor{Sepia}{$\Square$}}

\newcommand{\listTodoName}
{ \ifthenelse{\theTodoListCounter >1}
  {
    There are \textcolor{Sepia}{\theTodoListCounter} todo's
  }
  {
    There is \textcolor{Sepia}{\theTodoListCounter} todo
  }
}

\newlistof{todo}{tmp}{\listTodoName}

\providecommand{\newtodo}[2]
{
  \refstepcounter{todo}
  \noindent \rule{\linewidth}{1pt} \\
  \textbf{\large \theTodoListCounter} \textcolor{Sepia}{ TODO}: {\sc #2} \hfill \raisebox{-1ex}{\Huge #1}\\
  \rule{\linewidth}{1pt} \\
  \vspace{-1.5\baselineskip}
  \addcontentsline{tmp}{todo}{{\large #1} \textcolor{Sepia}{TODO \numberline{\thetodo:}}\textcolor{black}{#2}}
}

\newcommand{\listOfTodo}
{
  \ifthenelse{\theTodoListCounter > 0}
  {
    \clearpage
    \listoftodo
  }
  {
  }
}

\newcommand{\eval}[2]
{\ensuremath{#1\mathopen{}\left(#2\right)\mathclose{}}}

\newcommand{\Functional}[1]{E_{#1}}

\usepackage{xspace}

\algrenewcommand\algorithmicindent{1.0em}

\newcommand{\distortionFunction}
{\ensuremath{\eta\xspace}}

\newcommand{\qualityFunction}
{\ensuremath{\mathrm{q}\xspace}}

\newcommand{\mapName}
{\ensuremath{\bm{\phi}^*\xspace}}

\newcommand{\mapNameB}
{\ensuremath{\bm{\phi}\xspace}}

\newcommand{\elementName}
{\ensuremath{e\xspace}}

\newcommand{\domainName}
{\ensuremath{\Omega\xspace}}

\newcommand{\meshName}
{\ensuremath{\mathcal{M}\xspace}}

\newcommand{\pCoord}
{\ensuremath{\vec{x}}\xspace}

\newcommand{\iCoord}
{\ensuremath{\vec{y}}\xspace}

\newcommand{\distortion}[2][]
{\ifthenelse{\isempty{#2}}
	{\ensuremath{\distortionFunction_{#1}}}
	{\ensuremath{\distortionFunction_{#1}(#2)}}\xspace
}

\newcommand{\modDistortion}[2][]
{\ifthenelse{\isempty{#2}}
	{\ensuremath{\distortionFunction_{0}}}
	{\ensuremath{\distortionFunction_{0}(#2)}}\xspace
}

\newcommand{\quality}[2][]
{\ifthenelse{\isempty{#2}}
	{\ensuremath{\qualityFunction_{#1}}}
	{\ensuremath{\qualityFunction_{#1}(#2)}}\xspace
}

\newcommand{\modQuality}[2][]
{\ifthenelse{\isempty{#2}}
	{\ensuremath{\qualityFunction_{#1}^*}}
	{\ensuremath{\qualityFunction_{#1}^*(#2)}}\xspace
}

\newcommand{\objectiveFunction}[2][]
{\ifthenelse{\isempty{#2}}
	{\ensuremath{f_{#1}}}
	{\ensuremath{f_{#1}(#2)}}\xspace
}

\newcommand{\h}[1]
{\ifthenelse{\isempty{#1}}
	{\ensuremath{\determinant_\delta}}
	{\ensuremath{\determinant_\delta(#1)}}\xspace
}

\newcommand{\surface}[1]
{\ifthenelse{\isempty{#1}}
	{\ensuremath{\bm{\varphi}}}
	{\ensuremath{\bm{\varphi}(#1)}}\xspace
}

\newcommand{\curve}[1]
{\ifthenelse{\isempty{#1}}
	{\ensuremath{\bm{\gamma}}}
	{\ensuremath{\bm{\gamma}(#1)}}\xspace
}

\newcommand{\jacobian}[1][]
{\ifthenelse{\isempty{#1}}
	{\ensuremath{\mat{S}}}
	{\ensuremath{\mat{S}(#1)}}\xspace
}

\newcommand{\map}[2][]
{\ensuremath{\mapName
\ifthenelse{\isempty{#2}}
	{{}}
	{_{#2}}
\ifthenelse{\isempty{#1}}
	{{}}
	{(#1)}}\xspace
}

\newcommand{\mapB}[2][]
{\ensuremath{\mapNameB
\ifthenelse{\isempty{#2}}
	{{}}
	{_{#2}}
\ifthenelse{\isempty{#1}}
	{{}}
	{(#1)}}\xspace
}

\newcommand{\mapD}[2][]
{\ensuremath{\mapNameB_h
    \ifthenelse{\isempty{#2}}
    {{}}
    {_{#2}}
    \ifthenelse{\isempty{#1}}
    {{}}
    {(#1)}}\xspace
}

\newcommand{\determinant}[1][]
{\ifthenelse{\isempty{#1}}
	{\ensuremath{\sigma}}
	{\ensuremath{\sigma(#1)}}\xspace
}

\newcommand{\element}[1][]
{\ifthenelse{\isempty{#1}}
	{\ensuremath{\elementName}}
	{\ensuremath{\elementName_{#1}}}\xspace
}

\newcommand{\domain}[2][]
{\ensuremath{\domainName
	\ifthenelse{\isempty{#1}}
	{{}}
	{^{#1}}_{#2}}\xspace
}

\newcommand{\Gradient}[3][]
{\ensuremath{
	\nabla
	\ifthenelse{\isempty{#1}}
		{}
		{_{#1}}
	#2
	\ifthenelse{\isempty{#3}}
		{}
		{(#3)}
}\xspace
}

\newcommand{\Jacobian}[2][]
{\ifthenelse{\isempty{#1}}
	{\ensuremath{\textrm{\textbf{D}}#2}}
	{\ensuremath{\textrm{\textbf{D}}#2(#1)}}\xspace
}

\newcommand{\mesh}[2][]
{\ensuremath{\meshName^{{#1}}_{{#2}}}\xspace}

\newcommand{\projection}[2]
{\eval{\Pi_{#1}}{#2}}

\DeclareMathOperator*{\argmin}{arg\,min}

\newcommand{\node}[1]
{\ifthenelse{\isempty{#1}}
	{\ensuremath{v}}
	{\ensuremath{\vec x_{#1}}}\xspace
}

\newcommand{\scalarProduct}[3][]
{\ensuremath{\langle #2, #3\rangle
\ifthenelse{\isempty{#1}}
	{}
	{_{#1}}}\xspace
}

\newcommand{\trace}
{\ensuremath{\boldsymbol T}\xspace}

\graphicspath{{./figures/}}

\newcommand{\newtext}[1]
{#1}

\begin{document}

\begin{frontmatter}


\title{Automatic penalty and degree continuation for parallel pre-conditioned mesh curving on virtual geometry}



\author[bscaddress]{Eloi Ruiz-Giron\'es\corref{mycorrespondingauthor}}
\ead{eloi.ruizgirones@bsc.es}

\author[bscaddress]{Xevi Roca}
\ead{xevi.roca@bsc.es}

\cortext[mycorrespondingauthor]{Corresponding author}

\address[bscaddress]{
	Computer Applications in Science and Engineering,\\
	Barcelona Supercomputing Center - BSC, 08034 Barcelona, Spain}

\begin{abstract}
We present a distributed parallel mesh curving method for virtual geometry. The main application is to generate large-scale curved meshes on complex geometry suitable for analysis with unstructured high-order methods. Accordingly, we devise the technique to generate geometrically accurate meshes composed of high-quality elements. To this end, we advocate for degree continuation on a penalty-based second-order optimizer that uses global tight tolerances to converge the distortion residuals. To reduce the method memory footprint, waiting time, and energy consumption, we combine three main ingredients. First, we propose a matrix-free GMRES solver pre-conditioned with successive over-relaxation by blocks to reduce the memory footprint three times. We also propose an adaptive penalty technique, to reduce the number of non-linear iterations. Third, we propose an indicator of the required linear solver tolerance to reduce the number of linear iterations. On thousands of cores, the method curves meshes composed of millions of quartic elements featuring highly stretched elements while matching a virtual topology.
\end{abstract}



\end{frontmatter}


\section{Introduction}
\label{sec:introduction}

The capability to generate curved meshes on complex geometries is critical to perform high-fidelity simulations. These simulations demand computational methods featuring geometrical flexibility, high accuracy, and low numerical dissipation and dispersion. These requisites are fulfilled by many existent unstructured high-order methods. These methods also feature exponential convergence rates, and thus, they are faster than low-order methods in several applications  \cite{Sherwin-VSK:10, sherwin-CSKK:11a, Sherwin-CSKK:11b, Lohner:11, yano2012optimization, Kirby-KSC:12, huerta:isHDGCompetitive, Lohner:13,wang2013high}, especially in those problems where an implicit solver is required \cite{AA-HARP:13}. However, to keep their advantages, unstructured high-order methods need high-quality curved high-order meshes accurately approximating a complex boundary \cite{szabo:FEA91, schwab1998:p, deville2002:high, hesthaven:nodalDG, karniadakis2013:spectral}.

To generate curved high-order meshes approximating a complex boundary, there are many mesh curving methods \cite{dey1997:issuesPFEM, dey:curvilinearMeshing, luo:pMeshing, luo:automaticPMeshing, shephard:adaptiveCurvilinearMeshing, persson:curvedmesh, peiro:boundaryLayer, gargallo2015:tetOptimization, moxey2016high, fortunato2016winslow, eichstadt2018:programmingModel,sherwin:curvilinearMeshing,sevilla:HO3D,remacle:robustUntangling, karman2016:pointwiseHighOrder, stees2017:highorderWarp, panitanarak2018:lbwarpParallel, jiang2021:bichon}. These methods start with a linear mesh featuring elements of the desired shape and size, and then, the mesh boundary is curved to approximate the target geometry. This step may introduce low-quality and inverted elements that are repaired by the chosen curving technique. Nevertheless, there is still the need for parallel distributed mesh curving methods approximating a virtual geometry to generate large curved high-order meshes on complex models, as we detail in Section \ref{sec:relatedWork}. This capability is key to enable parallel unsteady (fine and graded meshes) or steady-state (fine meshes with stretched elements) high-fidelity flow simulations for large Reynolds numbers on highly complex geometries.

It is preferable to converge the mesh curving process with a tight tolerance for all the elements. A curved high-order mesh, obtained with a loose tolerance, might present oscillations in the element parameterization derivatives. Even when they are not visible, these oscillations might be sufficient to excite undesired spurious artifacts in the numerical solution. These oscillations decrease when the mesh curving uses a tighter tolerance for all the elements.

To achieve a tight convergence tolerance in all the mesh, we prefer to use a second-order optimization method. First-order methods feature a slower convergence rate. Nevertheless, their reduced memory footprint and computational cost per iteration are attractive. On the contrary, the memory footprint and computational cost per iteration of second-order methods are higher. Nevertheless, the faster convergence rate often reduces total optimization time when tight tolerances are needed. This time reduction is experienced when optimizing linear meshes, especially when dealing with highly graded and stretched elements.  We should expect similar results in mesh curving based on second-order optimization. Accordingly, we need to handle their higher memory footprint.

This high-memory footprint arises from the need to solve, at each non-linear iteration, a linear system of equations that involves all the mesh node coordinates as unknowns. The number of required mesh unknowns can be large in high-fidelity flow simulations. Thus, when using a second-order mesh curving optimization in all the mesh, we must solve the linear systems in a distributed fashion.

In our previous work \cite{ruiz2018:industrialHO, ruiz2019:pContinuation}, we proposed a distributed parallel mesh curving method for virtual geometries with a $p$-continuation technique. In \cite{ruiz2018:industrialHO, ruiz2019:pContinuation} we pose the mesh curving problem as a constrained optimization problem where we seek a minimum mesh distortion subject to a non-linear boundary condition. The constrained optimization problem is solved using a penalty method in which the boundary condition is introduced into the objective function using a penalty parameter. Then, several optimization problems are solved while increasing the penalty parameter to enforce the boundary condition. We devise the optimization process to favor deforming a valid mesh to a valid mesh at each optimization step. The $p$-continuation technique in \cite{ruiz2019:pContinuation} is critical to increasing the computational and energy efficiency of the curving process. However, mainly for large meshes, there is still the need to reduce the: memory footprint, waiting time, energy consumption, and total number of non-linear and linear iterations.

Herein, the first contribution is a new low memory pre-conditioned linear solver for curving. Specifically, we use a matrix-free GMRES linear solver combined with a specific-purpose pre-conditioner. Thus, we do not need to store the full sparse matrix. However, we still need a pre-conditioner to control GMRES iterations, especially for large graded meshes with stretched elements. The proposed pre-conditioner corresponds to a symmetric over-relaxation (SOR) method by sparse blocks. The blocks determine a $3 \times 3$ decomposition of the Hessian by groups of node coordinates. Using this pre-conditioner, we only need to store the diagonal blocks, and thus, we reduce the memory requirements three times.

The second contribution is a new methodology to adapt the penalty parameter during the optimization process. It reduces the required waiting time and energy consumption. Specifically, it reduces the number of non-linear problems to be solved. In our previous works \cite{ruiz2018:industrialHO, ruiz2019:pContinuation}, we increase the penalty parameter using a fixed factor. This fixed factor leads to additional intermediate non-linear problems if the increase factor is too small. Furthermore, it leads to non-linear problems that are difficult to solve if the increase factor is too high. We address this with our adaptive penalty parameter method. When the penalty method is in the convergence zone, we can increase the penalty parameter faster and predict the value needed to converge the curving process. Thus, we reduce the number of non-linear problems to be solved since we can increase the penalty parameter faster. Moreover, we can limit the penalty parameter since we can predict a value to converge the curving process. The latter is essential since as the penalty parameter increases, so does the condition number of the linear systems. Hence, large penalty parameters may lead to linear systems that cannot be solved. 

The third contribution is a new progress indicator for the penalty method devised to adapt the relative tolerance used to solve the linear systems. We deduce the required tolerance from the indicator. The indicator accounts for the different stages of the penalty optimization. In the first iterations of the penalty method, it is unnecessary to solve the linear systems with high accuracy because the current approximation is \emph{far} from the optimal solution. As the penalty method advances, it is necessary to solve the linear systems with higher accuracy. In the last iterations of the penalty method, we want to exploit Newton's method's quadratic convergence rate to meet a tight tolerance in all the mesh. Accordingly, we require the highest accuracy on the solution of the linear systems.

The rest of the paper is structured as follows. Section \ref{sec:relatedWork} reviews the literature related to the presented work. Section \ref{sec:problemStatement} presents the problem statement, and the input and output data. Section \ref{sec:meshCurving} presents the formulation of the proposed high order mesh curving methodology. Section \ref{sec:examples} presents several examples to show the capabilities of the proposed formulation. Section \ref{sec:discussion} discusses many aspects related to the proposed mesh curving process. Finally, Section \ref{sec:concludingRemarks} details the conclusions of this work.

\section{Related Work}
\label{sec:relatedWork}

Mesh curving methods can be divided into global and local methods. Global methods move all the nodes at the same time, while local methods move one node at a time. Global methods can be further divided into implicit methods that need to solve a sparse linear system, and explicit methods that move the nodes using an explicit formula. Note that herein, global (local) methods do not refer to obtain the global (local) minimum of an objective function.

One of the main bottlenecks in global implicit methods is the solution of a sparse linear problem to relocate all the mesh nodes at the same time. Therefore, efficient sparse linear solvers and pre-conditioners are necessary to curve high-order meshes composed of a large number of elements. In several works, the linear systems are solved using sparse direct solvers for morphing linear meshes \cite{staten2011:comparison}, and curving high-order meshes \cite{ruiz2016qualityDisparity}. Nevertheless, when the number of elements in the mesh increases, it is necessary to use iterative solvers. In general, the linear problems are solved either with MINRES, GMRES or conjugate gradients \cite{fortunato2016winslow, toulorge2016taylorDistance, panitanarak2018:lbwarpParallel, ruiz2019:augmentedlagrangian, dobrev2019:tmop}. To improve the computational efficiency, the linear problems can be pre-conditioned in different manners. For instance, diagonal pre-conditioners \cite{toulorge2016taylorDistance}, incomplete factorizations \cite{fortunato2016winslow, ruiz2019:augmentedlagrangian}, or especially designed pre-conditioners for the mesh curving problem \cite{garanzha2017:hyperelastic}. Global methods need to assemble a sparse matrix and solve the associated linear problem. Therefore, the parallelization of these methods need to assemble the matrix in parallel and apply a parallel sparse iterative linear solver, see \cite{panitanarak2018:lbwarpParallel, ruiz2018:industrialHO, dobrev2019:tmop}.

To avoid solving a sparse linear system, reference \cite{karman2018:viscousMeshing} relocates the nodes using a first order steepest descent minimization method. That is, the method relocates the nodes using a multiple of the objective function gradient. Although the convergence rate is lower than using Newton's method, the iterations are performed faster and the memory requirements are reduced.

Instead of solving a fully coupled linear problem to move all the nodes at the same time, local approaches move one node at a time. This approach is both applied for linear meshes \cite{sastry2014:logBarrierParallel, benitez2018:parallelPerformanceModel} and high-order meshes \cite{ruiz2016Hierarchical, peiro:boundaryLayer, eichstadt2018:programmingModel}. Although the local approach uses less memory and each iteration is faster than in the global approach, the convergence rate to the optimal solution can be hampered. In local approaches, nodes that do not belong to the same element can be relocated at the same time. Thus, several authors propose to color the nodes of the mesh in such a manner that nodes of the same color can be relocated at the same time \cite{sastry2014:logBarrierParallel, eichstadt2018:programmingModel, benitez2018:parallelPerformanceModel}. Moreover, in \cite{benitez2018:parallelPerformanceModel}, they propose a mesh partitioning method to ensure that the cost to relocate the nodes in each sub-domain is roughly the same.

In this work, we propose a distributed memory implementation that moves all the nodes at the same time. To accomplish this, we solve a sparse linear system of equations derived from Newton's method. Therefore, we expect faster convergence rates than first-order methods or local methods.

The main differences with the work presented in \cite{ruiz2019:pContinuation} is that we further reduce the waiting time and energy consumption while also reducing the memory footprint. Specifically, we propose an adaption of the penalty parameter to reduce the number of non-linear problems to be solved. We also propose a forcing term to reduce the number of iterations of the linear solver. Finally, we propose a block-SOR pre-conditioner that reduces the memory footprint three times.

\section{Problem Statement}
\label{sec:problemStatement}

\subsection{Input and Output}
\label{inputOutput}


\begin{figure*}[t!]
	\centering
	\hfill
	\begin{subfigure}[b]{0.25\textwidth}
		\includegraphics[width=\textwidth]{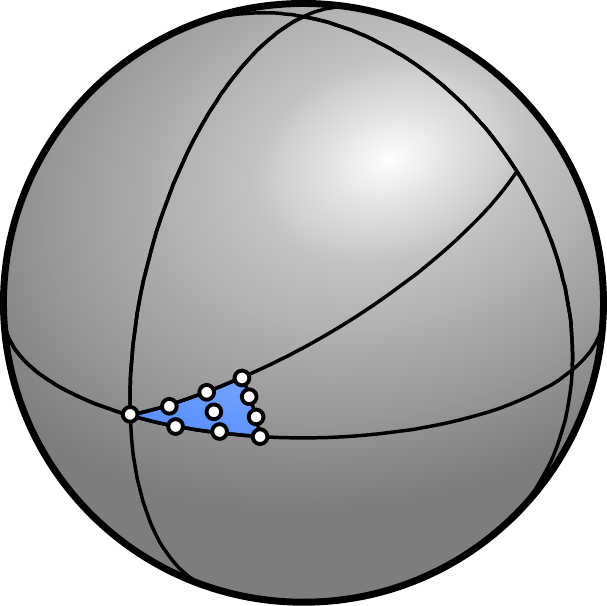}
		\caption{}
		\label{fig:sphereCAD_TriangleFaces}
	\end{subfigure}
	\hfill
	\begin{subfigure}[b]{0.25\textwidth}
		\includegraphics[width=\textwidth]{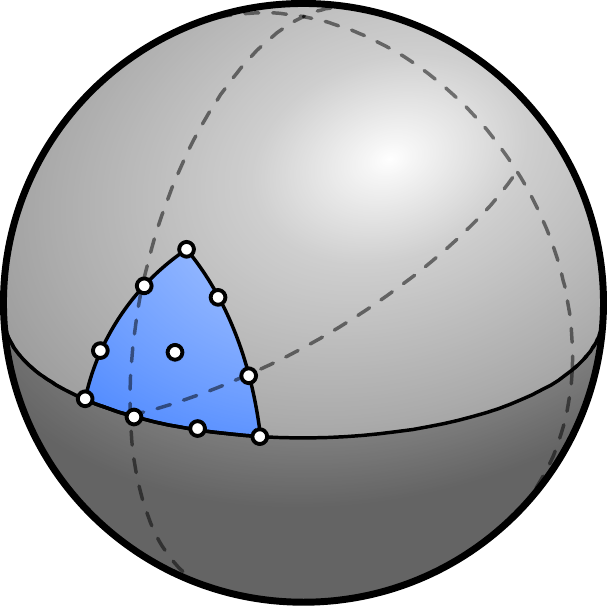}
		\caption{}
		\label{fig:sphereCAD_TriangleShells}
	\end{subfigure}
	\hfill\hspace{0cm}
	\caption{High order element approximating a geometric model of a sphere composed of:
	(a) nine surfaces and fourteen curves; and (b) two shells and one wire.}
	\label{fig:sphereCAD}
\end{figure*}

To perform the proposed mesh curving algorithm, we need several input data. The first input is a CAD model of the object of interest, in which several surfaces are grouped using a virtual geometry engine. The second input is a linear mesh composed of elements with the desired size and stretching. The output of our algorithm is an optimally converged curved high-order mesh that approximates the virtual model.

The virtual surfaces allow decoupling the topology of the original CAD model from the topology of the final mesh. Thus, we are able to avoid small angles and small surfaces that hamper the quality of the final mesh. For instance, Figure \ref{fig:sphereCAD_TriangleFaces} shows a CAD model of a sphere that contains a small angle. In blue, we depict a curved high-order element that approximates the sphere. The quality of this triangle is limited by the angle of the adjacent curves. In Figure \ref{fig:sphereCAD_TriangleShells}, we have grouped several surfaces into two virtual surfaces. In this case, the quality of the high-order triangle is improved because the triangle spans over several of the original surfaces.

The linear mesh approximates the virtual CAD model, and has the boundary triangles marked according to the virtual surface they approximate. The boundary marks allows us defining the appropriate boundary condition during the curving process to couple the curved high-order mesh and the virtual model.

\section{Parallel Mesh Curving}
\label{sec:meshCurving}

We first briefly summarize the proposed mesh curving solver on virtual geometry and the non-linear solver, as presented in \cite{ruiz2018:industrialHO}. Then, we detail the $p$-continuation technique introduced in \cite{ruiz2019:pContinuation}. Finally, we introduce the proposed techniques to increase the robustness and the efficiency of the curving method, and reduce the memory footprint.

\subsection{Mesh Curving Problem}
\label{sec:problem}

Given an initial linear mesh, \mesh[]{I}, we want to characterize a curved high-order one, \mesh[]{P}, in terms of a diffeomorphism \map{} \cite{gargallo2015:tetOptimization, ruiz2016Hierarchical}. The optimal diffeomorphism presents optimal point-wise distortion, and satisfies a prescribed boundary condition. That is, \map{} is the minimizer of
\begin{align}
	&\min_{\mapNameB \in \mathcal V} \eval{E}{\mapNameB} = \norm[\mesh{I}]{M\mapNameB}^2 \nonumber\\
	&\text{subject to:} \nonumber \\
	&\trace \mapNameB = \eval{\vec{g}_D}{\trace \mapB{}},
	\label{eqn:constrained}
\end{align}
\newtext{where
\[
	\norm[\mesh{I}]{f}^2 = \int_{\mesh{I}} f^2 \text{d}\domain{},
\]}
\trace is the trace operator, $\eval{\vec{g}_D}{\trace \mapB{}}$ is a non-linear Dirichlet boundary condition on $\partial\mesh{I}$ that depends on the values of \mapB{},  and
\[
\eval{M\mapB{}}{\iCoord} =
\eval{\eta}{\Jacobian{\eval{\mapB{}}{\iCoord}}}
= 
\frac{\norm[]{\eval{\Jacobian{\mapB{}}}{\iCoord}}^2}{n \eval{\determinant_0}{\eval{\Jacobian{\mapB{}}}{\iCoord}}^{2/n}}
\]
is a regularized point-wise distortion measure \cite{gargallo2015:tetOptimization} defined in terms of the shape distortion measure for linear simplices \cite{knupp:algebraicQuality}, where \norm{\cdot} is the Frobenius norm for matrices, and
\begin{equation}
	\determinant_0=\frac{1}{2}\left ( \determinant + |\determinant| \right),
	\label{eqn:pModDistortion}
\end{equation}
being \eval{\determinant}{\cdot} the determinant function. The regularized distortion measure takes a value of infinity when the determinant is negative or equal to zero, and takes finite values when the determinant is positive.

The non-linear boundary condition allows integrating a geometric model in the mesh curving process. In this work, we define the boundary condition as
\begin{equation}
	\eval{\vec g_D}{\trace \mapB[]{}} = 
	\sum_{i=1}^{N_b} \projection{}{\pCoord_i} N_i^b,
	\label{eqn:phiProjection}
\end{equation}
where $\pCoord_i$ are the coordinates of the mesh nodes, $N_b$ is the number of boundary nodes, $\{N^b\}_{i=1,\ldots,N_b}$ is a Lagrangian basis of shape functions \newtext{that are $\mathcal{C}^0-$}continuous between adjacent boundary faces, and \projection{}{\cdot} is a geometric orthogonal projection onto the CAD model. The boundary condition can be interpreted as an interpolation of the geometric model, in which the interpolation points are the projection of the boundary nodes. This boundary condition is non-linear and depends on the orthogonal projection of the boundary nodes.

\subsection{Mesh Curving Non-Linear Solver}
\label{sec:nonLinearSolver}

To solve the constrained optimization problem in \eqref{eqn:constrained}, we use a penalty approach, see \cite{nocedal:optimization}, in which we introduce the boundary constraint into the objective function in a weak sense as follows
\begin{equation}
	\min_{\mapB{} \in \mathcal V}
	\eval{\Functional{\mu}}{\mapB{}} = 
	\frac{\eval{\Functional{}}{\mapB{}}}{\norm[\mesh{I}]{1}^2} + 
	\mu \frac{\norm[{\partial \mesh{I}}]{\trace \mapB{} - \eval{\vec{g}_D}{\trace \mapB{}}}^2}{\norm[\partial \mesh{I}]{1}^2},
	\label{eqn:unconstrained}
\end{equation}
\newtext{where
\[
	\norm[\partial\mesh{I}]{f}^2 = \int_{\partial\mesh{I}} f^2 \text{d}\domain{},
\]}
$\mu$ is a penalty parameter that enforces the validity of the constraint when it tends to infinity. We have introduced the measures of the initial mesh and its boundary in order to balance the two contributions of the new functional.

The main idea is to solve several unconstrained optimization problems with increasing penalty parameter in order to enforce the boundary condition. Nevertheless, the boundary condition depends on the actual solution of the problem. Thus, we apply a fixed-point iteration as
\begin{align*}
	\vec g_D^{k} &= \eval{\vec g_D}{\trace \mapB[]{}^k}, \\
	\mapB{}^{k+1} &= \argmin_{\mapB{} \in \mathcal V} \eval{\Functional{\mu}}{\mapB{}; \vec g_D^k},
\end{align*}
being $k$ the $k$-th iteration of the proposed fixed-point solver.

We optimize each non-linear problem of the proposed penalty method using a backtracking line-search method in which the advancing direction is computed using Newton's method and the step-length is set using the Armijo's rule, see \cite{nocedal:optimization} for more details. We set a relative tolerance of 0.1 in the gradient to define the convergence criterion of Newton's method. Note that it is not necessary to fully converge all the non-linear problems, since the boundary condition and the penalty parameter will be modified. In this manner, we increase the efficiency of the curving process, and we allow a continuation in the solution with respect to the penalty parameter, and we fully converge the solution at the end of the curving process. In the last non-linear problem that has to be solved, we set the tolerance of Newton's method as the prescribed tolerance to finalize the penalty method. In this manner, we allow a Newton's method to converge the residual and obtain a fully converged mesh.

The convergence criterion of the penalty method is defined in terms of the gradient of the functional, and the value of the constraint as:
\[
	\norm[\partial{\mesh{I}}]{\mapB{}^k - \trace \mapB{}^k} / \norm[\partial{\mesh{I}}]{1}< \varepsilon^*,
	\qquad
	\norm[\infty]{\gradient{\eval{\Functional{\mu}}{\mapB{}^k; \vec g_D^k}}} < \omega^*.
\]
That is, we terminate the penalty method when both the boundary error and the residual are small enough. In our applications, we set $\omega^* = 10^{-8}$ and $\varepsilon^* = 10^{-12} \ell_c$, being $\ell_c$ a characteristic length of the model. 

\subsection{$p$-Continuation Technique}
\label{sec:pContinuation}

To improve the robustness of the proposed solver and to compute an initial condition for the non-linear solver, in \cite{ruiz2019:pContinuation}, we apply a $p$-continuation technique. Instead of directly computing the optimal mesh for a given polynomial degree, we iterate through the polynomial degrees and optimize them. The initial condition for each polynomial degree is the optimized mesh of the previous one. In our $p$-continuation technique, there are two main ingredients.

The first ingredient is an early termination criterion to stop the optimization of the initial polynomial degrees. The proposed early termination criterion reduces the computational cost of the full optimization process. To this end, we consider the mesh of the current polynomial degree, $\mapB{}^{p}$, and let $\mapB{}^{p+1}$ be the interpolation of $\mapB{}^{p}$ using element-wise polynomials of degree $p+1$. The early termination criterion is defined using the boundary condition error of both meshes as
\begin{equation}
	\alpha \norm[\partial{\mesh{I}}]{\trace \mapB{}^{p} - \eval{\vec g_D}{\trace \mapB{}^{p}}} < 
	\norm[\partial{\mesh{I}}]{\trace \mapB{}^{p+1} - \eval{\vec g_D}{\trace \mapB{}^{p+1}}}.
	\label{n:earlyTermination}
\end{equation}
That is, the optimization process of the current polynomial degree is finished when the error of the boundary condition is \emph{comparable} to the error of the boundary condition of the next polynomial degree. In this work, we take $\alpha=2$.

Note that $\eval{\vec g_D}{\trace \mapB{}^{p}}$ and $\eval{\vec g_D}{\trace \mapB{}^{p+1}}$ are in general terms different. Both terms are nodal projections on the virtual geometry of the current boundary state. However, they are different, since the the first one projects a nodal representation of degree p while the second one projects a different nodal representation of degree p+1.

\begin{algorithm*}[t!]
	\caption{Convergence criterion}
	\label{alg:convergence}
	\begin{algorithmic}[1]
		\Function{degreeConverged}{
			\NewParameterLine Map $\mapB{}_k$, Int $p$, Int $p_{\text{max}}$,
			\NewParameterLine Real $\varepsilon^*$, Real $\omega^*$}
		\If{$p = p_{\text{max}}$}
		\FirstLeftComment{Convergence criterion for $p_{\text{max}}$}
		\State $\varepsilon_k \gets \norm[\partial{\mesh{I}}]{\trace \mapB{}_{k} - \eval{\vec g_D}{\trace \mapB{}_{k}}} / \norm[\partial{\mesh{I}}]{1} $
		\State $\omega_k \gets \norm[\infty]{\gradient{\eval{\Functional{\mu}}{\mapB{}_k; \vec g_D^k}}}$
		\State converged $\gets \varepsilon_k < \varepsilon^*$ \AND $ \omega_k < \omega^*$
		\Else
		\FirstLeftComment{Early termination criterion}
		\State $\mapB{}^{p+1} \gets \Call{interpolate}{\mapB{}_{k}, p+1}$
		\State converged $\gets \alpha \norm[\partial{\mesh{I}}]{\trace \mapB{}_{k} - \eval{\vec g_D}{\trace \mapB{}_{k}}} < \norm[\partial{\mesh{I}}]{\trace \mapB{}^{p+1} - \eval{\vec g_D}{\trace \mapB{}^{p+1}}}$
		\EndIf  \label{lin:criterionEnd}
		\State \Return converged
		\EndFunction
	\end{algorithmic}
\end{algorithm*}

The second ingredient is the calculation of the initial penalty parameter for the optimization of each polynomial degree. A correct value of the penalty parameter facilitates the solution of the linear and non-linear problems and therefore, increases the robustness of the optimization process. For a sufficiently large penalty parameter, after optimizing the functional in Equation \eqref{eqn:unconstrained}, the Lagrange multipliers of the associated constraint are approximated as
\begin{equation}
	\lambda^{p} \simeq -2 \mu^p \frac{\trace \mapB{}^{p} - \eval{\vec g_D}{\trace \mapB{}^{p}}}{\norm[\partial \mesh{I}]{1}}.
	\label{eqn:multipliers}
\end{equation}
When the early termination criteria is satisfied, we interpolate $\mapB{}^p$ using polynomials of degree $p+1$, and compute the associated boundary condition of $\mapB{}^{p+1}$. Assuming that the associated Lagrange multipliers approximated using Equation \eqref{eqn:multipliers} are similar for both polynomial degrees, we set
\[
\lambda^{p+1} = \lambda^{p}.
\]
Therefore, by taking norms on both sides, we obtain that
\begin{equation}
	\mu^{p+1} = \mu^p 
	\frac{\norm[\partial \mesh{I}]{\trace \mapB{}^{p} - \eval{\vec g_D}{\trace \mapB{}^{p}}}}
	{\norm[\partial \mesh{I}]{\trace \mapB{}^{p+1} - \eval{\vec g_D}{\trace \mapB{}^{p+1}}}}.
	\label{eqn:nextDegreeMu}
\end{equation}

\begin{algorithm*}[t!]
	\caption{Penalty parameter estimation for new polynomial degree.}
	\label{alg:newDegreeMu}
	\begin{algorithmic}[1]
		\Function{firstIterationPenaltyParameter}{
			\NewParameterLine Map $\mapB{}^p$, Map $\mapB{}^{p+1}$,
			\NewParameterLine Real $\mu^{p}}$
		\State $\varepsilon^p \gets \norm[\partial{\mesh{I}}]{\trace \mapB{}^{p} - \eval{\vec g_D}{\trace \mapB{}^{p}}} / \norm[\partial{\mesh{I}}]{1} $
		\State $\varepsilon^{p+1} \gets \norm[\partial{\mesh{I}}]{\trace \mapB{}^{p+1} - \eval{\vec g_D}{\trace \mapB{}^{p+1}}} / \norm[\partial{\mesh{I}}]{1} $
		\State $\mu^{p+1} = \mu^{p} \varepsilon^{p} / \varepsilon^{p+1}$
		\State \Return $\mu^p$
		\EndFunction
	\end{algorithmic}
\end{algorithm*}

\subsection{Estimation of the penalty parameter}
\label{sec:penaltyParameter}

In this paper, we propose a novel methodology to adaptively increase the penalty parameter and reduce the number of non-linear problems to be solved, without hampering the robustness of the curving process. The selection of the penalty parameter has an important role in the robustness and efficiency of the curving process. If the penalty parameter is increased too fast, the linear and non-linear problems will be difficult to solve and therefore, the robustness of the method is reduced. On the contrary, if the penalty parameter is increased too slow, we will need to solve additional non-linear problems and therefore, the computational efficiency of the method is reduced.

To this end, we will use the same argument as in Section \ref{sec:pContinuation}. If the associated Lagrange multipliers of two consecutive iterations of the penalty method are similar,
\[
\lambda_{k} \simeq \lambda_{k-1},
\]
we obtain that
\[
\frac{\mu_{k}}{\mu_{k-1}} \simeq
\frac{\norm[\partial \mesh{I}]{\trace \mapB{}_{k-1} - \eval{\vec g_D}{\trace \mapB{}_{k-1}}}}
{\norm[\partial \mesh{I}]{\trace \mapB{}_{k} - \eval{\vec g_D}{\trace \mapB{}_{k}}}}.
\]

Therefore, if the ratio of two consecutive penalty parameters is well approximated by the ratio of two consecutive norms of the constraint error, we can assume that the penalty parameter of the current iteration is high enough. For this reason, we can estimate the penalty parameter to obtain the given tolerance of the constraint, $\varepsilon^*$, as
\begin{equation}
	\mu^{*} = \mu_k m^* = \mu_k 1.01  \frac{\varepsilon^*}{\norm[\partial \mesh{I}]{\trace \mapB{}_{k} - \eval{\vec g_D}{\trace \mapB{}_{k}}} / \norm[\partial \mesh{I}]{1}},
	\label{eqn:optimalMu}
\end{equation}
where $m^*$ is the optimal increase factor. If the penalty process is in the convergence region, the optimal increase factor would reduce the boundary condition norm exactly to the prescribed tolerance $\varepsilon^*$. For this reason, we have increased the optimal factor by 1.01 in order to obtain a boundary condition norm slightly below the prescribed tolerance.

Nevertheless, we need to limit the increase of the penalty parameter when the Lagrange multipliers are not well approximated. The main idea is to allow a larger increase of the penalty parameter when the Lagrange multipliers are better approximated. When the Lagrange multipliers are in the convergence zone, an increase of the penalty parameter by a given factor leads to a decrease of the norm of the boundary condition by the same factor. Therefore, we define the ratio
\[
s_k = \frac{\mu_{k-1}}{\mu_k}
\frac{\norm[\partial \mesh{I}]{\trace \mapB{}_{k-1} - \eval{\vec g_D}{\trace \mapB{}_{k-1}}}}
{\norm[\partial \mesh{I}]{\trace \mapB{}_{k} - \eval{\vec g_D}{\trace \mapB{}_{k}}}}.
\]
The ratio $s_k$ checks the reduction of the boundary condition norm against the increase of the penalty parameter for two consecutive iterations. In the convergence region, the ratio $s_k$ should be close to one. Thus, we define a penalty convergence indicator as
\[
e_k = \max\{ s_k, 1/s_k \} - 1.
\]
Using this indicator, we define a factor to increase the penalty parameter as
\begin{equation}
	m_k = \max\{ 10, 1/e_k \}.
	\label{eqn:mFactor}
\end{equation}
The factor $m_k$ is larger when the error estimator is smaller. Therefore, it will lead to a larger increase of the penalty parameter. If the convergence indicator is not sufficiently small, we set a default factor of 10. The penalty parameter for the next iteration is defined using the optimal penalty parameter in Equation \eqref{eqn:optimalMu}, and the factor $m_k$ in Equation \eqref{eqn:mFactor} as
\begin{equation}
	\mu_{k+1} = \min\{m^* \mu_k, m_k\mu_k \}.
	\label{eqn:nextMu}
\end{equation}

If the Lagrange multipliers are not well approximated, we will not increase the penalty parameter by a large factor. On the contrary, if the Lagrange multipliers are well approximated, we will be able to increase the penalty parameter by a large factor. Moreover, the penalty parameter will not get larger than the optimal one. Thus, we will not increase the penalty parameter in excess and therefore, we will not increase the condition number of the linear systems.

\begin{algorithm*}[t!]
	\caption{Penalty parameter adaption}
	\label{alg:muAdaption}
	\begin{algorithmic}[1]
		\Function{penaltyParameterAdaption}{
			\NewParameterLine Map $\mapB{}_k$, Map $\mapB{}_{k-1}$,
			\NewParameterLine Real $\mu_{k-1}$, Real $\mu_k$, Real $\varepsilon^*$}
		\FirstLeftComment{Compute constraint norms}
		\State $\varepsilon_{k-1} \gets \norm[\partial \mesh{I}]{\trace \mapB{}_{k-1} - \eval{\vec g_D}{\trace \mapB{}_{k-1}}} / \norm[\partial \mesh{I}]{1}$
		\State $\varepsilon_{k} \gets \norm[\partial \mesh{I}]{\trace \mapB{}_{k-1} - \eval{\vec g_D}{\trace \mapB{}_{k}}} / \norm[\partial \mesh{I}]{1}$
		
		\LeftComment{Compute optimal increase factor}
		\State $m^* = 1.01 \varepsilon^* / \varepsilon_{k}$
		
		\LeftComment{Compute error indicator}
		\State $s_k = (\mu_{k-1}/\mu_k)/(\varepsilon_{k-1}/\varepsilon_k)$
		\State $e_k = \max\{s_k,1/s_k\} - 1$
		\State $m_k = \max\{10, 1/e_k\}$
		
		\LeftComment{Compute next penalty parameter}
		\State $\mu_{k+1} = \min\{ m^* \mu_k, m_k \mu_k\}$
		\State \Return $\mu_{k+1}$
		\EndFunction
	\end{algorithmic}
\end{algorithm*}

\subsection{Sparse Matrix Pre-Conditioned Linear Solver}
\label{sec:linearSolve}

At each iteration of the non-linear solver we solve the following linear system derived from Newton's method:
\[
\hessian{\eval{\Functional{\mu}}{\vec u}} \delta \vec u =
- \gradient{\eval{\Functional{\mu}}{\vec u}}.
\]
Since in this work we are dealing with large-scale meshes, we assemble and solve the linear system in a parallel framework. We use the FEniCS software to automatically generate the derivative expressions required to evaluate the Hessian and the gradient. Since we solve this problem in parallel, we need to distribute the mesh elements and degrees of freedom. To this end, we use the ParMETIS library. The resulting distribution allows computing the element contributions in a distributed manner. Specifically, we use the FEniCS software, where each elemental contribution to the residual and the residual Jacobian are computed in parallel. To this end, non-blocking communication is performed to send and receive the required data from adjacent processors. Then, the contributions of the inner processor elements are computed and overlapped with the previously non-blocking communication. Then, the contributions from the processor boundary elements are computed. Finally, the residual and the residual Jacobian are assembled.

We have used the linear solvers and pre-conditioners implemented in the PETSc library \cite{petsc4py}. We solve the linear systems using a GMRES method restarted every 20 iterations. To reduce the number of iterations of the linear solver, we use a restricted additive Schwarz pre-conditioner with one overlap level. The local problems are approximated using two iterations of a symmetric successive over-relaxation method with $\omega=1$.

\subsection{Reduced memory: parallel and pre-conditioned linear solver}
\label{sec:matrixFree}

When solving a linear system using the GMRES method, we need to perform several matrix-vectors products. This operation can be performed without storing the matrix in memory. Specifically, we perform local operations at each element and then, assemble the resulting local vector into the global distributed vector. In this manner, the memory footprint of the linear solver is reduced. However, we need to properly apply a pre-conditioner to be able to solve the linear system in a reasonable number of iterations. Note that we are not able to apply the pre-conditioner proposed in Section \ref{sec:linearSolve} since we do not have the system matrix stored in memory.

The preconditioning strategy that we propose only stores a part of the full matrix and therefore, we are able to reduce the memory footprint of using classical linear solvers. To this end, we first reorder the unknowns of the linear system of equations $\vec H \vec x = \vec b$ to obtain a matrix and right hand side as
\[
	\vec H = \begin{pmatrix}
		\vec H_{1,1} & \vec H_{1,2} & \vec H_{1,3} \\
		\vec H_{2,1} & \vec H_{2,2} & \vec H_{2,3} \\
		\vec H_{3,1} & \vec H_{3,2} & \vec H_{3,3} 
		\end{pmatrix}
	,
	\vec b =
	\begin{pmatrix}
		\vec b_1 \\
		\vec b_2 \\
		\vec b_3
	\end{pmatrix},
\]
where $\vec H_{i,j}$ and $\vec b_i$ are associated to the dimensional components $i$ and $j$, for $i$ and $j$ from 1 to 3. Using this block decomposition, we will derive a block-SOR pre-conditioner. First, we split the original matrix as a sum of three contributions as
\[
	\vec H = \vec H_l + \vec H_d + \vec H_u =
	\begin{pmatrix}
		\vec 0       & \vec 0       & \vec 0 \\
		\vec H_{2,1} & \vec 0       & \vec 0 \\
		\vec H_{3,1} & \vec H_{3,2} & \vec 0 
	\end{pmatrix}
	+
	\begin{pmatrix}
		\vec H_{1,1} & \vec 0       & \vec 0 \\
		\vec 0       & \vec H_{2,2} & \vec 0 \\
		\vec 0       & \vec 0       & \vec H_{3,3} 
	\end{pmatrix}
	+
	\begin{pmatrix}
		\vec 0       & \vec H_{1,2} & \vec H_{1,3} \\
		\vec 0       & \vec 0       & \vec H_{2,3} \\
		\vec 0       & \vec 0 & \vec 0 
	\end{pmatrix}.
\]
Inserting this into the original linear system and reorganizing terms, we obtain the following equality:
\[
	(\vec H_l + \vec H_d) \vec x = \vec b - \vec H_u \vec x.
\]
This equality allows proposing an iterative linear solver method in the following manner
\[
	(\vec H_l + \vec H_d) \vec x^{l+1} = \vec b - \vec H_u \vec x^l.
\]

Specifically, at each iteration, we solve a block triangular linear system of the form
\[
	\begin{pmatrix}
		\vec H_{1,1} & \vec 0       & \vec 0 \\
		\vec H_{2,1} & \vec H_{2,2} & \vec 0 \\
		\vec H_{3,1} & \vec H_{3,2} & \vec H_{3,3}
	\end{pmatrix} 
	\begin{pmatrix}
		\vec x^{l+1}_1 \\
		\vec x^{l+1}_2 \\
		\vec x^{l+1}_3 
	\end{pmatrix}
	=
	\begin{pmatrix}
		\vec b_1 \\
		\vec b_2 \\
		\vec b_3 
	\end{pmatrix}
	-
	\begin{pmatrix}
		\vec 0       & \vec H_{1,2} & \vec H_{1,3} \\
		\vec 0       & \vec 0       & \vec H_{2,3} \\
		\vec 0       & \vec 0       & \vec 0      
	\end{pmatrix} 
	\begin{pmatrix}
		\vec x^{l}_1 \\
		\vec x^{l}_2 \\
		\vec x^{l}_3 
	\end{pmatrix}.
\]

To solve this linear system in an effective manner, we use a block-based forward substitution. Thus, we solve three linear systems in which the system matrix contain three times less unknowns. Specifically, the three linear systems that we solve at each iteration of the block-SOR pre-conditioner are:
\[
	\vec H_{i,i} \vec x^l_i = \vec b_i - \sum_{j=1}^{i-1} \vec H_{i,j} \vec x^{l+1}_j - \sum_{j=i+1}^3 \vec H_{i,j} \vec x^{l}_j, \qquad \text{for $i=1,2,3$}.
\]

To solve the second linear system, we need the solution of the first one, and to solve the third linear system we need the solution of the first and second ones. Each linear system is solved using the strategy presented in Section \ref{sec:linearSolve}. That is, we use a GMRES linear solver preconditioned with the restricted additive Schwarz in which the local problems are approximated with two iterations of the SSOR. Note that we only need to store the diagonal blocks since the other blocks of the matrix are only used to perform a matrix-vector product that we perform in a matrix-free manner. Therefore, the memory footprint is reduced by a factor of three.

The pre-conditioner that we propose for the matrix-free GMRES is one iteration of the block-SOR iterative method. Therefore, at each iteration of the GMRES method, we solve three linear systems with three times less unknowns. To solve these problems, we use the same relative tolerance as the original linear problem.

\subsection{Linear solver tolerance, forcing term}
\label{sec:forceTerm}


During the optimization process, it is not necessary to solve all the linear systems with the same tolerance. In the first non-linear iterations of the curving process, it is possible to select a loose tolerance to solve the linear system since we are \emph{far} to the optimal solution. In the last iterations of the process, we want to obtain the quadratic convergence of Newton's method and therefore, we need to use a tight tolerance in order to accurately solve the linear system. Using appropriate tolerances, we can avoid unnecessary iterations of the linear solver and therefore, increase the computational efficiency of the code.

The main idea is to use the error of the boundary condition as an indicator of the evolution of the penalty method. The main reason is that the residual of the non-linear problem at each iteration of the penalty method is dominated by the error of the constraint. When a non-linear iteration of the penalty method is completed, we have a converged residual up to a given tolerance. Then, in the next iteration of the penalty method, we increase the penalty parameter and the residual is dominated by the error of the constraint. Thus, the error of the constraint is a good indicator of the evolution of the curving process.

Let $\varepsilon_k = \norm[\partial \mesh{I}]{\trace \mapB{}_k - \eval{\vec g_D}{\trace \mapB{}_k}} / \norm[\partial \mesh{I}]{1}$ be the norm of the constraint at the $k$-th iteration of the penalty method. At the $k$-th iteration, if the penalty parameter has been increased by a factor $m_k$, we can assume that
\[
	\varepsilon_{k+1} \approx \varepsilon_k / m_k.
\]
That is, when we solve the non-linear problem associated to the $k$-th iteration of the penalty parameter, we can expect a decrease of the constraint norm by a factor of $m_k$. To quantify the progress on approximating the boundary condition, we define a progress variable as:
\[
	t_k = \frac
	{\eval{\log}{\frac{\varepsilon_0}{\varepsilon_k / m_k}}}
	{\eval{\log}{\frac{\varepsilon_0}{\varepsilon^*}}},
\]
where, $\varepsilon^*$ is the prescribed tolerance for the constraint error. The variable $t_k$ estimates the progress on approximating the boundary condition at the next iteration of the penalty method in logarithmic scale. That is, $t_k$ tracks how close is the boundary condition error to the prescribed tolerance in terms of orders of magnitude.

If the norm of the constraint is estimated to be near the initial constraint norm, then $t_k$ is near zero. On the contrary, when the constraint norm is estimated to be near the prescribed tolerance, $\varepsilon^*$, $t_k$ is near one. Using the progress indicator $t_k$, we define the relative tolerance to solve the linear systems of the $k$-th iteration of the penalty parameter as
\begin{equation}
	\delta = {\delta_{\text{max}}}^{1-t_k} \cdot {\delta_{\text{min}}}^{t_k},
	\label{eqn:forcingTerm}
\end{equation}
where $\delta_{\text{max}}$ and $\delta_{\text{min}}$ are two parameters that control the maximum and minimum allowed tolerances relative to the initial residual of the linear system. In our case, we use $\delta_{\text{max}} = 10^{-3}$, and $\delta_{\text{min}} = 10^{-8}$.

\begin{algorithm*}[t!]
	\caption{Calculation of the forcing term}
	\label{alg:forcingTerm}
	\begin{algorithmic}[1]
		\Function{computeForcingTerm}{
			\NewParameterLine Real $\varepsilon_k$, Real $\varepsilon_0$, Real $\varepsilon^*$,
			\NewParameterLine Real $m_k$,
			\NewParameterLine Real $\delta_{\text{max}}$, Real $\delta_{\text{min}}$}
			\State $t_k \gets \frac
			{\eval{\log}{\frac{\varepsilon_0}{\varepsilon_k / m_k}}}
			{\eval{\log}{\frac{\varepsilon_0}{\varepsilon^*}}}$
			
			\State $\delta_k \gets {\delta_{\text{max}}}^{1-t_k} \cdot {\delta_{\text{min}}}^{t_k}$
			\State \Return $\delta_k$
		\EndFunction
	\end{algorithmic}
\end{algorithm*}

\subsection{Algorithm Description}
\label{sec:algorithm}

Algorithm \ref{alg:penaltyMethod} describes the proposed penalty method with $p$-continuation technique for mesh curving. The input of the algorithm is an initial linear mesh \mesh[]{I}, a virtual topology model pointing to a CAD model to compute the boundary condition, the final polynomial degree, $p_{\text{max}}$, and the required tolerances for the boundary condition and the non-linear problem, $\varepsilon^*$ and $\omega^*$, respectively. In Lines \ref{lin:initStart}--\ref{lin:initEnd}, we set the initial polynomial degree, and we initialize the quadratic mesh, $\mapB{}^2$, to the identity mapping. That is, we introduce the additional nodes on the edges of the mesh, and we keep the straight-edged elements. The identity mapping is optimal with respect of the mesh quality however, it does not satisfy the boundary condition. The initial penalty parameter, $\mu$, and the increase factor, $m$, are set to $10$. In Line \ref{lin:pLoop} we start the $p$-continuation loop, in which we iterate through all the polynomial degrees. Then, in Line \ref{lin:penaltyLoop}, we start the penalty method for the given polynomial degree. In Lines \ref{lin:fixPointStart} and \ref{lin:fixPointEnd} we perform the fixed-point iteration. First, we update the boundary condition and then, we optimize the functional in Equation \ref{eqn:unconstrained} to compute the new approximation of the optimal mesh. In Line \ref{lin:convergence} we check the convergence of the current polynomial degree. If the convergence check passes and we are optimizing the last polynomial degree, the algorithm ends, Line \ref{lin:finished}. If the convergence check passes and we are not in the last polynomial degree, we compute the new penalty parameter, Line \ref{lin:computeMu}, according to Equation \eqref{eqn:nextMu}, and we perform the optimization of the next polynomial degree. If the convergence check fails and we are optimizing the last polynomial degree, we adapt the increase factor of the penalty parameter, Line \ref{lin:mudAdaption}. Finally, in Line \ref{lin:muIncrease}, we increase the penalty parameter for the next iteration of the penalty method.

\begin{algorithm*}[t!]
	\caption{$p$-continuation penalty method for mesh curving.}
	\label{alg:penaltyMethod}
	\begin{algorithmic}[1]
		\Function{meshOptimization}{
			\NewParameterLine Mesh \mesh{I}, CAD $\Omega$, Int $p_{\text{max}}$, Real $\varepsilon^*$, Real $\omega^*$}
			\FirstLeftComment{Variable initialization}
			\State $p \gets 2$ \label{lin:initStart}
			\State $\mapB{}^p \gets \mat{Id}$
			\State $\mu \gets 10$, $m \gets 10$, $\alpha \gets 2$
			\State $\varepsilon_0 = \norm[\partial \mesh{I}]{\trace \mapB{}^p - \eval{\vec g_D}{\trace \mapB{}^p}} / \norm[\partial \mesh{I}]{1}$ \label{lin:initEnd} 
			\LeftComment{$p$-continuation loop} \label{lin:pLoop}
			\For{$p \in \{2,\ldots,p_{\text{max}}\}$}
				\State converged $\gets$ \textbf{false}
				\LeftComment{Penalty method loop}
				\While{\textbf{not} converged} \label{lin:penaltyLoop}
					\State $\delta \gets \Call{computeForcingTerm}{\mapB{}^p,  \varepsilon_0, \varepsilon^*, m}$
					\LeftComment{Fixed-point iteration}
					\State $\vec g_D^p \gets  \eval{\vec g_D}{\trace \mapB{}^p_{k-1}}$ \label{lin:fixPointStart}
					\State $\mapB{}^p_{k} \gets \Call{optimize}{\eval{\Functional{\mu}}{\mapB{}^p_{k-1},\vec g_D^p},\delta}$ \label{lin:fixPointEnd}
					
					\LeftComment{Check convergence}
					\State converged $\gets \Call{degreeConverged}{\mapB{}^p_{k}, p, p_{\text{max}}, \varepsilon^*, \omega^*}$ \label{lin:convergence}
					
					\If{converged}
						\If{$p = p_{\text{max}}$}
							\FirstLeftComment{Algorithm has finished}
							\State \Return $\mapB{}^{p}_{k}$ \label{lin:finished}
						\Else
							\FirstLeftComment{Interpolate solution and compute next penalty parameter}
							\State $\mapB{}^{p+1}_k \gets \Call{interpolate}{\mapB{}^p_{k}, p+1}$
							\State $\mu_{k} \gets \Call{firstIterationPenaltyParameter}{\mapB{}^p_{k}, \mapB{}^{p+1}_k, \mu_k}$ \label{lin:computeMu}
						\EndIf
					\Else
						\If{$p = p_{\text{max}}$}
							\FirstLeftComment{In last polynomial degree, adapt the penalty parameter}
							\State $m \gets \Call{penaltyParameterAdaption}{\mapB{}^p_{k}, \mapB{}^p_{k-1}, \mu_{k}, \mu_{k-1}}$ \label{lin:mudAdaption}
						\EndIf
						\State $\mu_{k+1} \gets m \mu_k$ \label{lin:muIncrease}
					\EndIf
				\EndWhile
			\EndFor
		\EndFunction
	\end{algorithmic}
\end{algorithm*}

\section{Examples}
\label{sec:examples}

\newcommand{\spheresWidth}{0.375\textwidth}

We present several examples that show the capabilities of the proposed high-order mesh curving method. To generate the initial linear meshes, we have used Pointwise \cite{pointwise}. The mesh curving solver has been implemented in Python \cite{python} using the FEniCS \cite{alnaes:fenics} and the petsc4py \cite{petsc4py} libraries. To project the boundary high-order nodes we have used both the geode \cite{geode} and the Open CASCADE \cite{opencascade} libraries interfaced with an in-house python wrapper developed using swig \cite{swig}.

All the meshes have converged with a residual tolerance of $10^{-8}$ in the infinity norm. That is, all the components of the residual vector in absolute value are less or equal than $10^{-8}$. Thus, the obtained meshes in these examples are curved with a tight residual tolerance and therefore, are fully converged. \newtext{Recall that to converge the curving process, we also impose a constraint norm less than $10^{-12} \ell_c$, where $\ell_c$ is a characteristic length of the model. Although at the end of the curving process the nodes do not lie exactly on the CAD surfaces, they are at a distance twelve orders of magnitude smaller than the chosen characteristic length.}

\newtext{Our formulation ensures the mesh validity at the integration points during the curving process. This validity is so since our functional goes to infinity when the mesh becomes invalid. Nevertheless, at the end of the curving process, we check that the Jacobian determinant is positive at the integration points to validate the results of our implementation.}

The optimization process has been performed in the MareNostrum4 super-computer located at the Barcelona Supercomputing Center. It is composed of 3456 nodes, connected using an Intel Omni-Path network. Each node contains two Intel Xeon Platinum 8160 CPU with 24 cores, each at 2.10 GHz, and 96 GB of RAM memory. To obtain the total energy consumed by all tasks of the optimization job, we have used the \texttt{sacct} command of the SLURM Workload Manager.

The mesh visualization has been performed using Paraview 5.5.2 \cite{ahrens2005:paraview} in parallel in the MareNostrum4 super-computer. We have used the high-order mesh visualization implementation of Paraview that subdivides each element in a given number of sub-elements. Note that the mesh partition to perform the visualization does not need to coincide with the the one used in optimization. In general, for visualization purposes, less cores are needed since no global matrices are assembled and no linear systems are solved.

We compute the elemental quality relative to the initial mesh as in \cite{gargallo2015:tetOptimization}
\[
q_{\element[P]} = \frac{1}{\eta_{\element[P]}},
\quad \text{where} \quad
\eta_{\element[P]} = \left(
\frac
{\displaystyle \int_{\element_{I}} (M\mapB[]{})^2\ \text{d}\domain{}}
{\displaystyle \int_{\element_{I}} 1 \ \text{d}\domain{}}
\right)^{1/2}.
\]
The relative element quality takes values in the interval $[0,1]$. An ideal element has quality equal to one, and an inverted or tangled element has a quality of zero. In all the examples, we color the elements according to  $1-q_{\element[P]}$ in logarithmic scale to check how close is the element quality to one. Thus, lower values denote higher quality elements.

\subsection{Influence of the solver improvements: external uniform mesh of a sphere}


\begin{figure*}[t!]
	\centering
	\begin{subfigure}[b]{0.66\textwidth}
		\includegraphics[width=\textwidth]{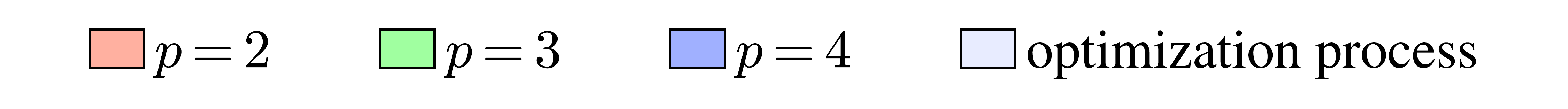}
	\end{subfigure}
	\\
	\hfill
	\begin{subfigure}[b]{0.33\textwidth}
		\includegraphics[width=\textwidth]{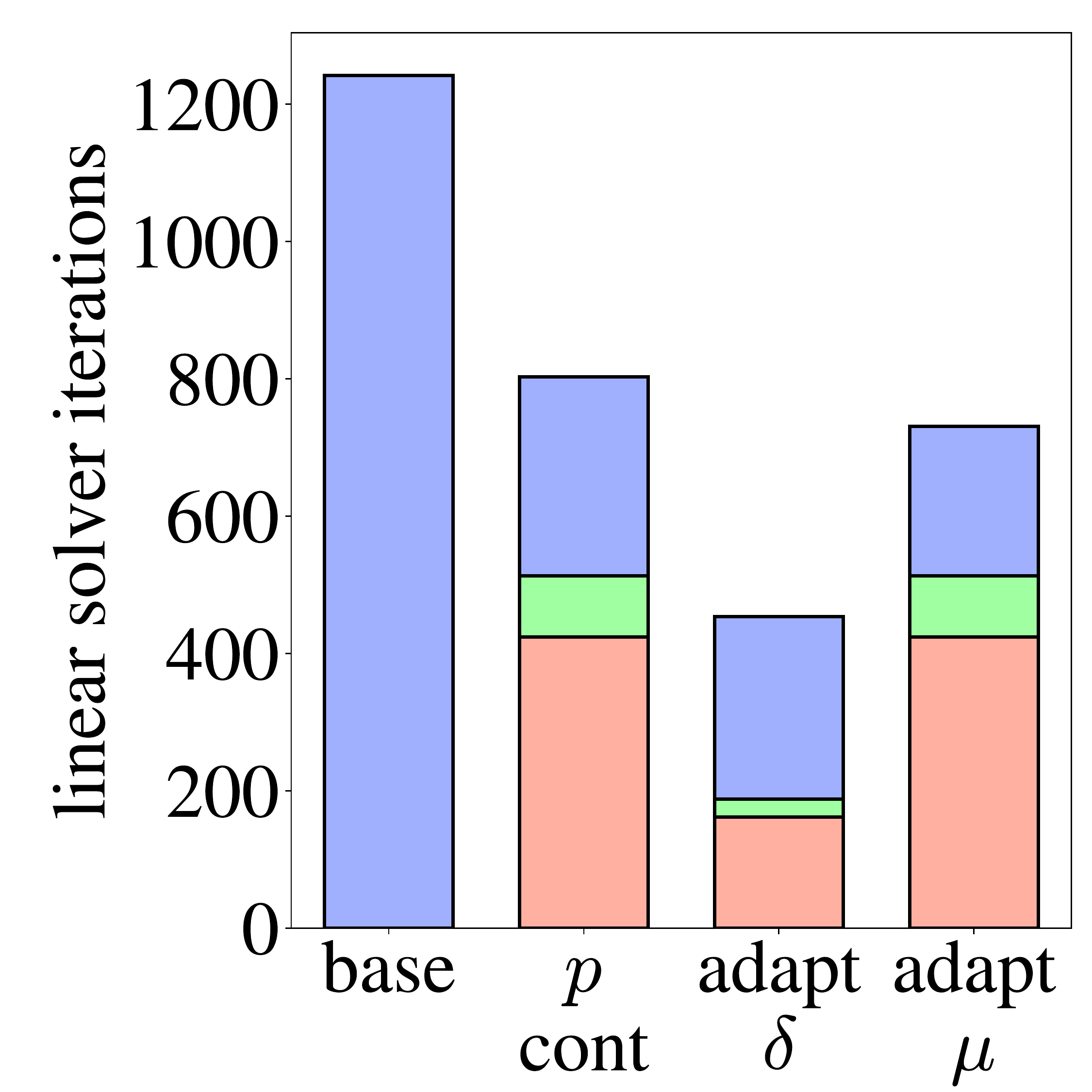}
		\caption{}
		\label{fig:comparison_NMF_Time}
	\end{subfigure}
	\hspace{0.05\textwidth}
	\begin{subfigure}[b]{0.22\textwidth}
		\includegraphics[width=\textwidth]{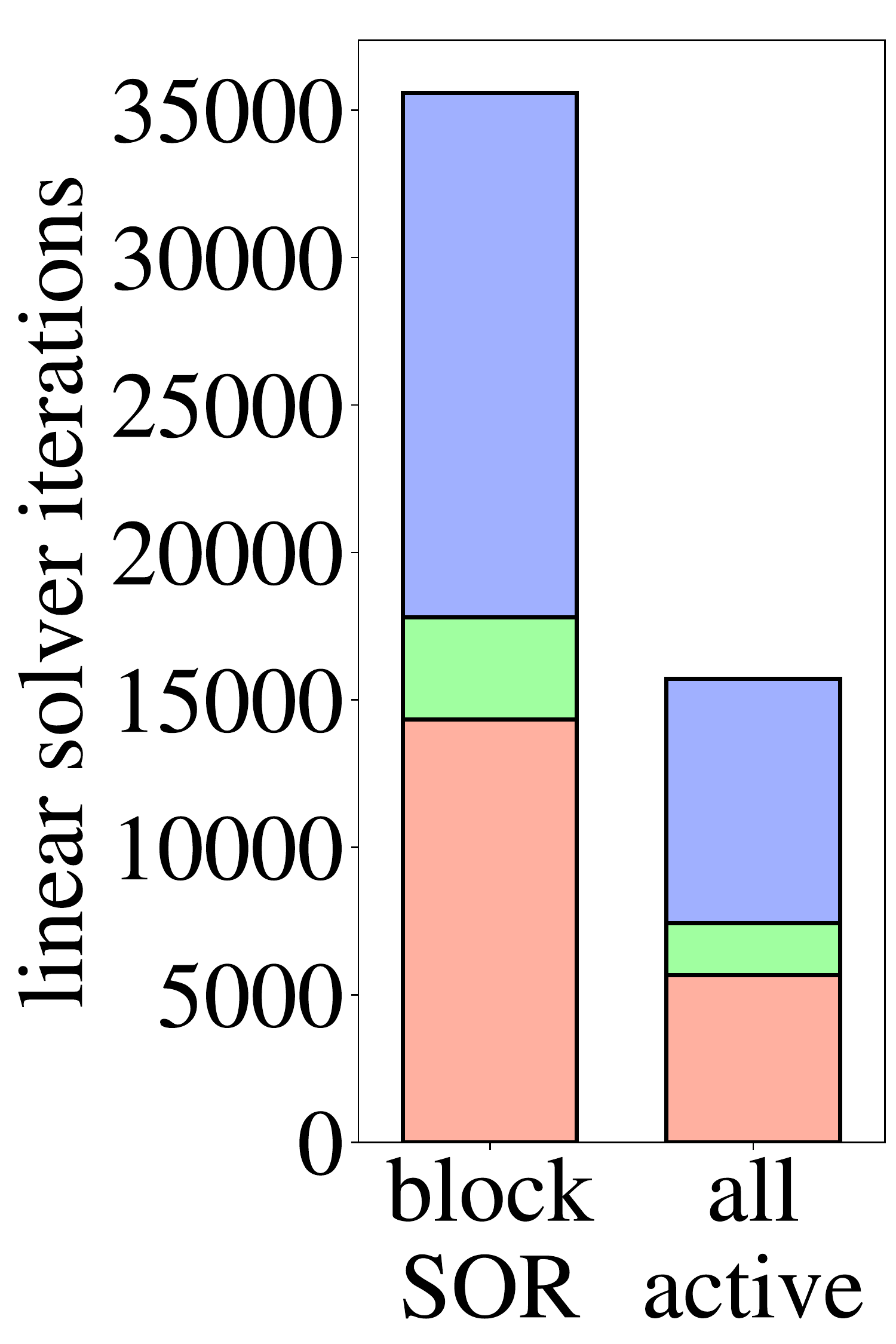}
		\caption{}
		\label{fig:comparison_MF_Iterations}
	\end{subfigure}
	\hfill\hspace{0cm}
	\\
	\hfill
	\begin{subfigure}[b]{0.44\textwidth}
		\includegraphics[width=\textwidth]{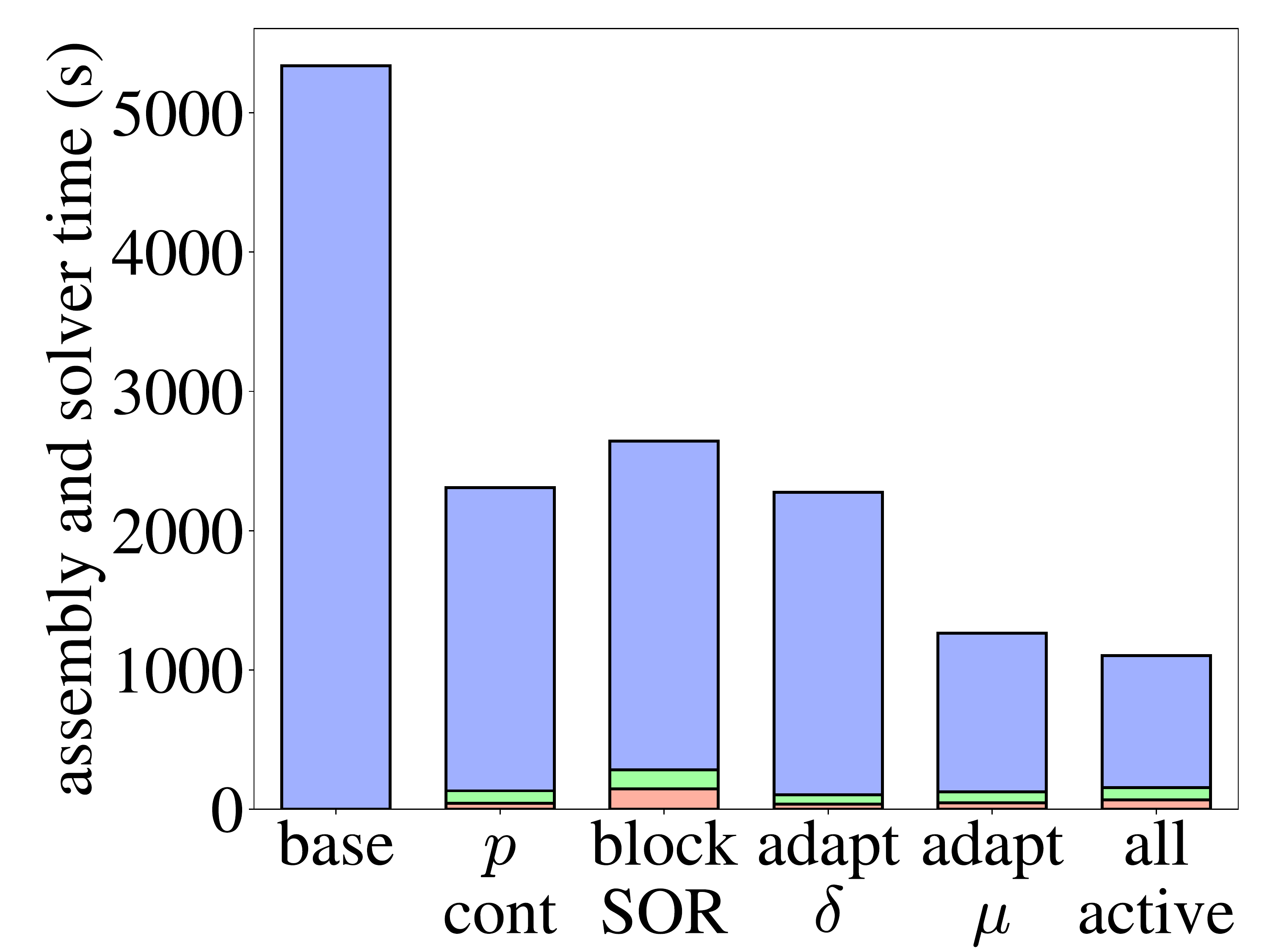}
		\caption{}
		\label{fig:comparison_TimeScaling}
	\end{subfigure}
	\hfill
	\begin{subfigure}[b]{0.44\textwidth}
		\includegraphics[width=\textwidth]{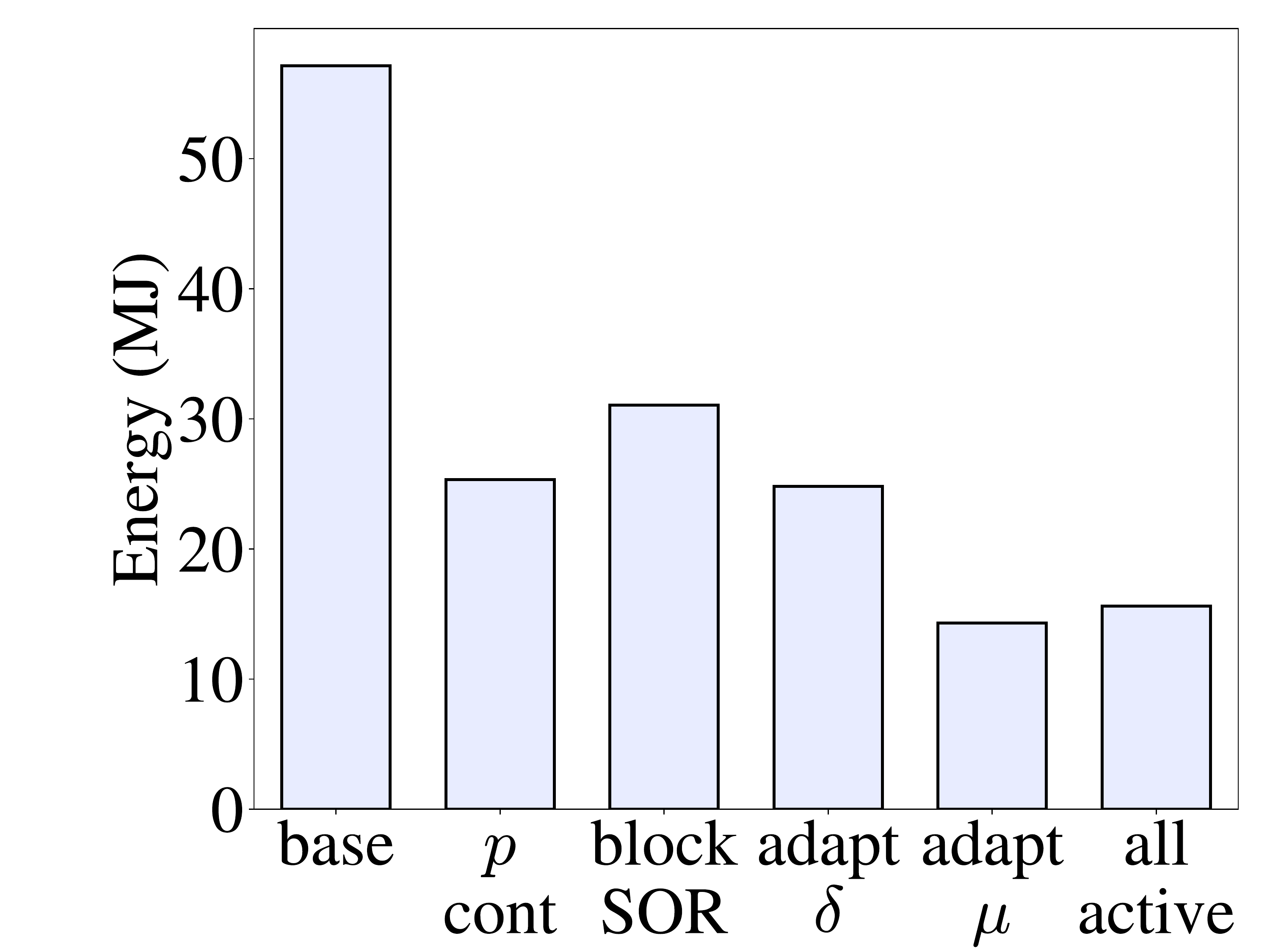}
		\caption{}
		\label{fig:comparison_Energy}
	\end{subfigure}
	\hfill\hspace{0cm}
	\caption{Analysis of the proposed improvements:
	(a) and (b) number of linear iterations;
	(c) assembly and solver time; and
	(d) energy consumption.}
	\label{fig:comparisonScaling}
\end{figure*}

To illustrate the influcence of each solver improvement, we consider a curved high-order mesh of polynomial degree four composed of 1.44 million isotropic elements of constant size. The domain is a sphere of radius four with a spherical hole of radius one.

To quantify the influence of each improvement in the optimization, in Figure \ref{fig:comparisonScaling} we analyze the time to assemble and solve the linear systems, the number of iterations of the iterative linear solver, and the energy consumption of generating the curved mesh. We use the red, green and blue colors to denote the quantities of interest for the meshes of polynomial degree two, three and four when using the $p$-continuation technique. In all the cases, we have used 768 processors to perform the optimization process and we have obtained the same curved mesh up to errors smaller than the non-linear solver tolerance. Although the convergence criterion in the residual norm is $10^{-8}$, the obtained meshes have converged with residual norms in the range of $4.63 \cdot 10^{-10}$ and $1.68 \cdot 10^{-9}$.

The first case consists on generating the mesh without any of the proposed improvements nor the $p$-continuation technique proposed in \cite{ruiz2019:pContinuation}. In Figure \ref{fig:comparisonScaling}, this case is labeled ``base''. In all the other cases, we use the $p$-continuation technique. In the second case, we only use the $p$-continuation technique (labeled ``$p$-cont''). In the third case, we use the proposed matrix-free linear solver (labeled ``block SOR''). In the fourth case, we use the proposed technique to adapt the tolerance of the linear solvers (labeled ``adapt $\delta$''). In the fifth case, we use the proposed technique to adapt the penalty parameter (labeled ``adapt $\mu$''). Finally, in the sixth case, we activate all the proposed improvements (labeled ``all active'').

The $p$-continuation technique reduces the wall-clock time, the number of linear iterations, and the energy consumption. Specifically, the time to assemble and solve the linear systems, and the energy consumption is reduced by a factor greater than two. For this reason, we will compare the improvements proposed in this paper against using only the $p$-continuation technique.

When using the block-SOR pre-conditioner, the memory footprint has been reduced by a factor of three. On top of that, the time to assemble and solve the linear systems, and the energy consumption are only slightly larger than using the classical linear solver.

The cost of assembling the small matrices and performing a linear iteration is less than the cost of assembling and performing a linear iteration using the full sparse matrix. This is so since the matrices involved in the pre-conditioner have three times less unknowns and nine times less non-zero entries than the system matrix. Although the number of linear iterations to apply the pre-conditioner is larger than the number of iterations to solve the linear system using the classical sparse matrix solver, each linear iteration is performed faster. Note that the number of iterations of the block-SOR pre-conditioner is obtained by adding the iterations of the three linear systems involved in the pre-conditioner.

When we apply the proposed technique to adapt the tolerance to solve the linear systems, we reduce the total number of linear iterations. However, this reduction is more apparent when curving the quadratic and cubic meshes. The main reason is that the tolerance to solve the linear systems is looser at the start of the curving process and it becomes tighter when the curving process advances. Thus, when curving the mesh of polynomial degree four, the curving process has advanced sufficiently to demand the quadratic convergence of Newton's method. Therefore, we need to solve the linear systems with a tight tolerance.

\begin{figure*}[t!]
	\centering
	\hfill
	\begin{subfigure}[b]{0.45\textwidth}
		\includegraphics[width=\textwidth]{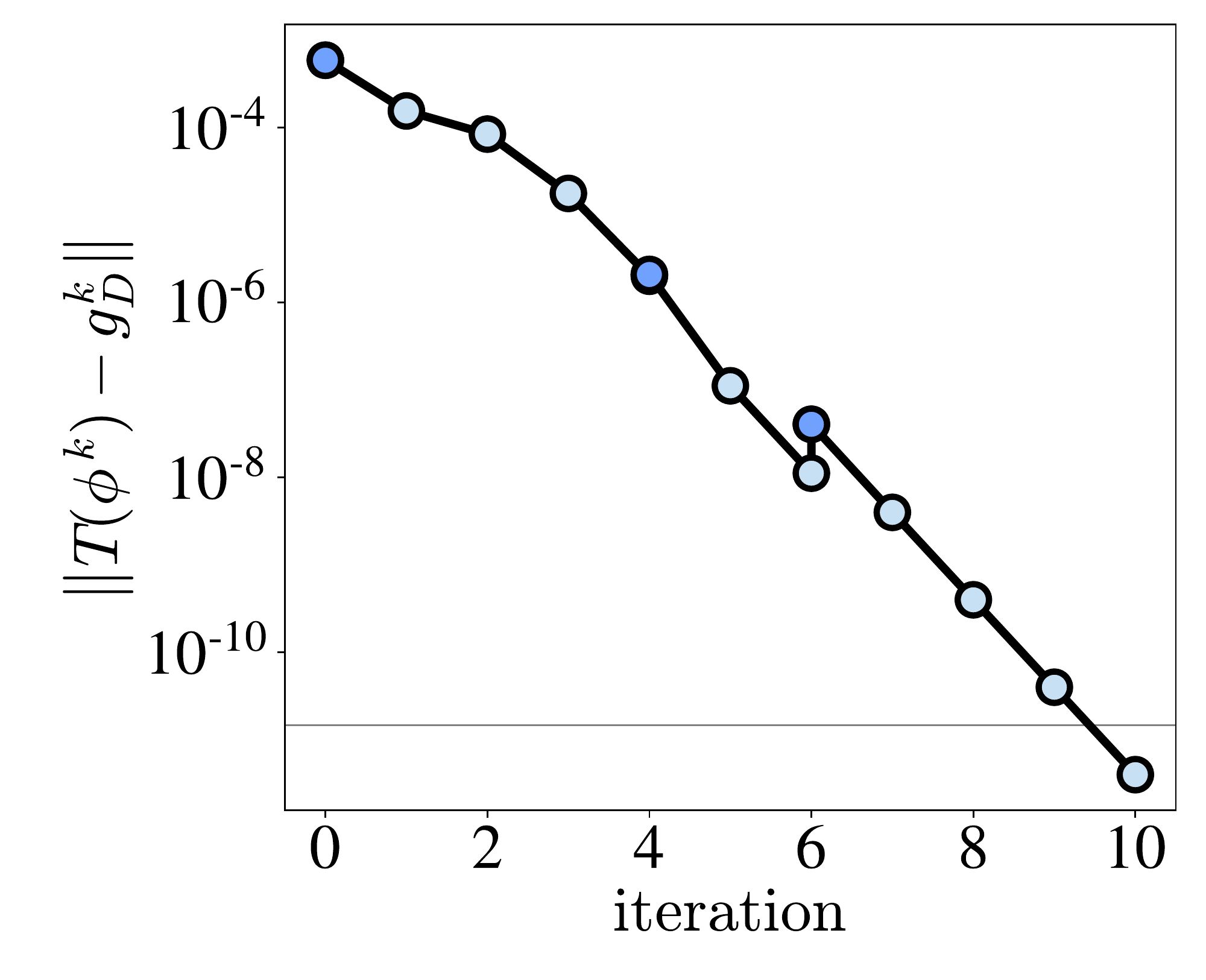}
		\caption{}
		\label{fig:constraintEvolution_NAM}
	\end{subfigure}
	\hfill
	\begin{subfigure}[b]{0.45\textwidth}
		\includegraphics[width=\textwidth]{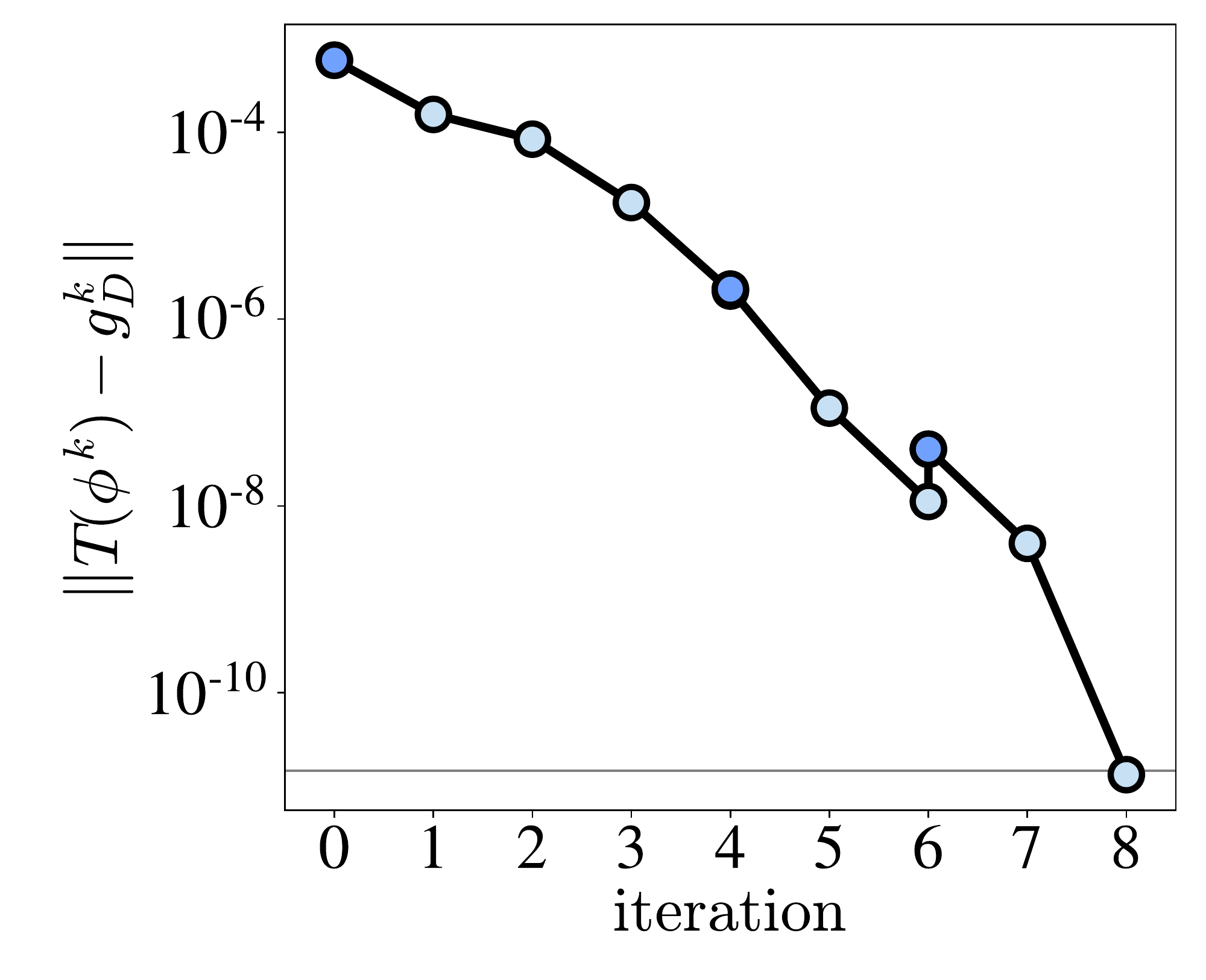}
		\caption{}
		\label{fig:constraintEvolution_AM}
	\end{subfigure}
	\hfill\hspace{0cm}
	\caption{Evolution of the norm of the constraint through the optimization process for the mesh optimized:
		(a) without $\mu$ adaptation; and
		(b) with $\mu$ adaptation.}
	\label{fig:constraintEvolution_comparison}
\end{figure*}

The adaption of the penalty parameter reduces the number of non-linear problems to solve and therefore, reduces the time to assemble and solve the linear systems, the linear iterations, and the energy consumption. Specifically, the number of non-linear problems solved during the curving process of the polynomial degree four is reduced in half. This leads to a reduction of the time to assemble and solve the linear systems, and the energy consumption almost by a factor of two.

To show the effect of the penalty parameter adaption, we plot in Figure \ref{fig:constraintEvolution_comparison} the evolution of constraint norm over the iterations of the penalty method. Specifically, Figure \ref{fig:constraintEvolution_NAM} plots the evolution of the constraint norm when the penalty parameter is not adapted, and Figure \ref{fig:constraintEvolution_AM} plots the evolution of the constraint norm when the penalty parameter is adapted. In both figures, dark blue circles denote the starting iteration of each polynomial degree, and the thin black line denotes the target tolerance. The evolution of the constraint norm for the polynomial degrees two and three is the same for both cases, since the penalty parameter adaption is not yet activated. When we perform the curving process for the polynomial degree four and the penalty parameter is not adapted, the norm of the constraint decreases linearly in logarithm space. The main reason is that the penalty method is in the convergence region and thus, an increase of the penalty parameter by a given factor induces a decrease of the constraint norm by the same factor. For this reason, it takes four iterations of the penalty method to converge the polynomial degree four. Instead, when we adapt the penalty parameter, it takes only two iterations to converge the polynomial degree four, since the adaption of the penalty parameter is able to estimate correctly the optimal penalty parameter to converge the curving process at the second iteration.

Finally, when we activate all the improvements, we obtain a solver that reduces the memory footprint by a factor of three. Moreover, the linear solver time and energy consumption is two times smaller than for the p-continuation solver. The block-SOR solver provides the memory reduction, at the price of slightly increasing the number of linear iterations, the linear solver time, and energy consumption. The tolerance adaption helps to reduce the number of linear iterations without a significant influence in the the linear solver time and energy consumption. The adaption of the penalty parameter reduces the number of non-linear problems to solve and therefore, it leads to a significant reduction of the linear solver time and the energy consumption. The p-continuation provides significant improvements when compared to the base case, that multiply by the corresponding factor the influence of the rest of  improvements.

\begin{figure*}[t!]
	\centering
	\hfill
	\begin{subfigure}[b]{\spheresWidth}
		\includegraphics[width=\textwidth]{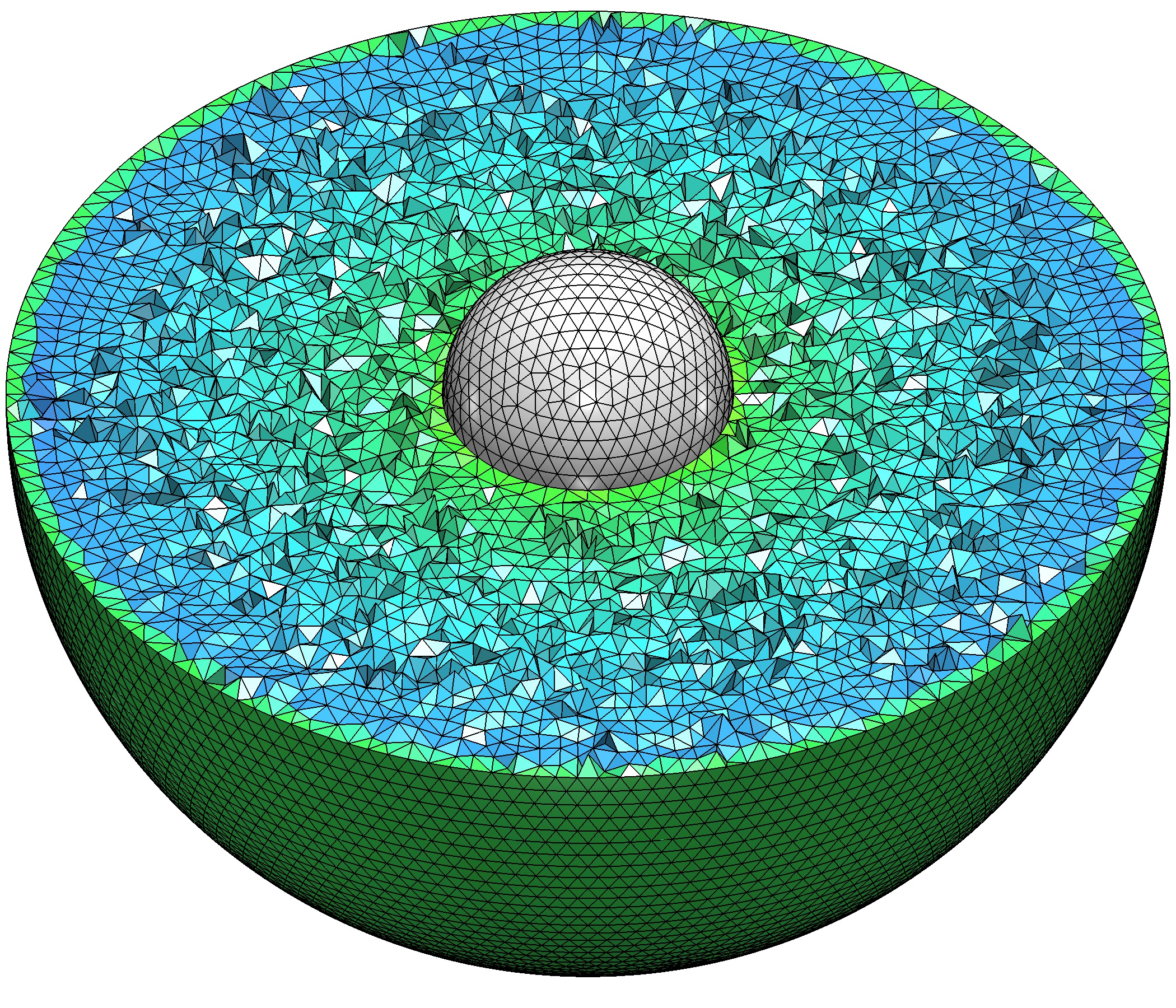}
		\caption{}
		\label{fig:hScalingLevel1}
	\end{subfigure}
	\hfill
	\begin{subfigure}[b]{\spheresWidth}
		\includegraphics[width=\textwidth]{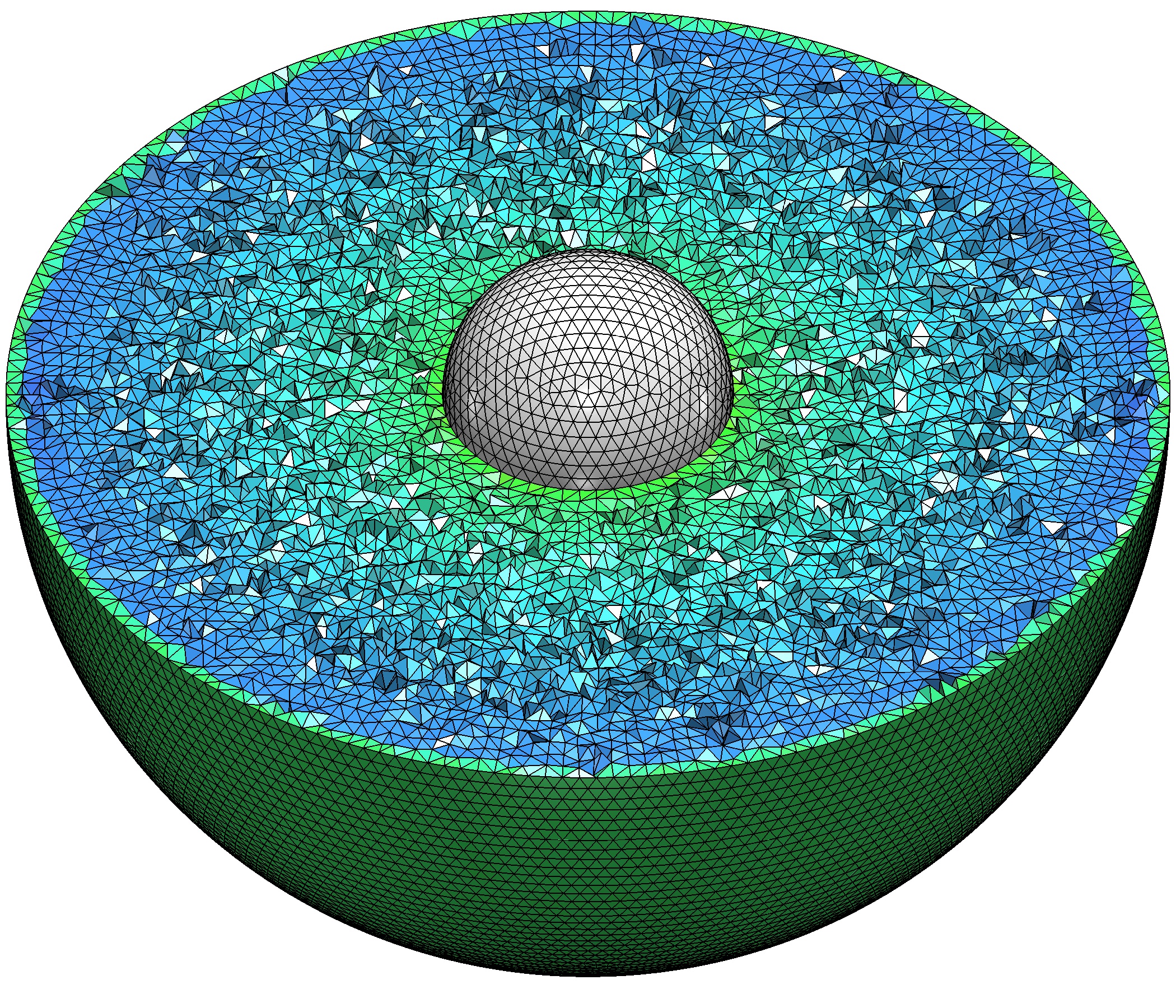}
		\caption{}
		\label{fig:hScalingLevel2}
	\end{subfigure}
	\hfill\hspace{0pt}
	\\
	\hfill
	\begin{subfigure}[b]{\spheresWidth}
		\includegraphics[width=\textwidth]{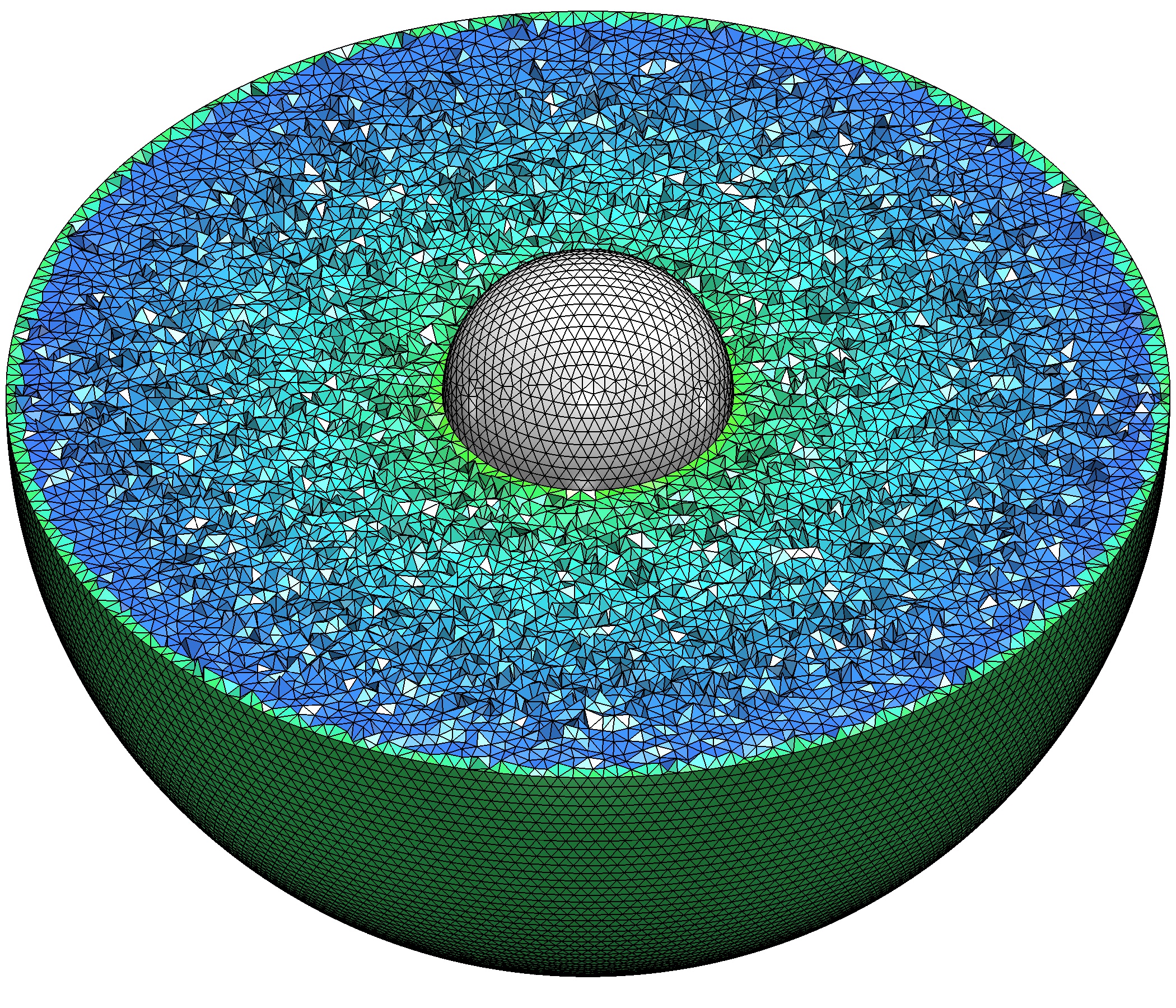}
		\caption{}
		\label{fig:hScalingLevel3}
	\end{subfigure}
	\hfill
	\begin{subfigure}[b]{\spheresWidth}
		\includegraphics[width=\textwidth]{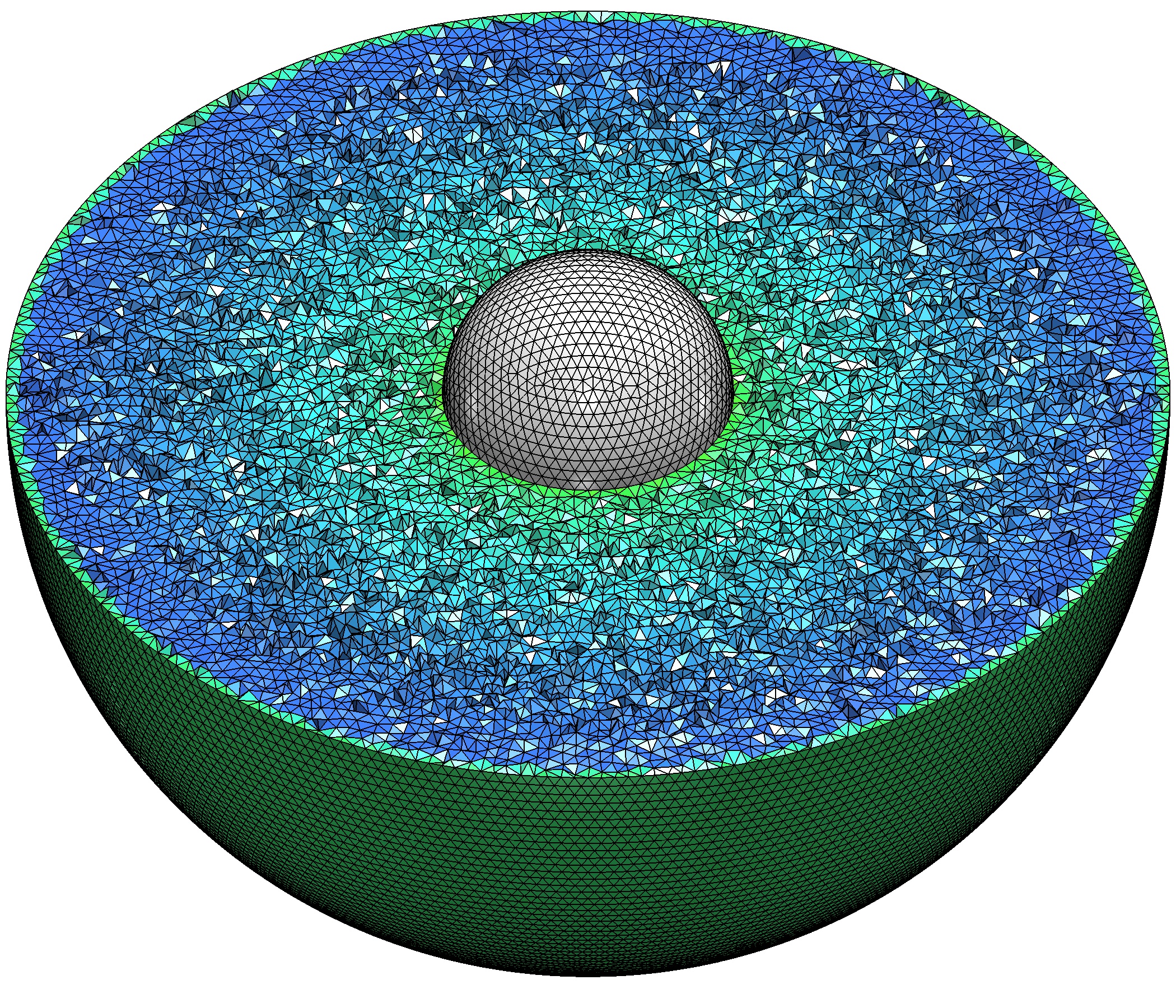}
		\caption{}
		\label{fig:hScalingLevel4}
	\end{subfigure}
	\hfill\hspace{0pt}
	\\
	\begin{subfigure}[b]{\spheresWidth}
		\includegraphics[width=\textwidth]{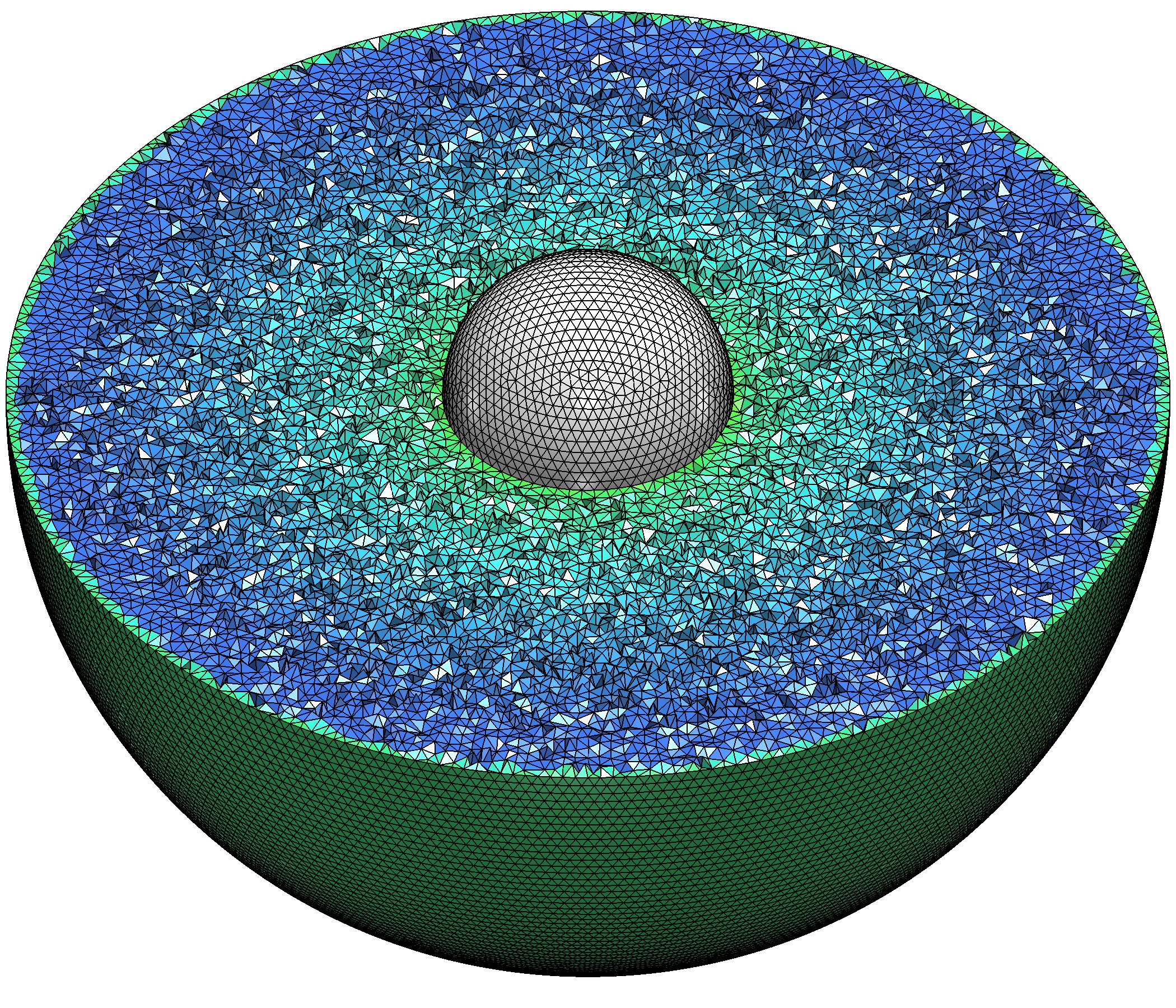}
		\caption{}
		\label{fig:hScalingLevel5}
	\end{subfigure}
	\begin{subfigure}[b]{0.75\textwidth}
		\includegraphics[width=\textwidth]{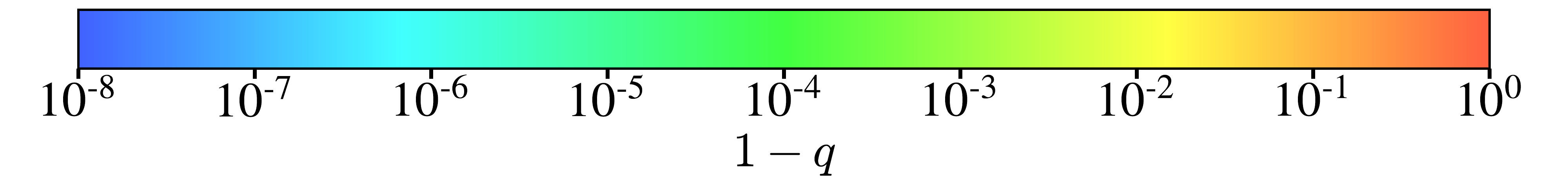}
	\end{subfigure}
	\caption{Optimized meshes of polynomial degree four using the proposed mesh curving solver, approximately composed of:
		(a) $0.72 \cdot 10^6$ elements on 480 cores;
		(b) $1.44 \cdot 10^6$ elements on 960 cores;
		(c) $2.16 \cdot 10^6$ elements on 1440 cores;
		(d) $2.88 \cdot 10^6$ elements on 1920 cores; and
		(e) $3.60 \cdot 10^6$ elements on 2400 cores.}
	\label{fig:hScalingMeshes}
\end{figure*}

\subsection{Influence of uniform mesh size}

\begin{figure*}[t!]
	\centering
	\hfill
	\begin{subfigure}[b]{0.32\textwidth}
		\includegraphics[width=\textwidth]{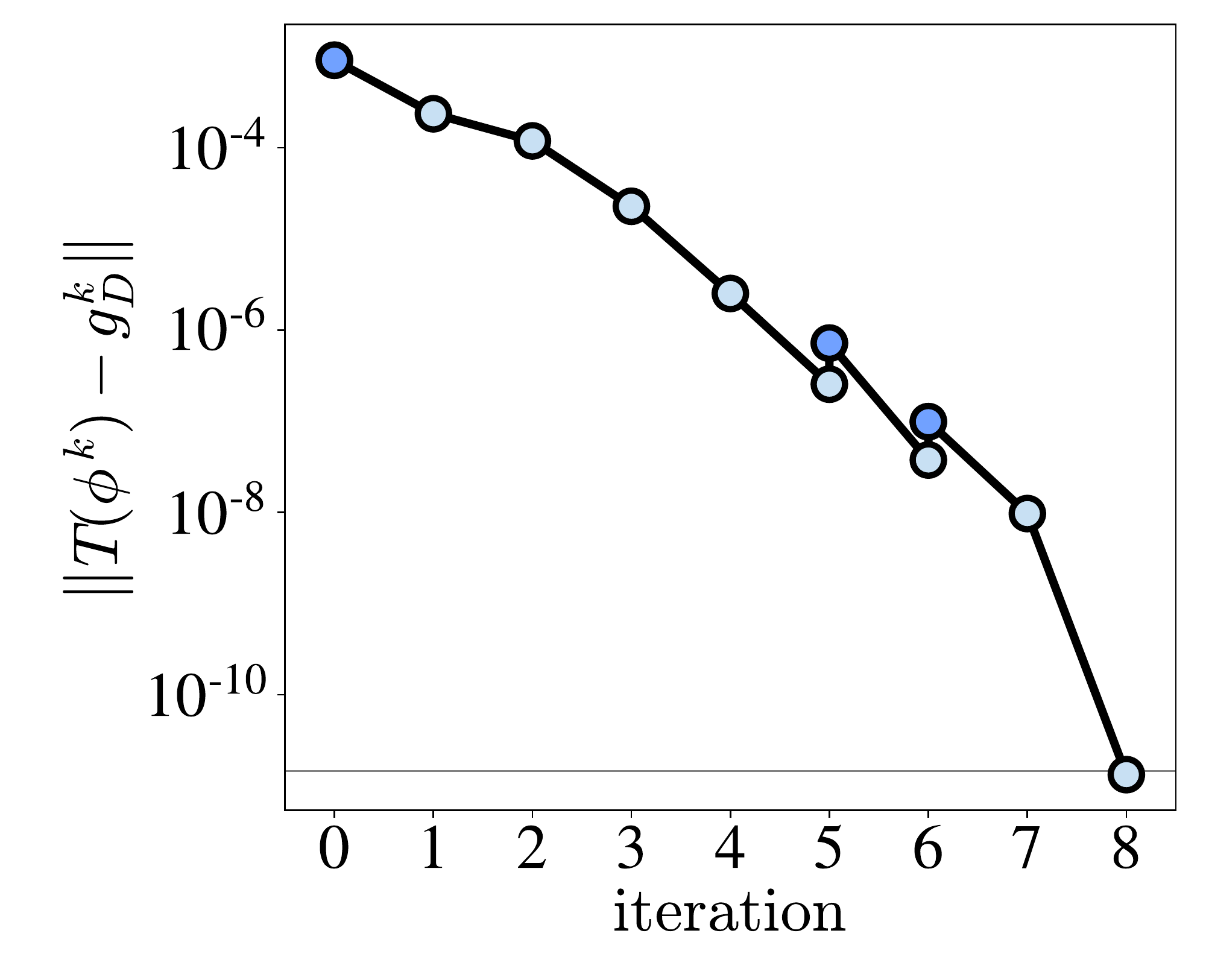}
		\caption{}
		\label{fig:spheres1_MF_AM_ATP4_Constraint}
	\end{subfigure}
	\hfill
	\begin{subfigure}[b]{0.32\textwidth}
		\includegraphics[width=\textwidth]{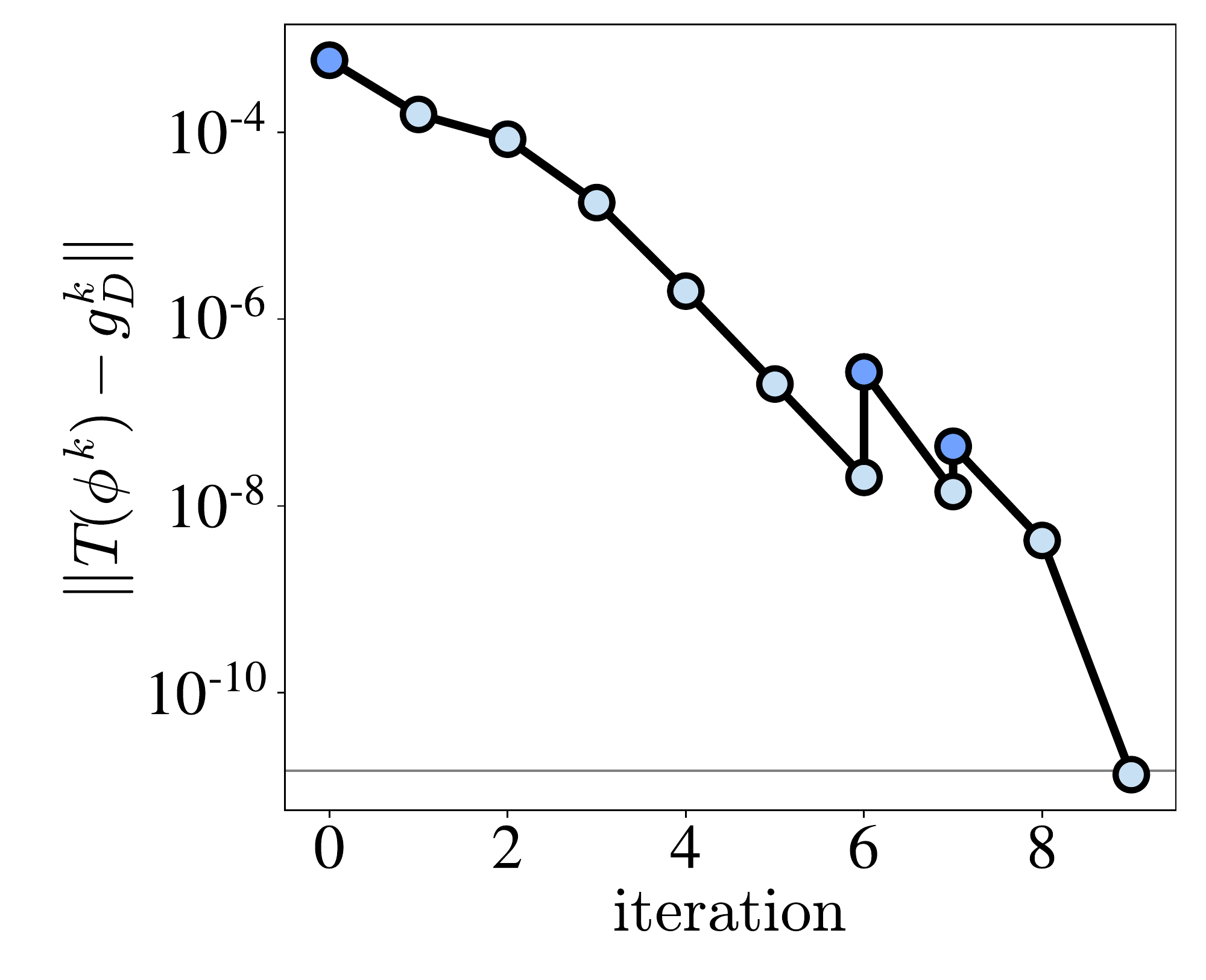}
		\caption{}
		\label{fig:spheres2_MF_AM_ATP4_Constraint}
	\end{subfigure}
	\hfill
	\begin{subfigure}[b]{0.32\textwidth}
		\includegraphics[width=\textwidth]{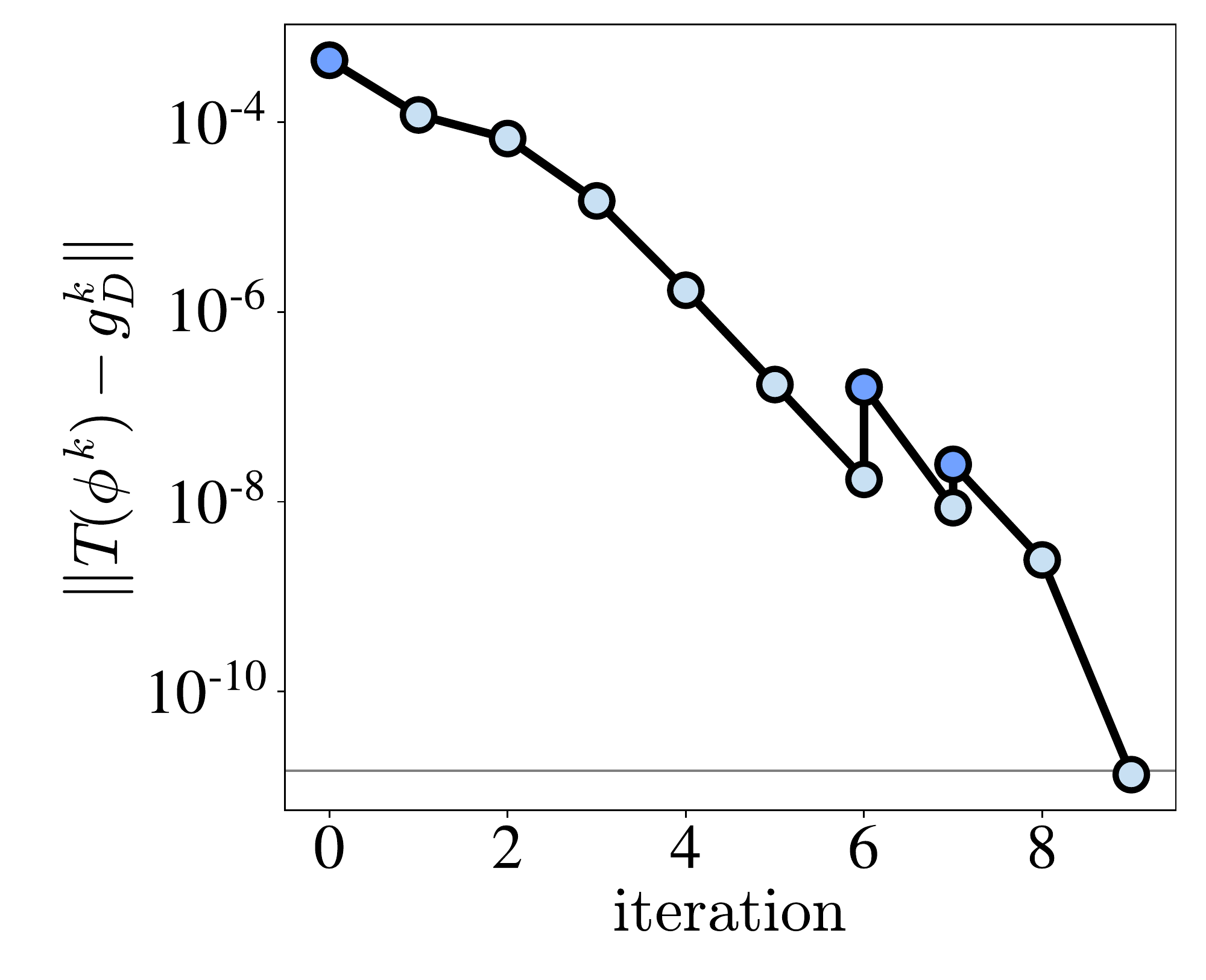}
		\caption{}
		\label{fig:spheres3_MF_AM_ATP4_Constraint}
	\end{subfigure}
	\hfill\hspace{0cm}
	\\
	\hfill
	\begin{subfigure}[b]{0.32\textwidth}
		\includegraphics[width=\textwidth]{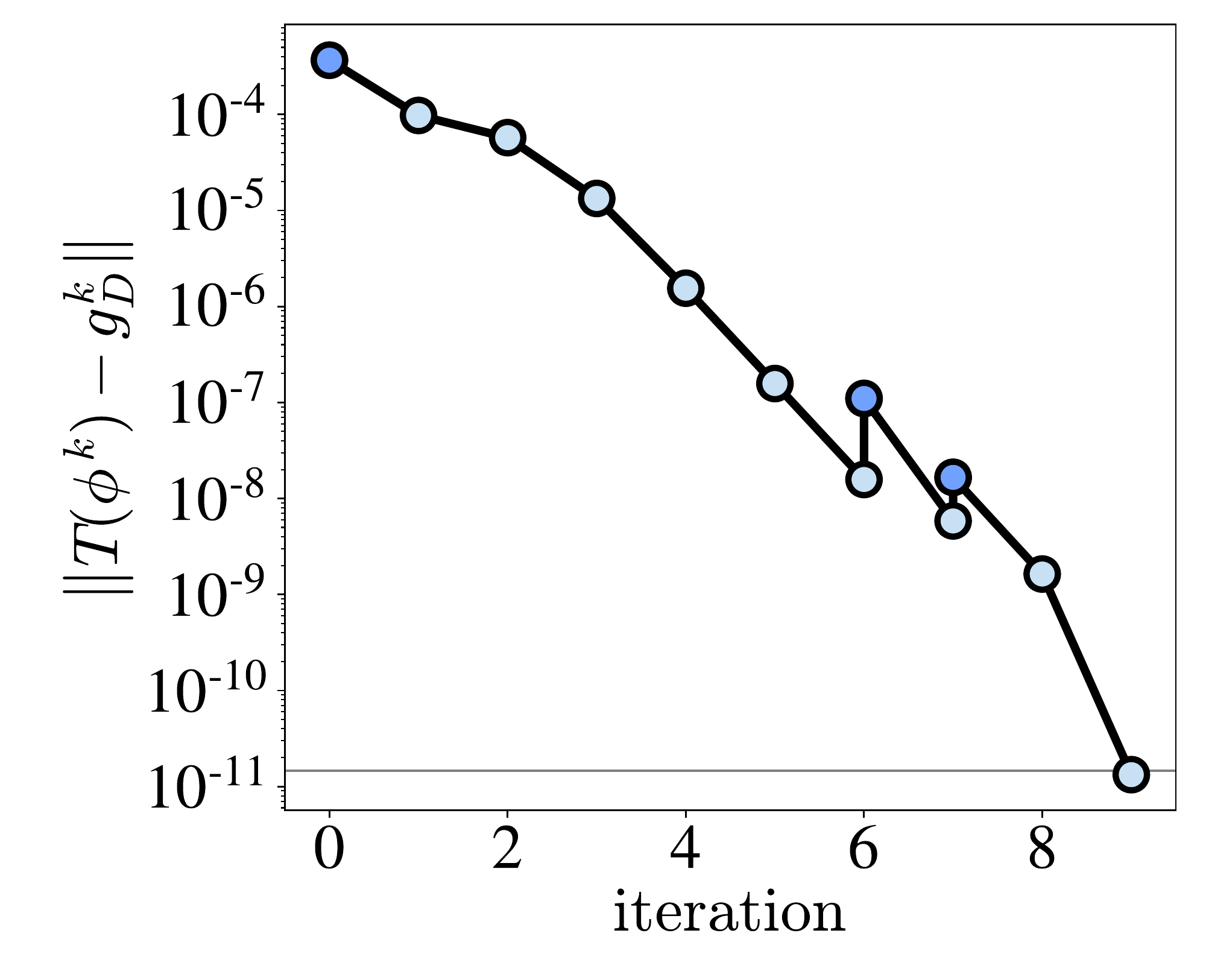}
		\caption{}
		\label{fig:spheres4_MF_AM_ATP4_Constraint}
	\end{subfigure}
	\hfill
	\begin{subfigure}[b]{0.32\textwidth}
		\includegraphics[width=\textwidth]{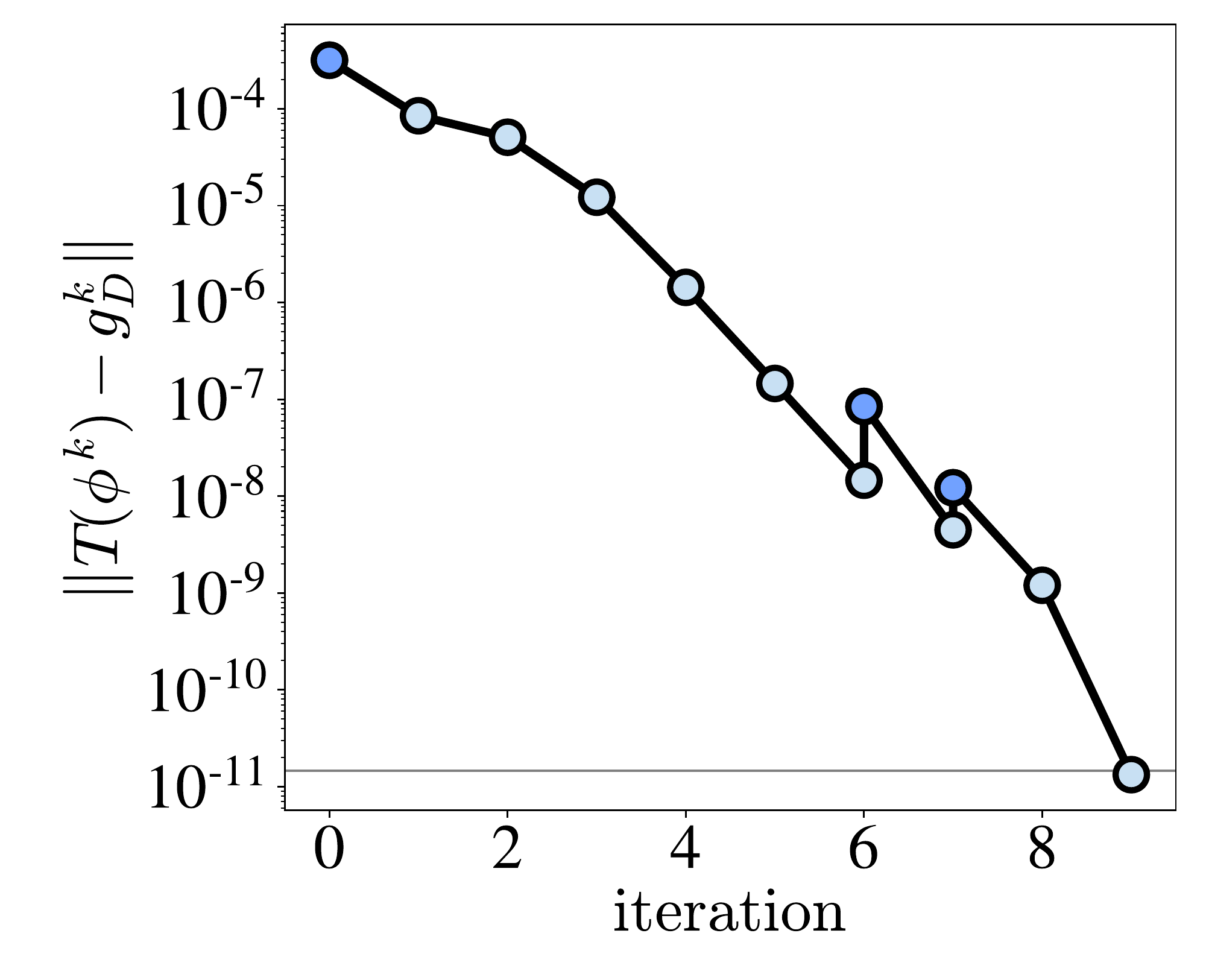}
		\caption{}
		\label{fig:spheres5_MF_AM_ATP4_Constraint}
	\end{subfigure}
	\hfill\hspace{0cm}
	\caption{Evolution of the norm of the constraint through the optimization process for the mesh approximately composed of:
		(a) $0.72 \cdot 10^6$ elements on 480 cores;
		(b) $1.44 \cdot 10^6$ elements on 960 cores;
		(c) $2.16 \cdot 10^6$ elements on 1440 cores;
		(d) $2.88 \cdot 10^6$ elements on 1920 cores; and
		(e) $3.60 \cdot 10^6$ elements on 2400 cores.}
	\label{fig:spheresConstraint}
\end{figure*}

\begin{figure*}[t!]
	\centering
	\begin{subfigure}[b]{0.66\textwidth}
		\includegraphics[width=\textwidth]{comparison_LabelsScaling}
	\end{subfigure}
	\\
	\hfill
	\begin{subfigure}[b]{0.32\textwidth}
		\includegraphics[width=\textwidth]{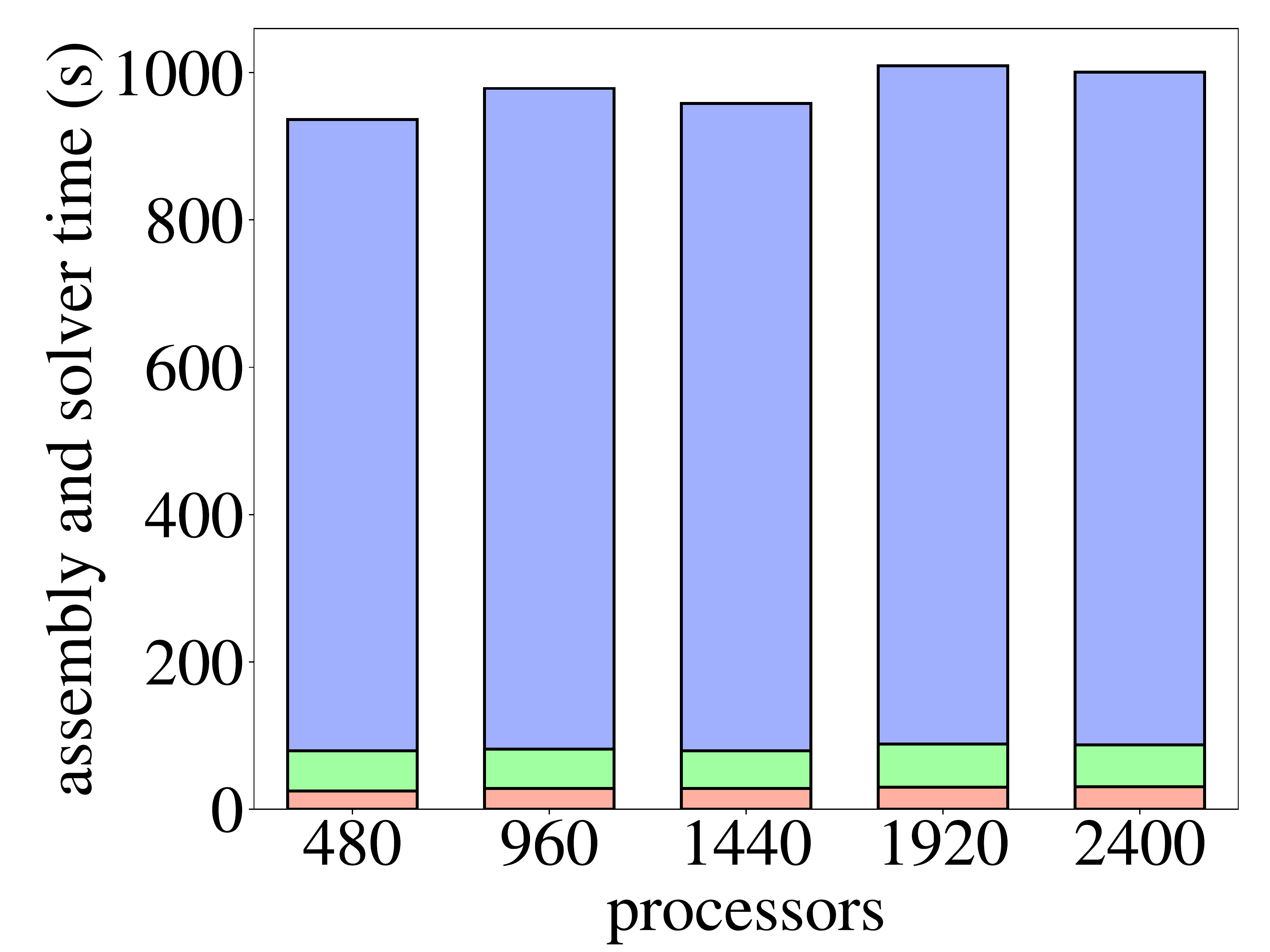}
		\caption{}
		\label{fig:scalingH_NMF_TimeScaling}
	\end{subfigure}
	\hfill
	\begin{subfigure}[b]{0.32\textwidth}
		\includegraphics[width=\textwidth]{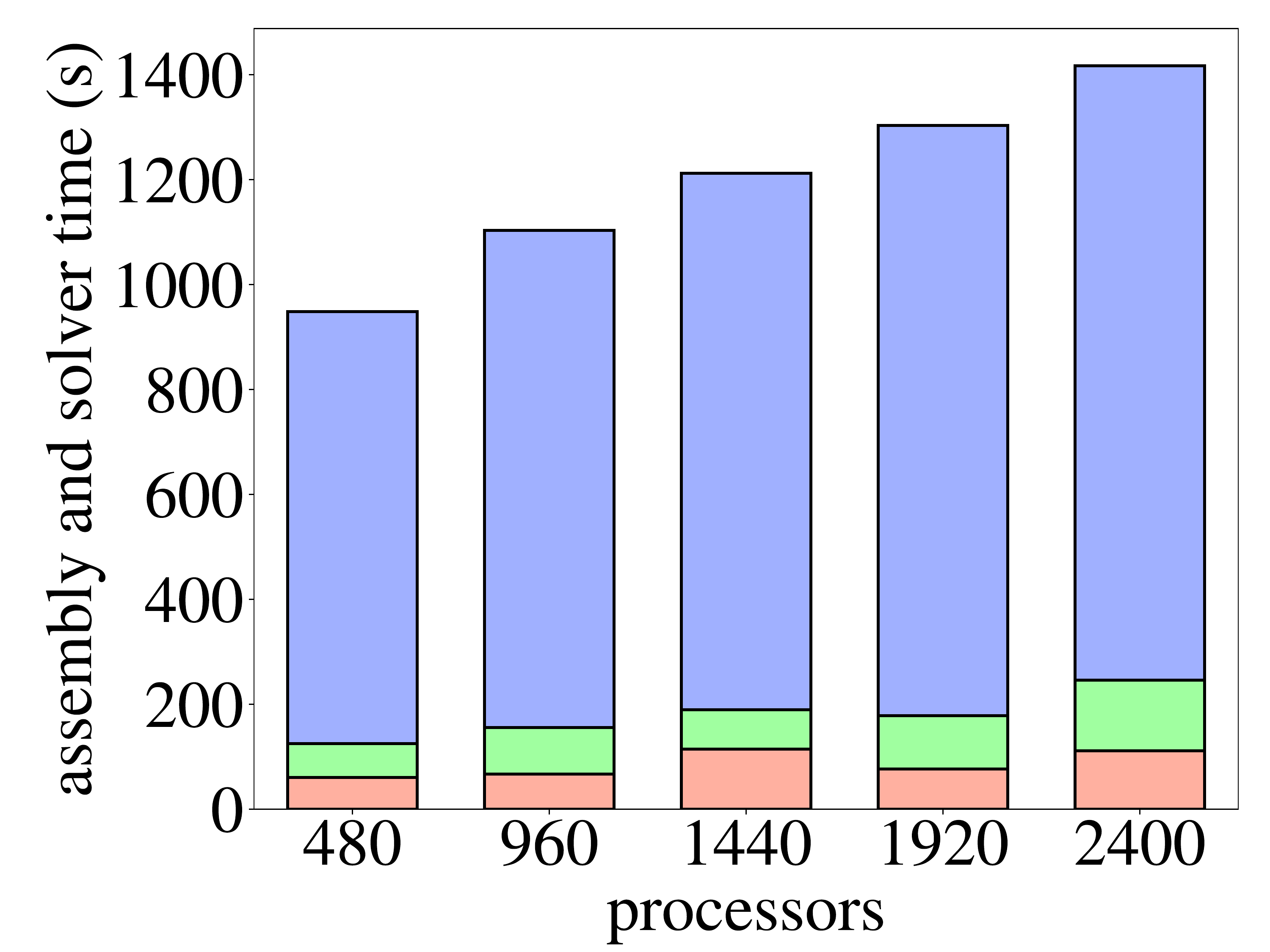}
		\caption{}
		\label{fig:scalingH_MF_TimeScaling}
	\end{subfigure}
	\hfill
	\begin{subfigure}[b]{0.32\textwidth}
		\includegraphics[width=\textwidth]{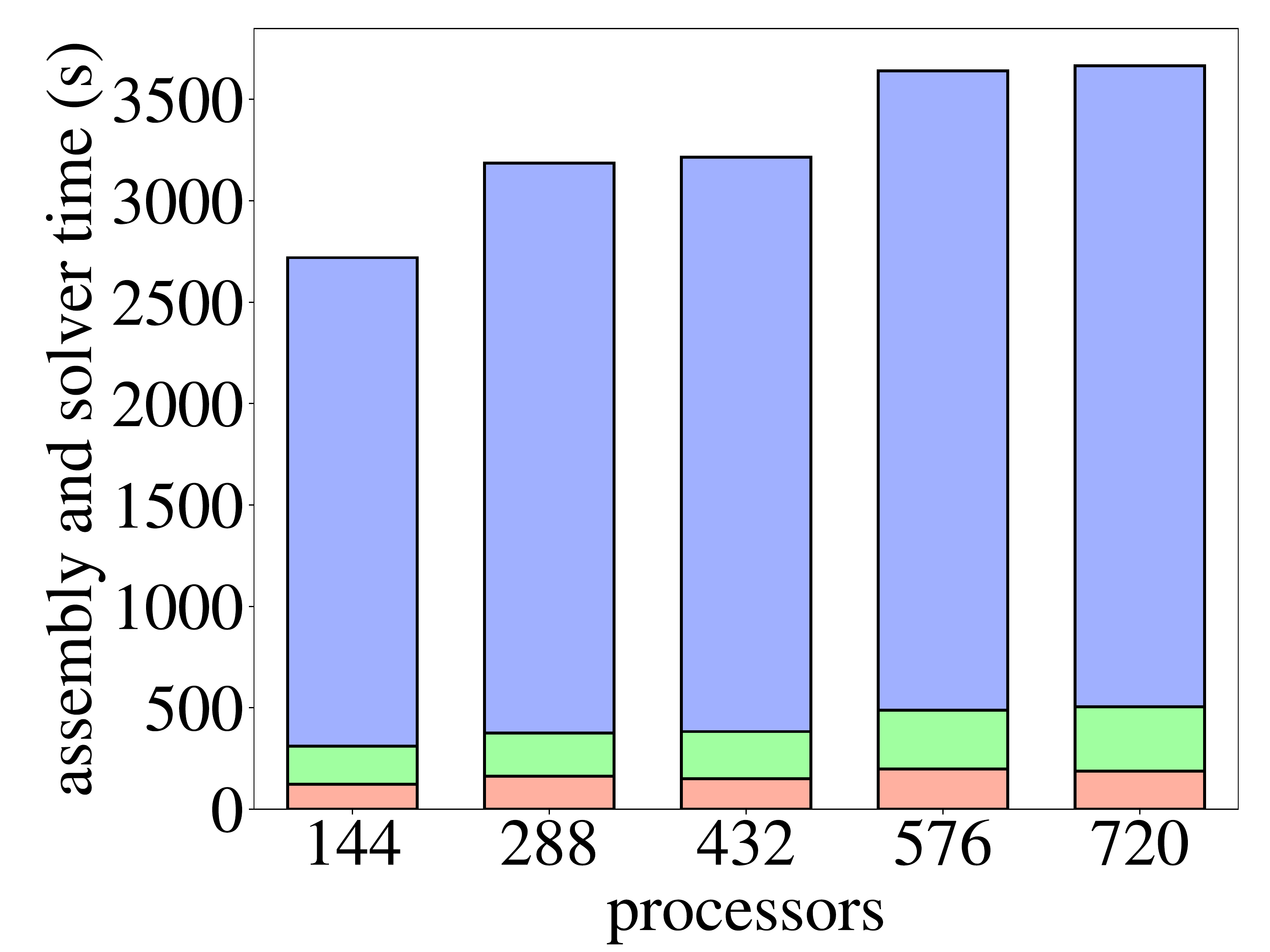}
		\caption{}
		\label{fig:scalingH_MF_LP_TimeScaling}
	\end{subfigure}
	\hfill\hspace{0cm}
	\\
	\hfill
	\begin{subfigure}[b]{0.32\textwidth}
		\includegraphics[width=\textwidth]{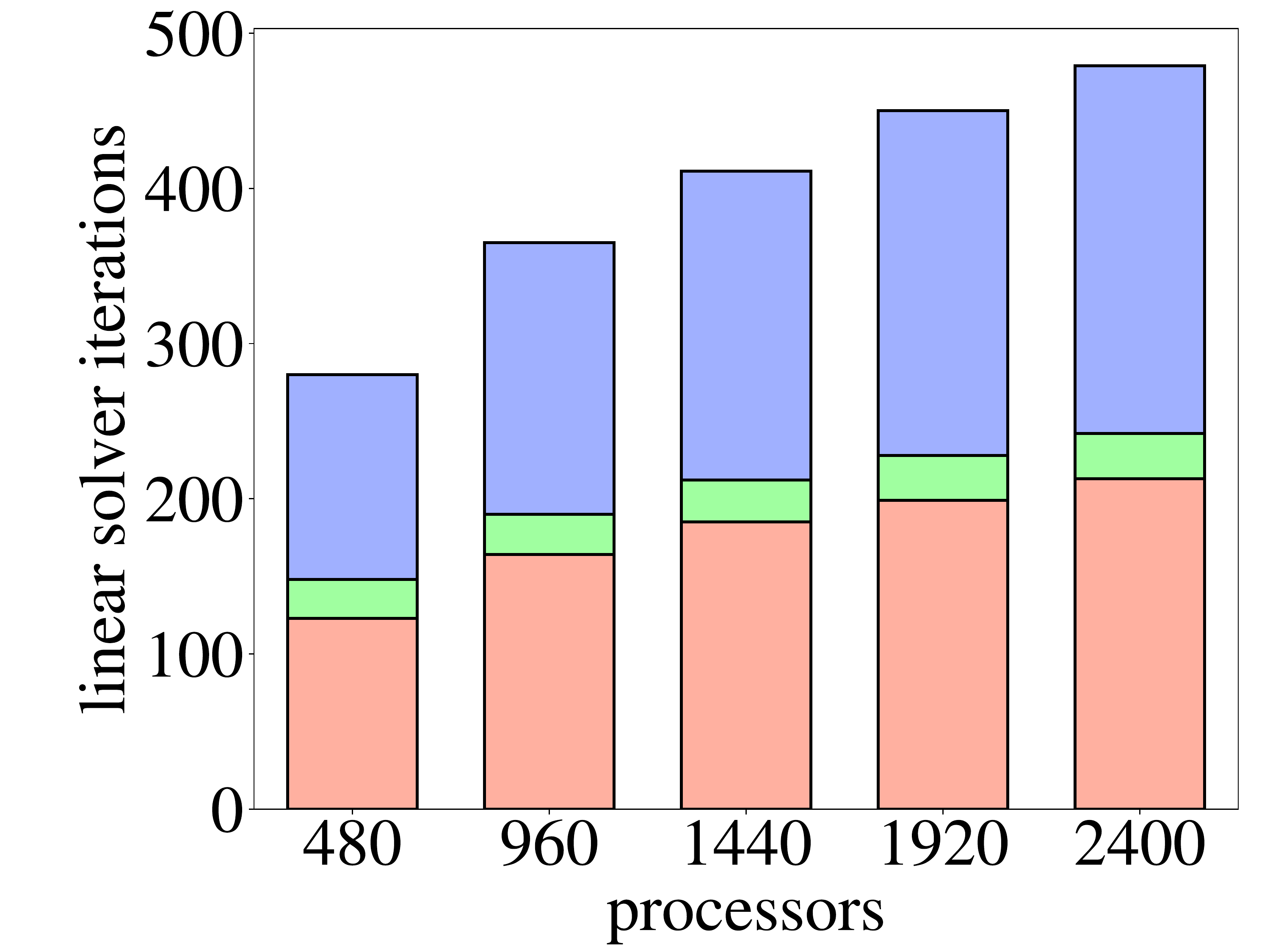}
		\caption{}
		\label{fig:scalingH_NMF_IterationsScaling}
	\end{subfigure}
	\hfill
	\begin{subfigure}[b]{0.32\textwidth}
		\includegraphics[width=\textwidth]{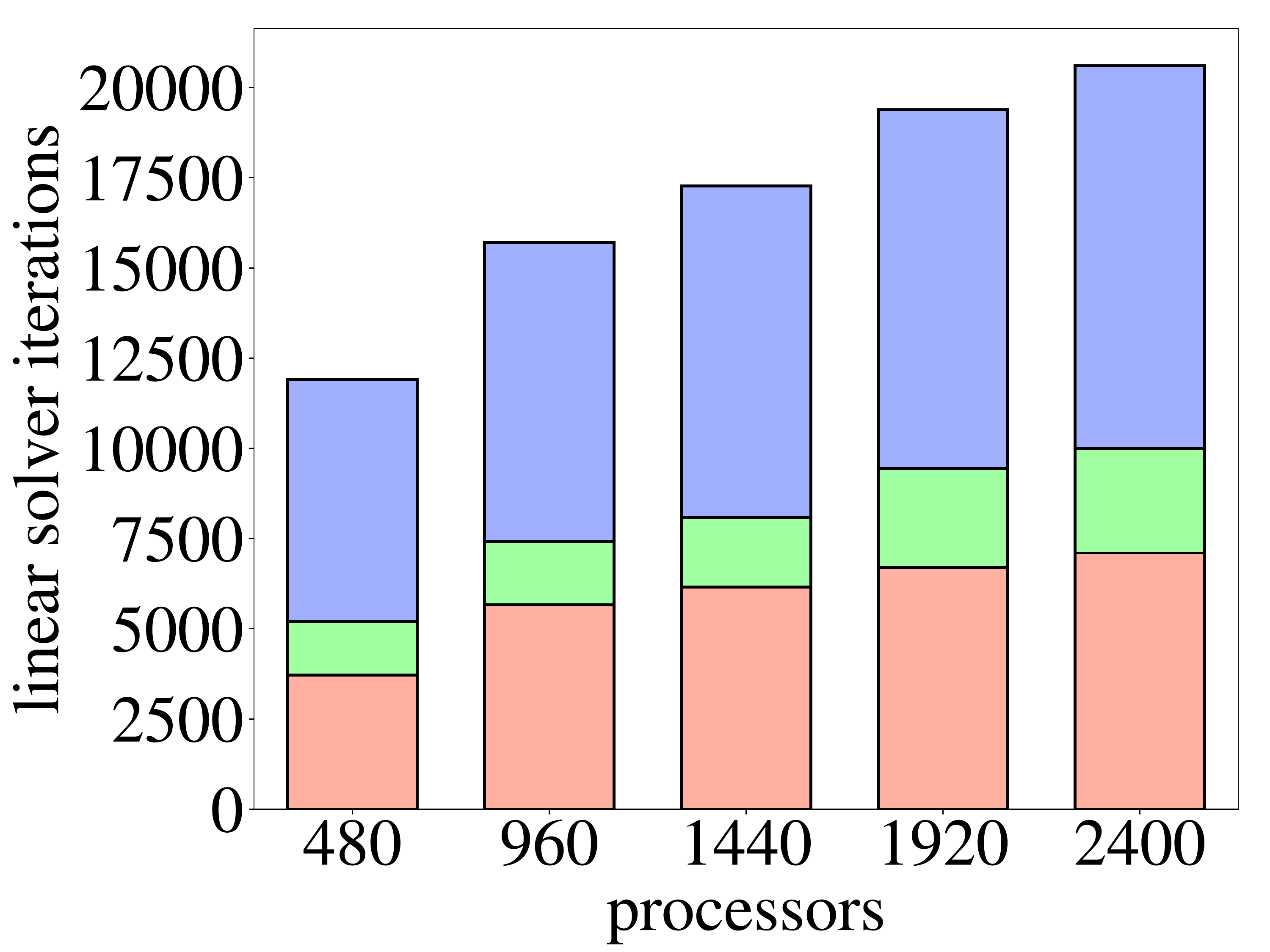}
		\caption{}
		\label{fig:scalingH_MF_IterationsScaling}
	\end{subfigure}
	\hfill
	\begin{subfigure}[b]{0.32\textwidth}
		\includegraphics[width=\textwidth]{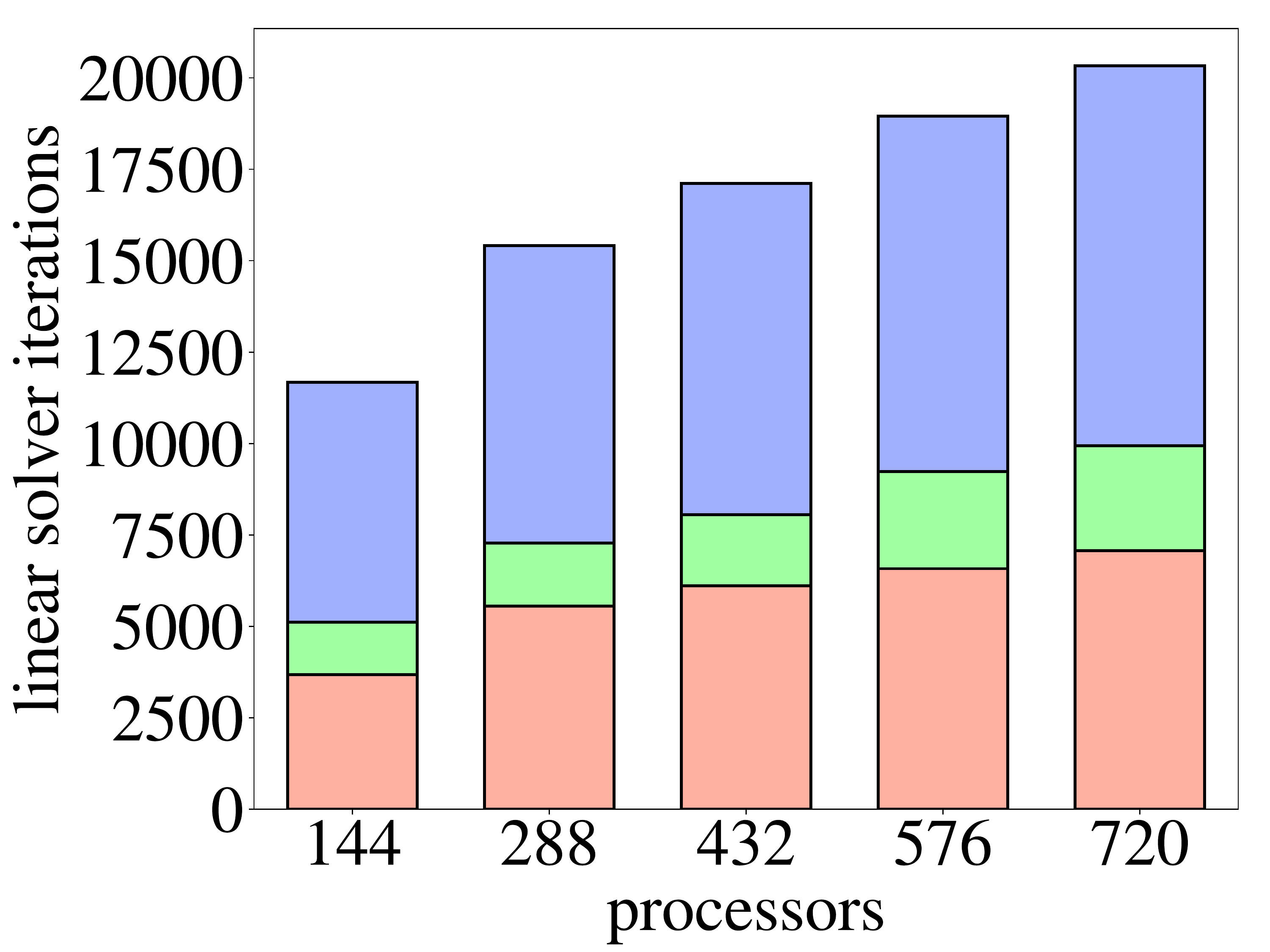}
		\caption{}
		\label{fig:scalingH_MF_LP_IterationsScaling}
	\end{subfigure}
	\hfill\hspace{0cm}
	\\
	\hfill
	\begin{subfigure}[b]{0.32\textwidth}
		\includegraphics[width=\textwidth]{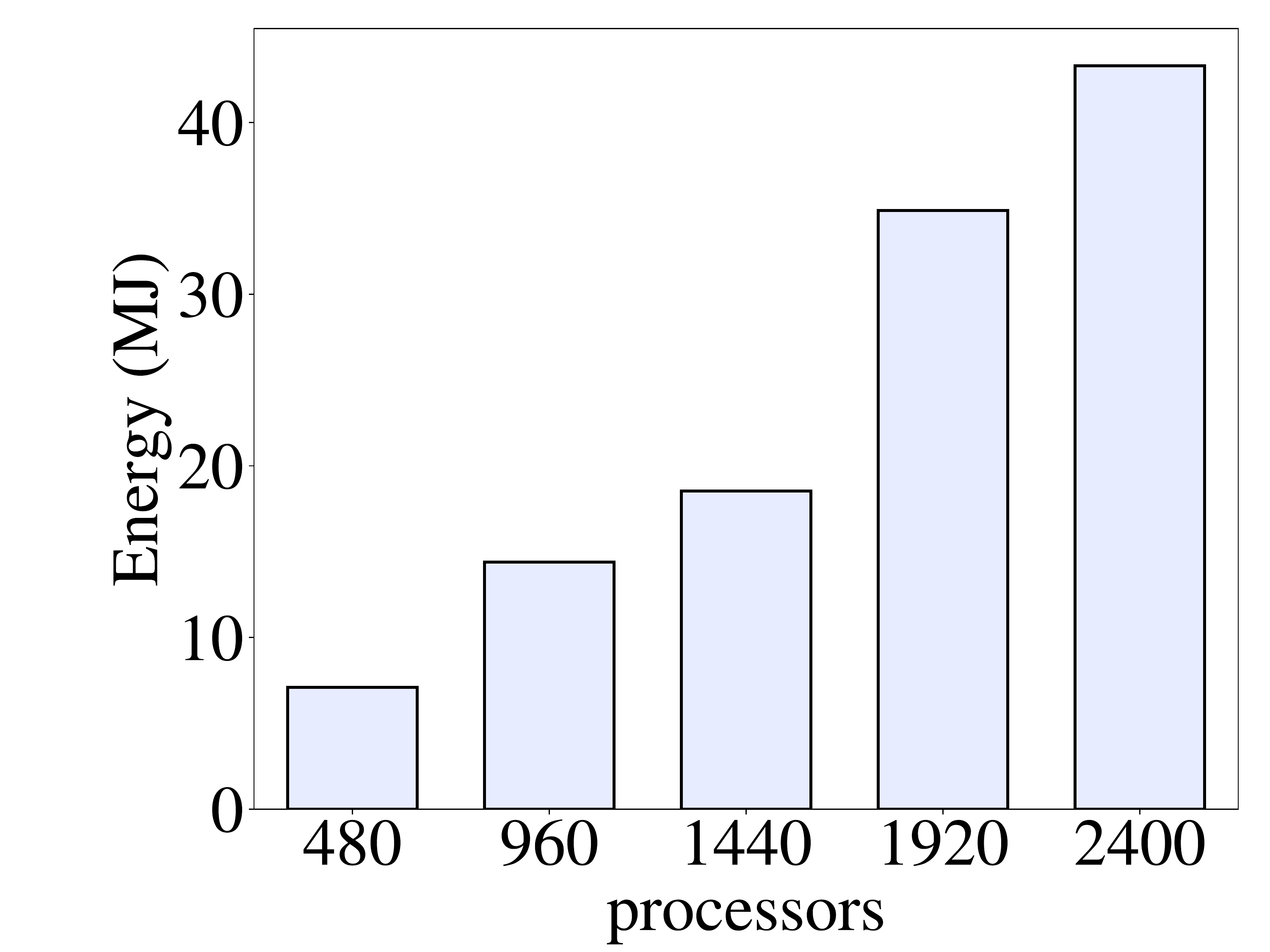}
		\caption{}
		\label{fig:scalingH_NMF_Energy}
	\end{subfigure}
	\hfill
	\begin{subfigure}[b]{0.32\textwidth}
		\includegraphics[width=\textwidth]{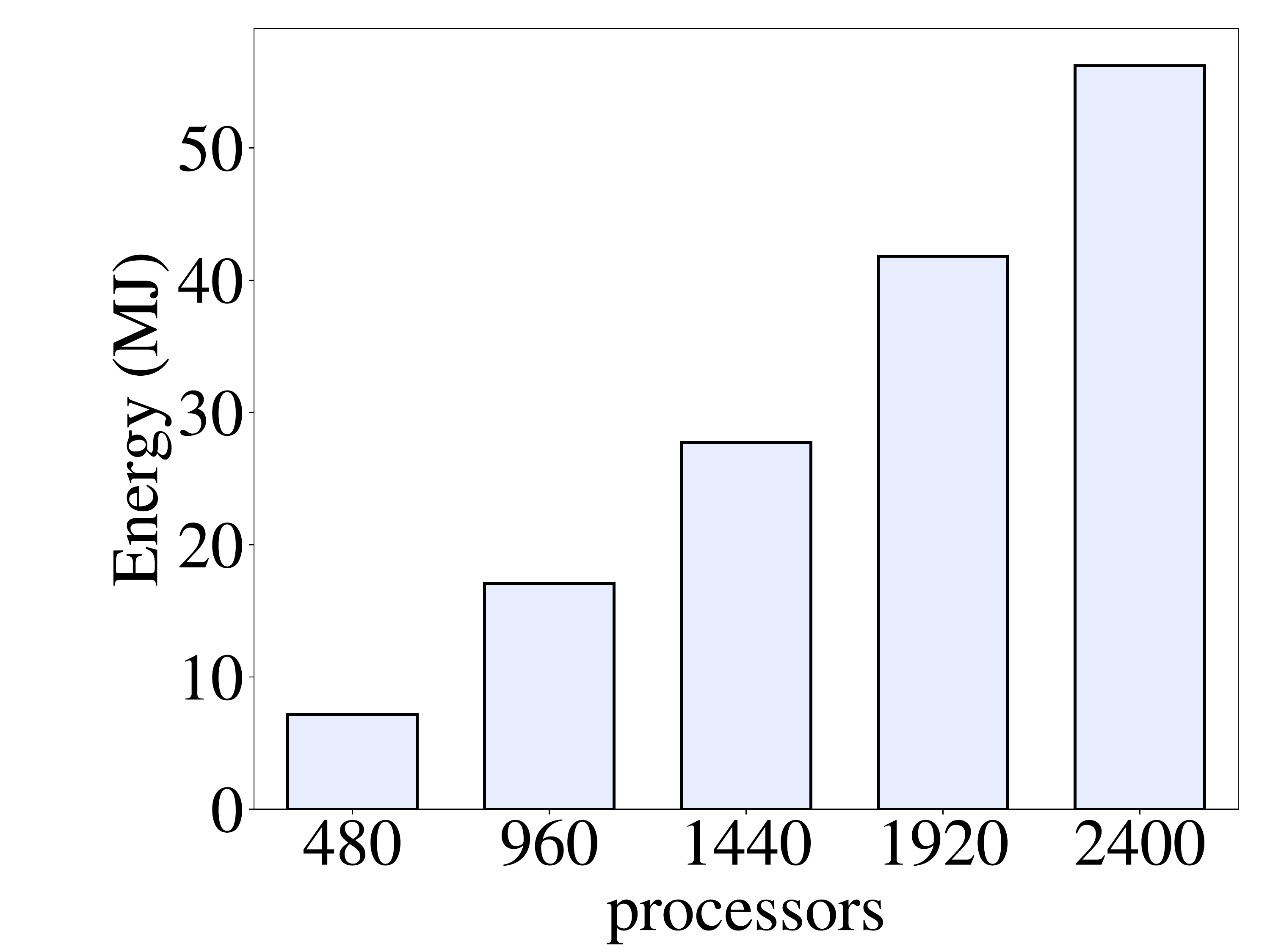}
		\caption{}
		\label{fig:scalingH_MF_Energy}
	\end{subfigure}
	\hfill
	\begin{subfigure}[b]{0.32\textwidth}
		\includegraphics[width=\textwidth]{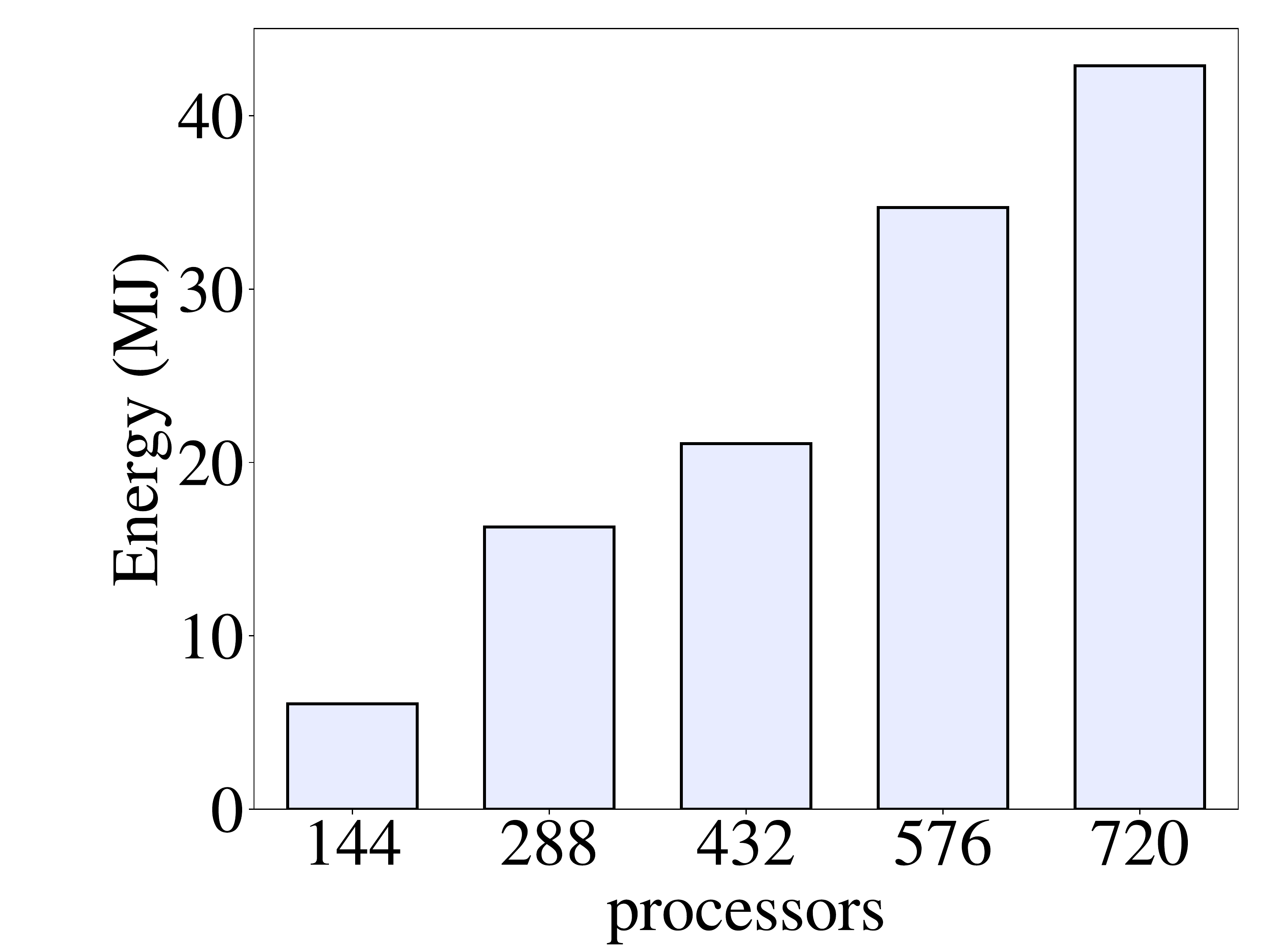}
		\caption{}
		\label{fig:scalingH_MF_LP_Energy}
	\end{subfigure}
	\hfill\hspace{0cm}
	\\
	\hfill
	\begin{subfigure}[b]{0.32\textwidth}
		\includegraphics[width=\textwidth]{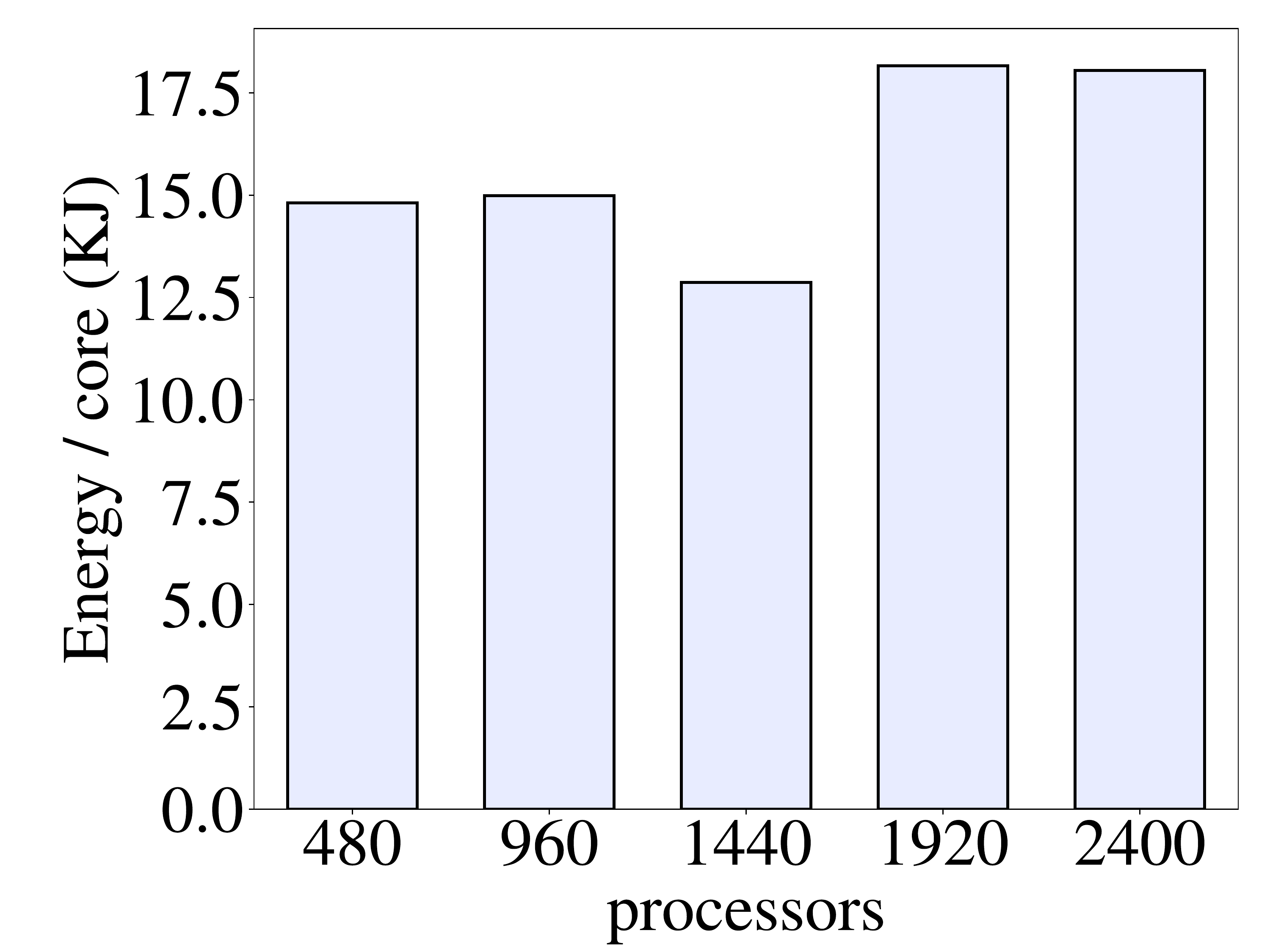}
		\caption{}
		\label{fig:scalingH_NMF_EnergyPerCore}
	\end{subfigure}
	\hfill
	\begin{subfigure}[b]{0.32\textwidth}
		\includegraphics[width=\textwidth]{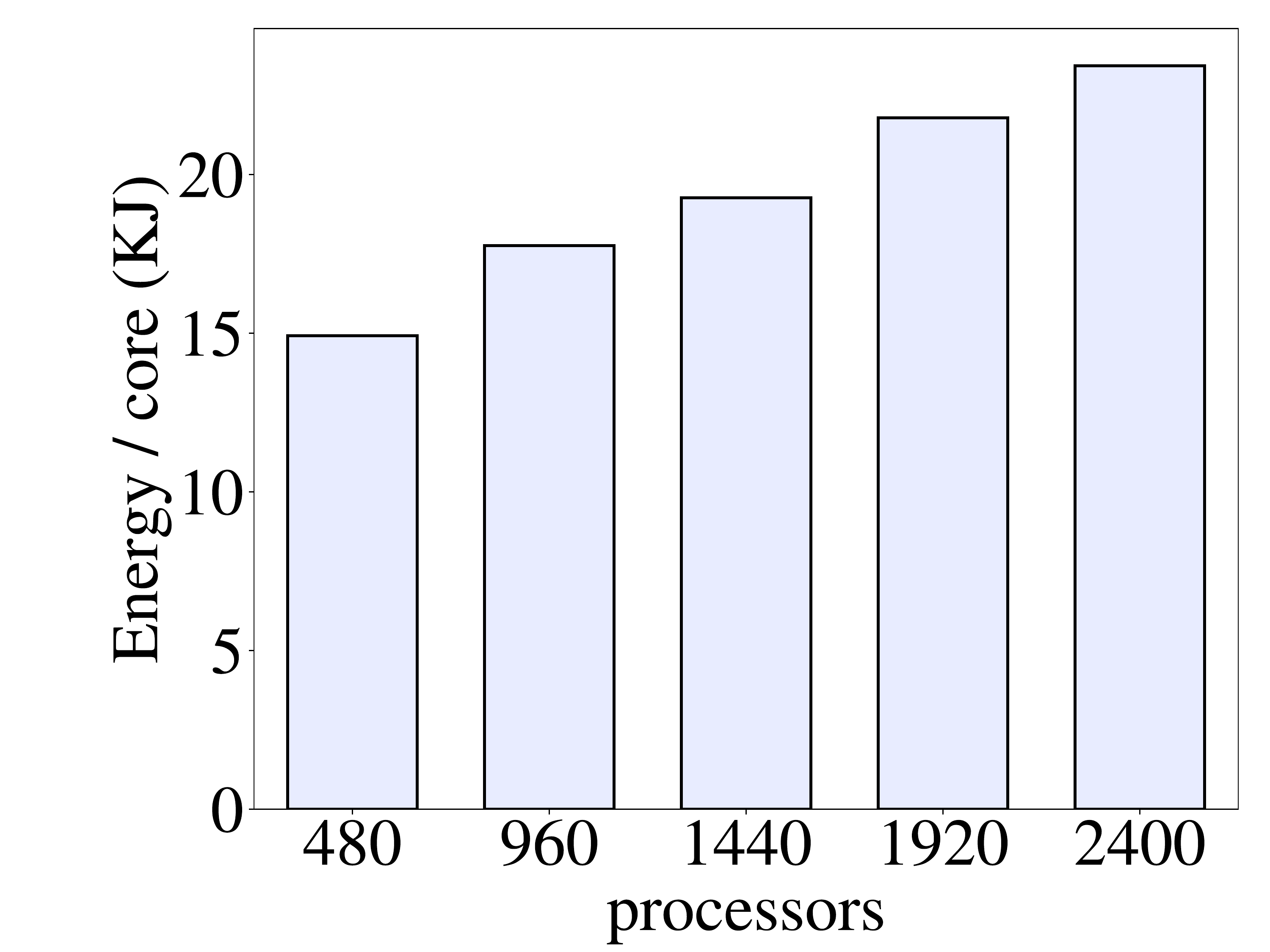}
		\caption{}
		\label{fig:scalingH_MF_EnergyPerCore}
	\end{subfigure}
	\hfill
	\begin{subfigure}[b]{0.32\textwidth}
		\includegraphics[width=\textwidth]{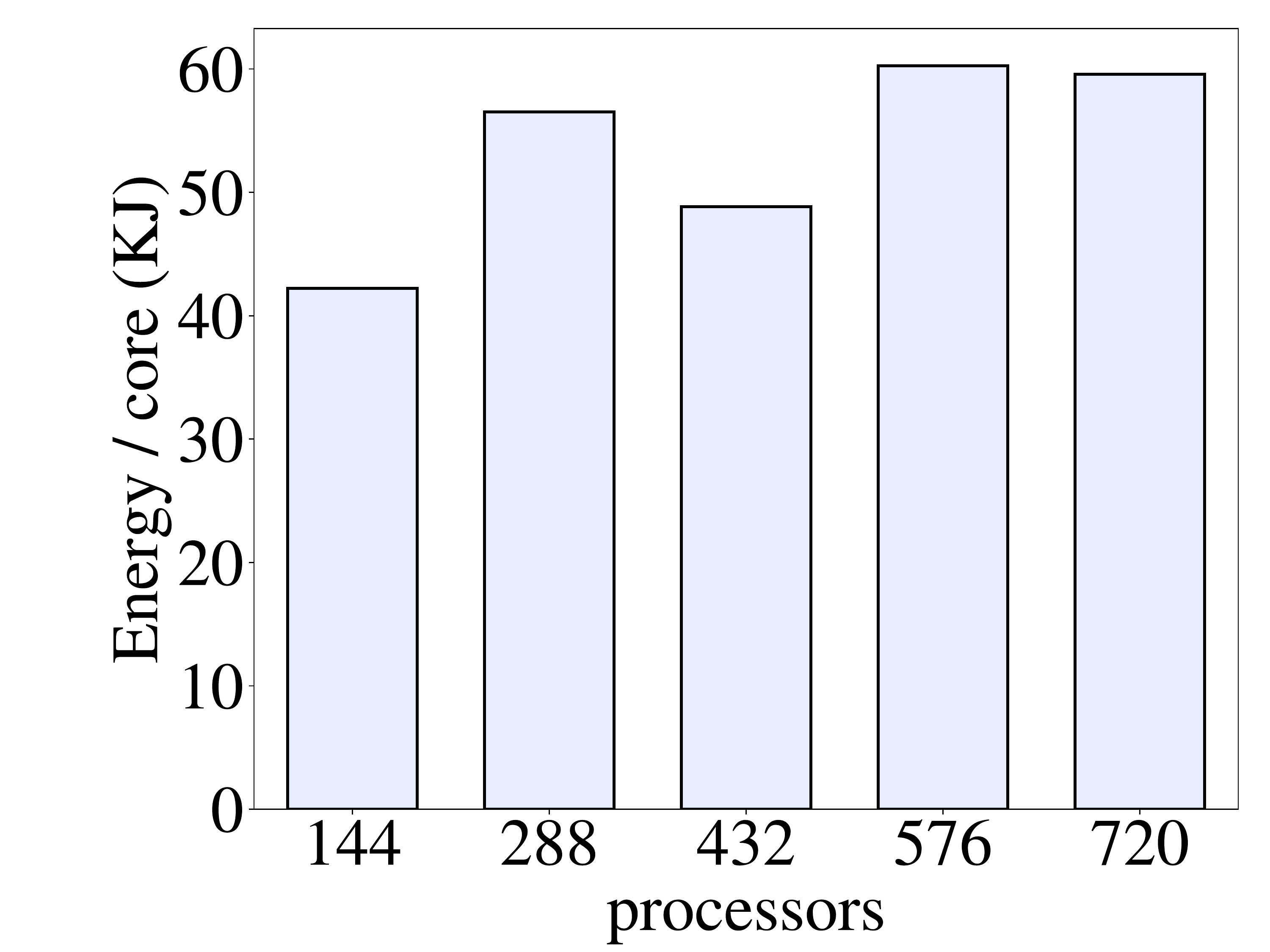}
		\caption{}
		\label{fig:scalingH_MF_LP_EnergyPerCore}
	\end{subfigure}
	\hfill\hspace{0cm}
	\caption{In columns, different pre-conditioners: 
		(a), (d) (g) and (j) ASDD(1)/SOR(2); (b), (e), (h) and (k) block-SOR(1)/ASDD(1)/SOR(2); (c), (f), (i) and (l) block-SOR(1)/ASDD(1)/SOR(2) with less processors .
		In rows, (a), (b) and (c) linear solver time;
		(d), (e) and (f) linear solver iterations; 
		(g), (h) and (i) total energy consumption; and
		(j), (k) and (l) energy consumption per core.}
	\label{fig:spheresH_Scalings}
\end{figure*}

In this example, we perform an analysis on the element size for isotropic elements. We generate five isotropic meshes increasing the number of elements and the used cores in such a way that the number of elements per core remains constant. The domain is a sphere of radius four with a spherical hole in the center of radius one. The element sizes are chosen such that there are around 1500 elements per core, and we have used 480, 960, 1440, 1920 and 2400 cores. This leads to meshes that are composed of $0.72\cdot10^6$, $1.44\cdot10^6$, $2.16\cdot10^6$, $2.88\cdot10^6$ and $3.60\cdot10^6$ elements of polynomial degree four. Since the proposed block-SOR pre-conditioner reduces the memory footprint of the curving method, we also generate the same five meshes with a number of cores that leads to 5000 elements per core. Specifically, we have used 144, 288, 432, 576 and 720 cores. Figures \ref{fig:hScalingLevel1} to \ref{fig:hScalingLevel5} show the five curved high-order meshes generated for this example. The obtained meshes have converged with residual norms in the range of $1.90 \cdot 10^{-12}$ and $5.05 \cdot 10^{-12}$.

Figure \ref{fig:spheresConstraint} shows the evolution of the constraint norm during the iterations of the non-linear optimization for the five meshes. Note that we obtain the same evolution with all the linear solvers analyzed in this example. At each non-linear iteration, we optimize the functional in Equation \eqref{eqn:unconstrained} and the penalty parameter is increased in order to enforce the boundary condition. Dark blue circles denote the initial iteration of each polynomial degree in the $p$-continuation technique. The evolution of the constraint norm in all cases follows a similar pattern, even though the boundary mesh is not the same.

We observe that the proposed formulation presents a mesh independent behavior at the non-linear level. Apart of the first case, the number of iterations to perform the whole optimization process is the same in all the cases and moreover, the number of non-linear iterations at each polynomial degree is also the same. During the first iterations of the quadratic mesh, the constraint norm decreases \emph{slowly}. Nevertheless, from iteration three onward, the constraint norm decreases geometrically with the non-linear iterations. When the constraint norm is of the same order as the constraint norm of the next polynomial degree, the early-termination criterion is activated. Thus, the norm of the boundary condition is similar to the norm of the boundary condition of the next polynomial degree. Then, we compute the new penalty parameter for the next polynomial degree. Note that the constraint norm also decreases geometrically with the non-linear iterations. Therefore, this shows that we have correctly selected the value of the penalty parameter. Finally, when curving the quartic mesh, we activate the adaption of the penalty parameter and we finalize the curving process with two iterations.

In Figure \ref{fig:spheresH_Scalings}, we present the time to assemble and solve the linear problems, the total number of linear solver iterations, the energy consumption, and the energy consumption per core for the matrix-free and the sparse matrix linear solvers. When using the matrix-free linear solver with the block-SOR pre-conditioner, the time to assemble and solve the linear systems for the finer meshes is larger than the time to curve the coarser meshes, even when the number of elements per core is the same. For this reason, as the meshes become finer, the energy consumption per core also increases because each core is working more time. In the case of using the sparse matrices, the time to assemble and solve the linear systems is more or less the same for all the meshes, and so is the energy consumption per core.

In all cases, the number of linear iterations increases as the meshes become finer. Most of the linear iterations are performed when curving the polynomial degrees two and four. However, the time to assemble and solve the linear systems for the quadratic and cubic meshes is almost negligible compared to the total time to assemble and solve all the linear systems of the curving process.

In this example, the energy consumption is slightly larger when using the proposed matrix-free pre-conditioner. Nevertheless, the main advantage of this pre-conditioner is the reduction of the memory footprint by a factor of three. Thus, we are able to reduce the number of processors by three. When using less processors, the execution time increases because there is more work per core. Although we have reduced the number of cores by a factor greater than three, the time to assemble and solve the linear systems increases by a factor less than three. The number of liner solver iterations is more or less the same as when curving with a large number of cores. Nevertheless, since there are less processors, the communication workload is reduced and therefore, each iteration is performed faster. Moreover, the energy consumption is slightly reduced when  we use less processors. The energy consumption per core has increased because the execution time has also increased. Thus, we have shown that using the proposed block-SOR pre-conditioner, we can generate curved meshes with less computational resources. When reducing the number of cores, although the time to generate the mesh increases, the energy consumption remains similar.

\begin{figure*}[t!]
	\centering
	\hfill
	\begin{subfigure}[b]{\spheresWidth}
		\includegraphics[width=\textwidth]{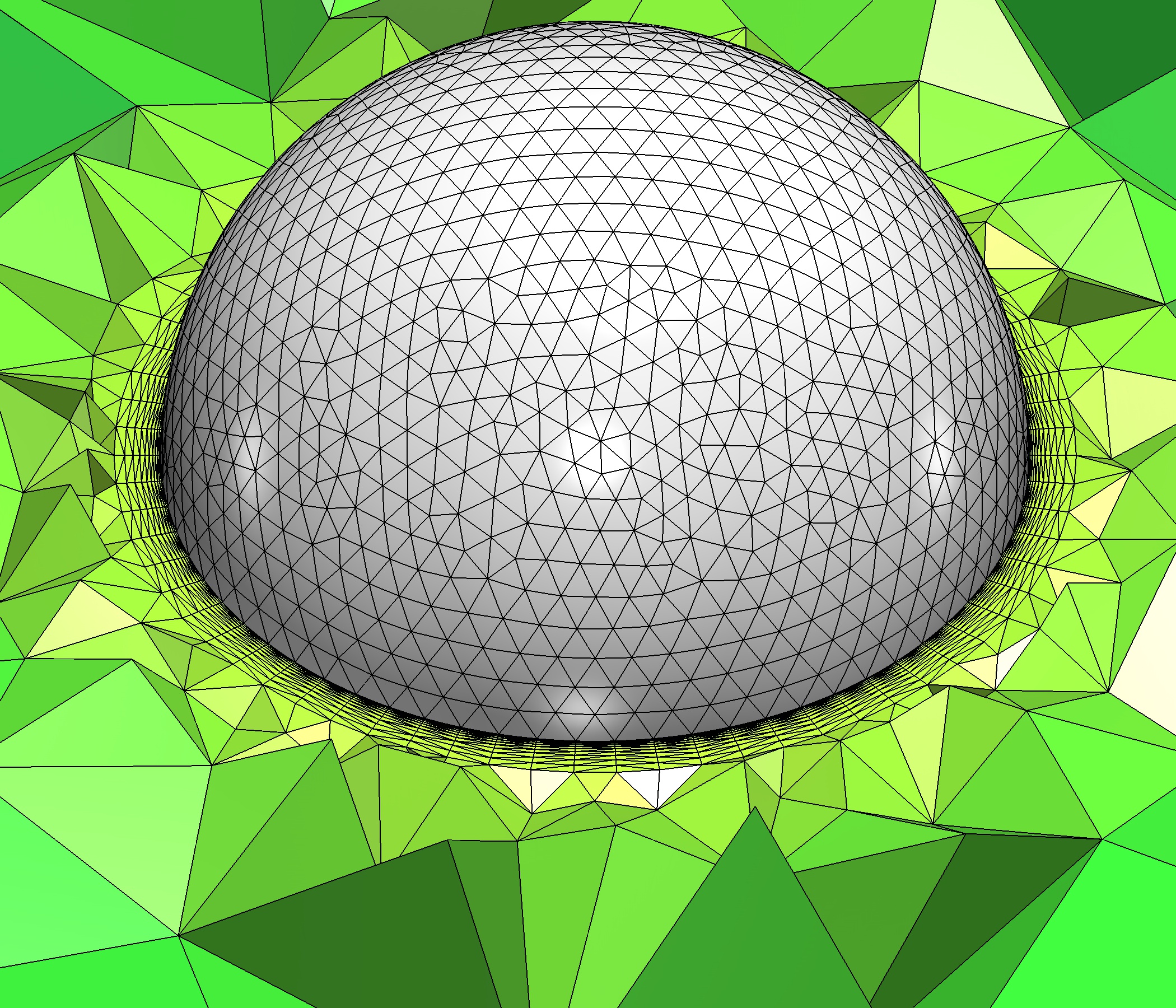}
		\caption{}
		\label{fig:blScalingLevel1}
	\end{subfigure}
	\hfill
	\begin{subfigure}[b]{\spheresWidth}
		\includegraphics[width=\textwidth]{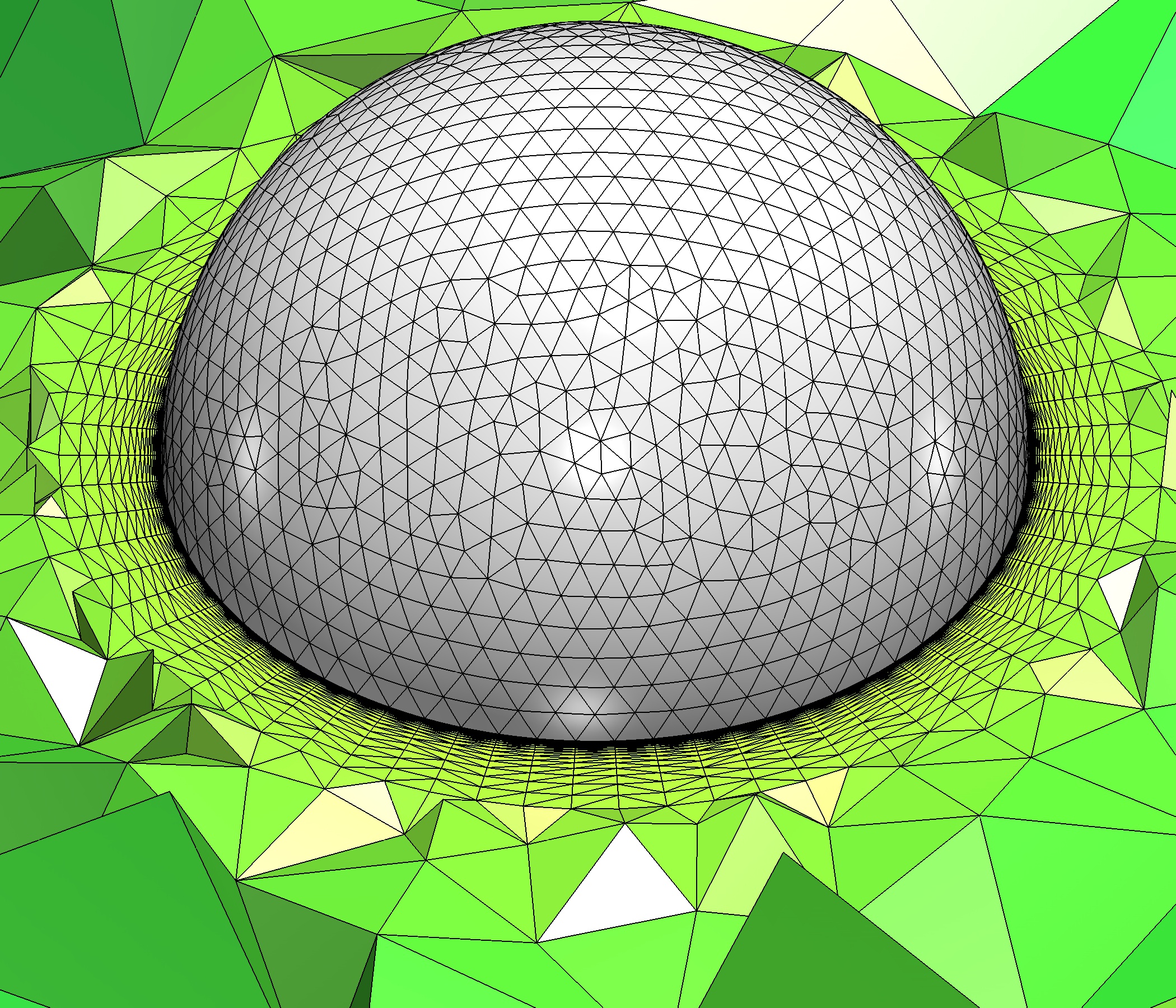}
		\caption{}
		\label{fig:blScalingLevel2}
	\end{subfigure}
	\hfill\hspace{0pt}
	\\
	\hfill
	\begin{subfigure}[b]{\spheresWidth}
		\includegraphics[width=\textwidth]{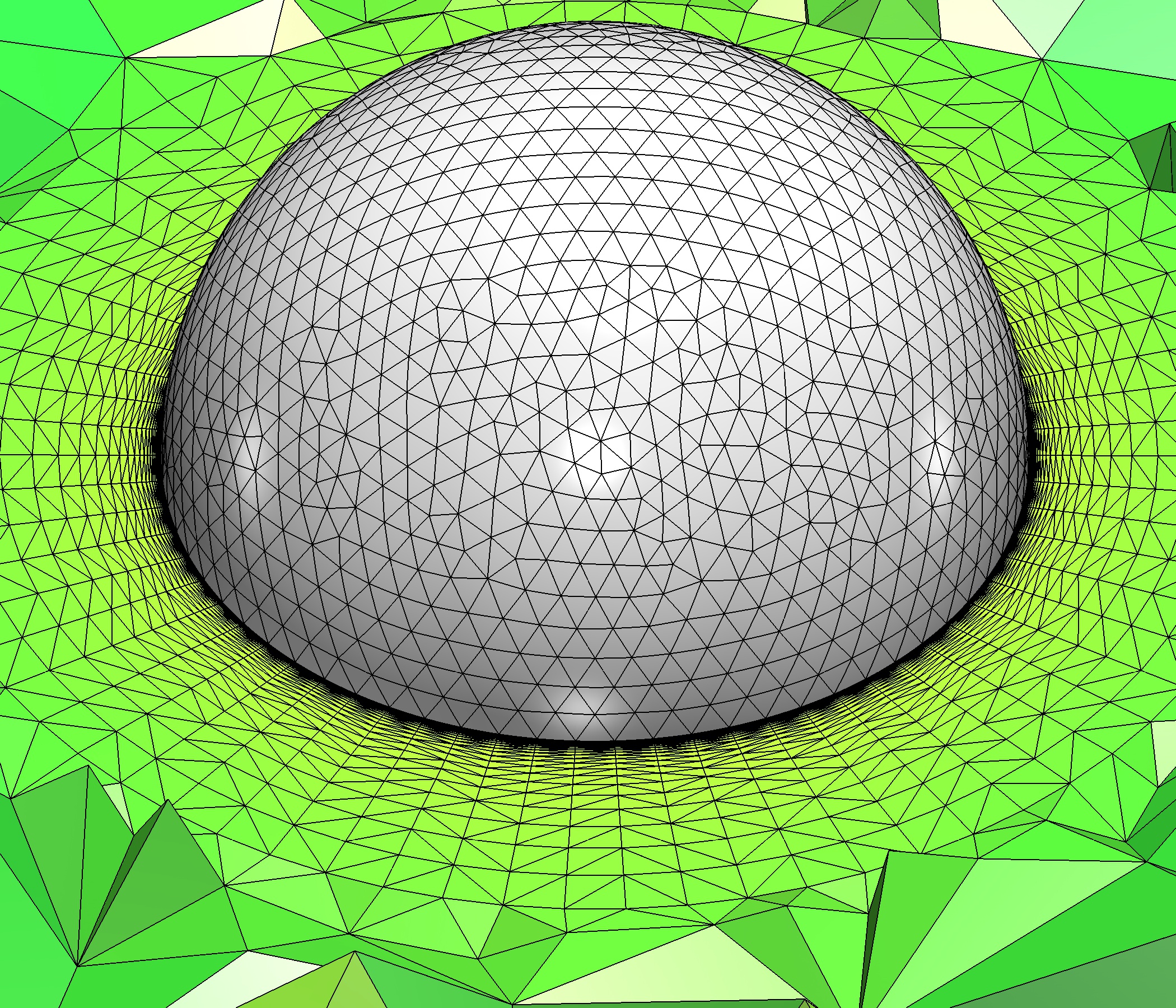}
		\caption{}
		\label{fig:blScalingLevel3}
	\end{subfigure}
	\hfill
	\begin{subfigure}[b]{\spheresWidth}
		\includegraphics[width=\textwidth]{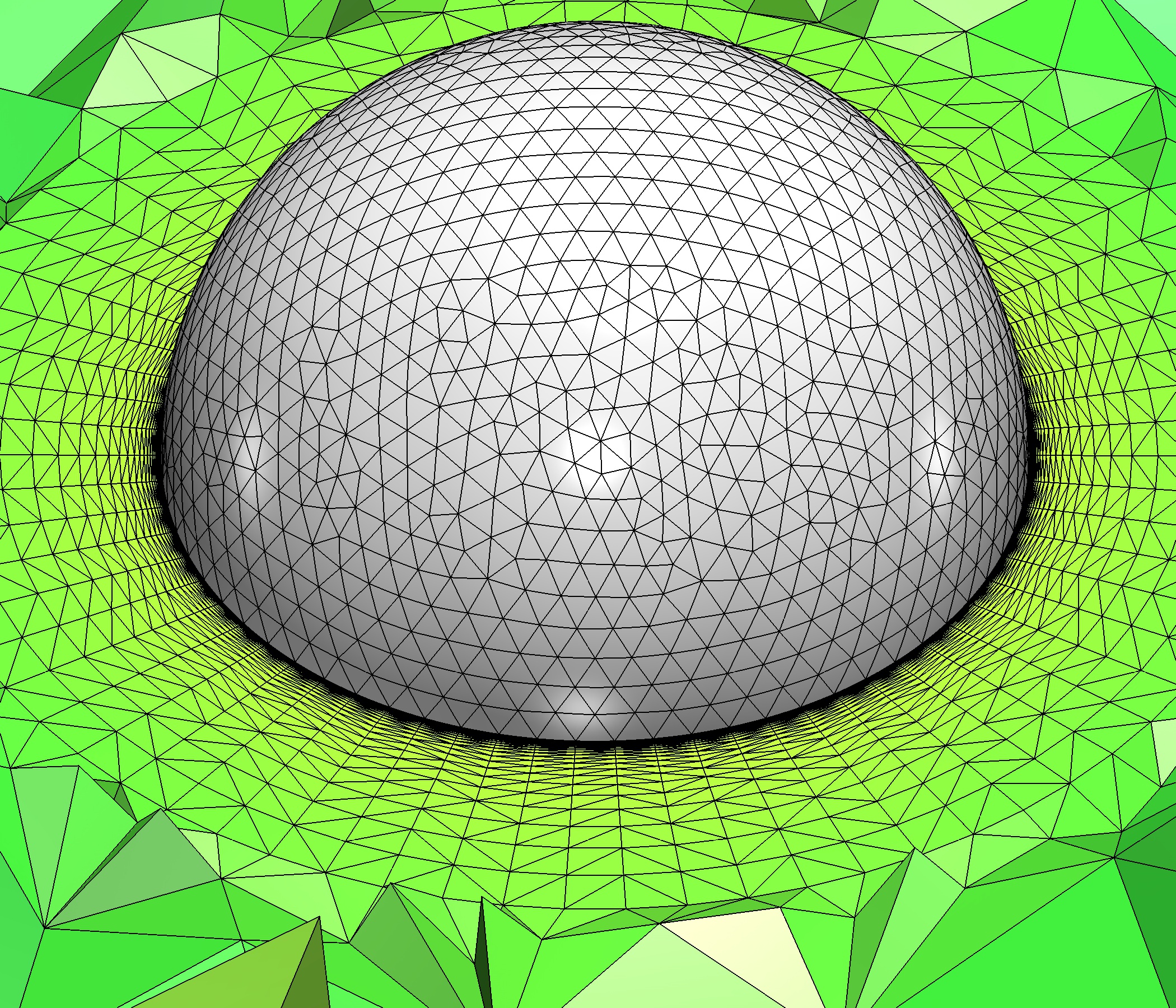}
		\caption{}
		\label{fig:blScalingLevel4}
	\end{subfigure}
	\hfill\hspace{0pt}
	\\
	\begin{subfigure}[b]{\spheresWidth}
		\includegraphics[width=\textwidth]{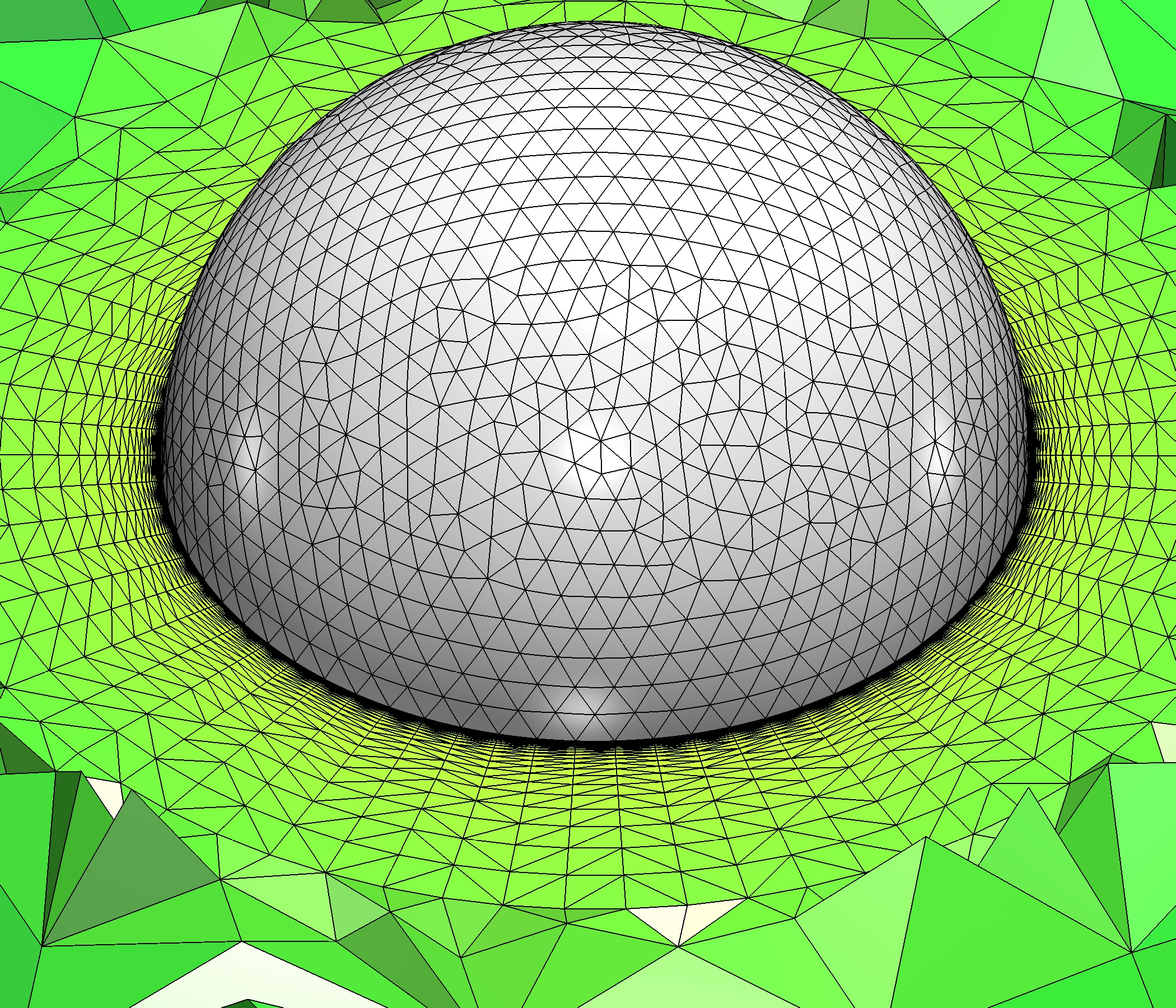}
		\caption{}
		\label{fig:blScalingLevel5}
	\end{subfigure}
	\begin{subfigure}[b]{0.75\textwidth}
		\includegraphics[width=\textwidth]{qualityLegend}
	\end{subfigure}
	\caption{Optimized meshes using the proposed mesh curving solver with boundary layer stretching of:
		(a) $1:1\cdot 10^1$;
		(b) $1:1\cdot 10^2$;
		(c) $1:1\cdot 10^3$;
		(d) $1:1\cdot 10^4$; and
		(e) $1:1\cdot 10^5$.}
	\label{fig:blScalingMeshes}
\end{figure*}

\begin{figure*}[t!]
	\centering
	\hfill
	\begin{subfigure}[b]{0.32\textwidth}
		\includegraphics[width=\textwidth]{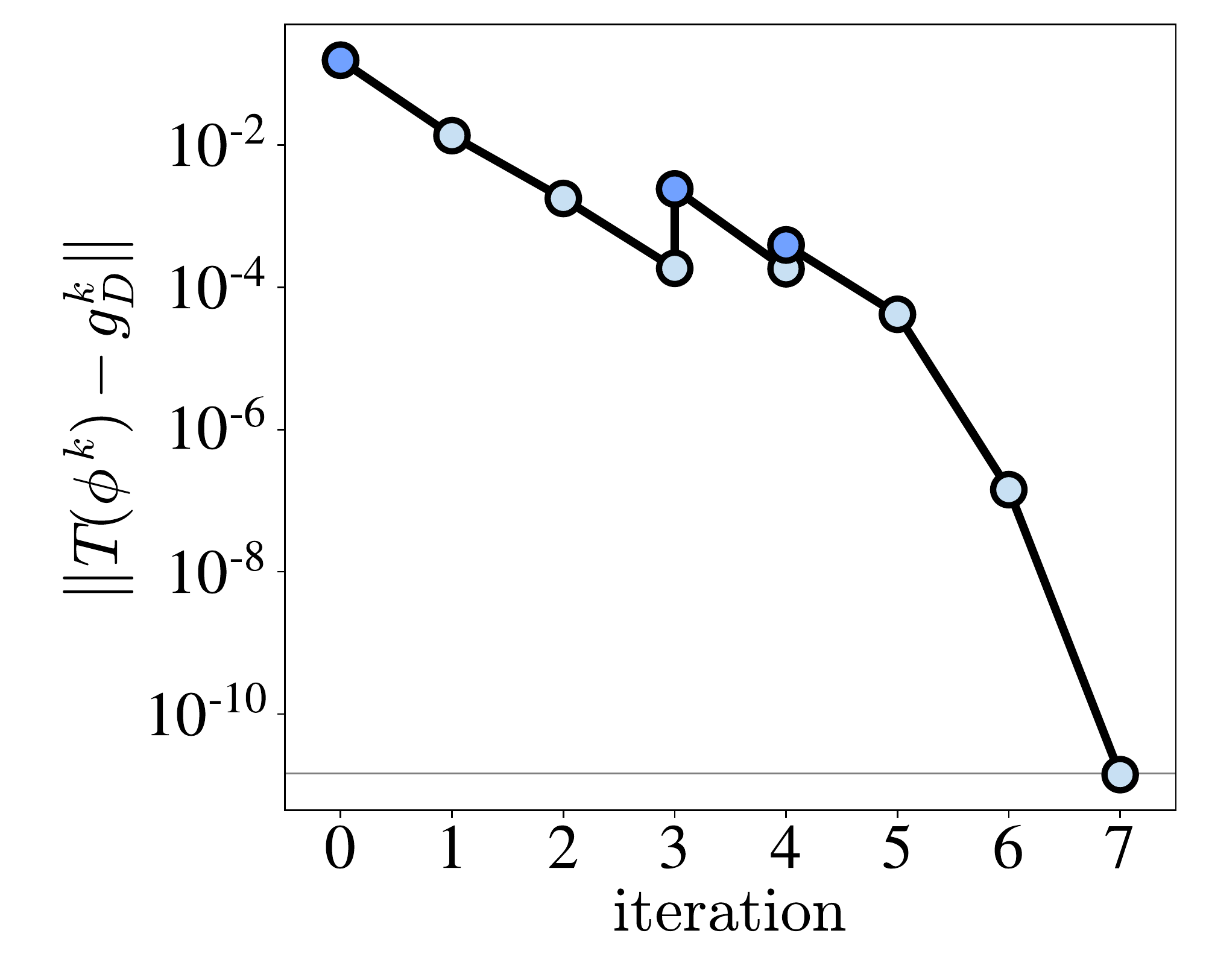}
		\caption{}
		\label{fig:spheresBL1_MF_AM_ATP4_Constraint}
	\end{subfigure}
	\hfill
	\begin{subfigure}[b]{0.32\textwidth}
		\includegraphics[width=\textwidth]{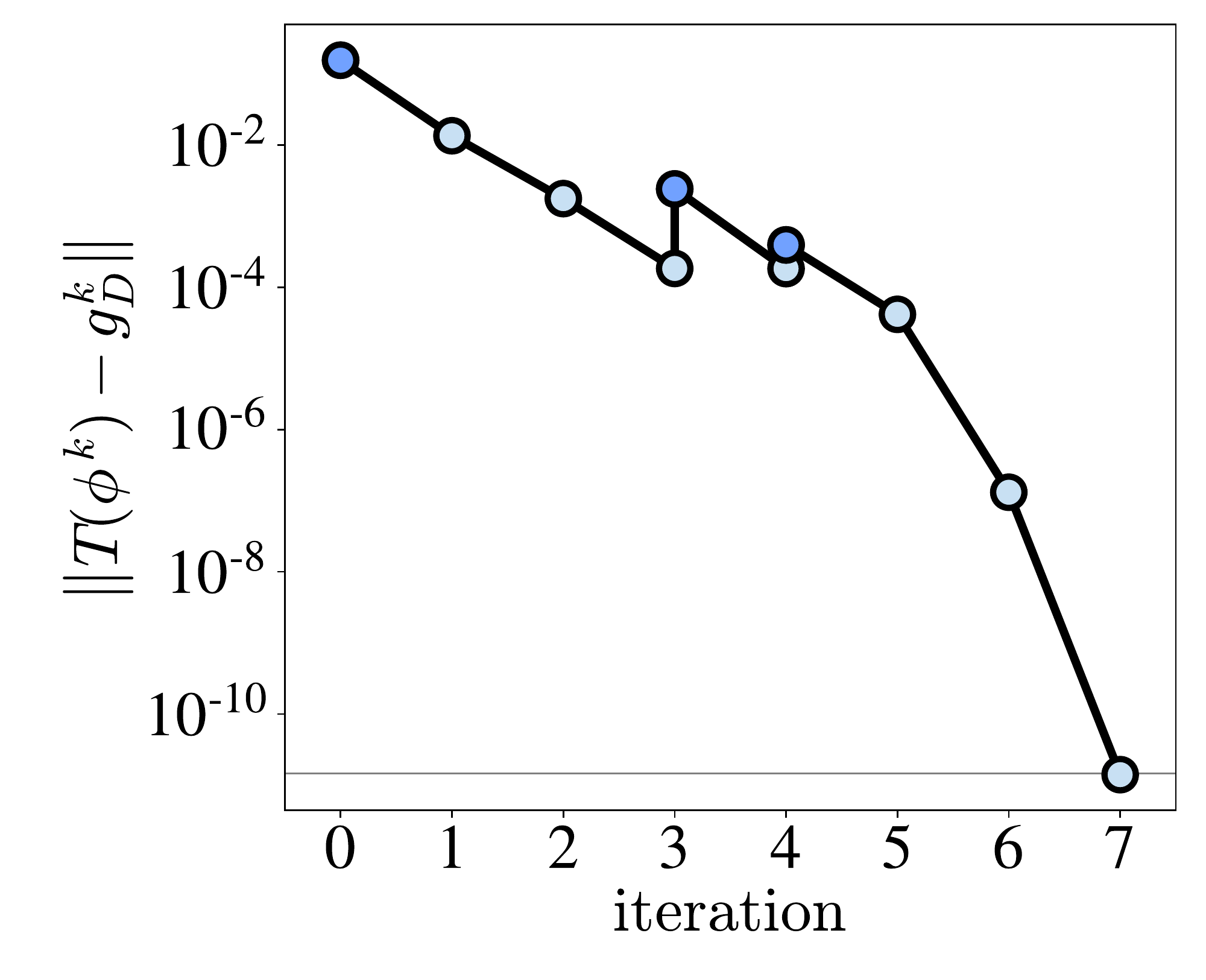}
		\caption{}
		\label{fig:spheresBL2_MF_AM_ATP4_Constraint}
	\end{subfigure}
	\hfill
	\begin{subfigure}[b]{0.32\textwidth}
		\includegraphics[width=\textwidth]{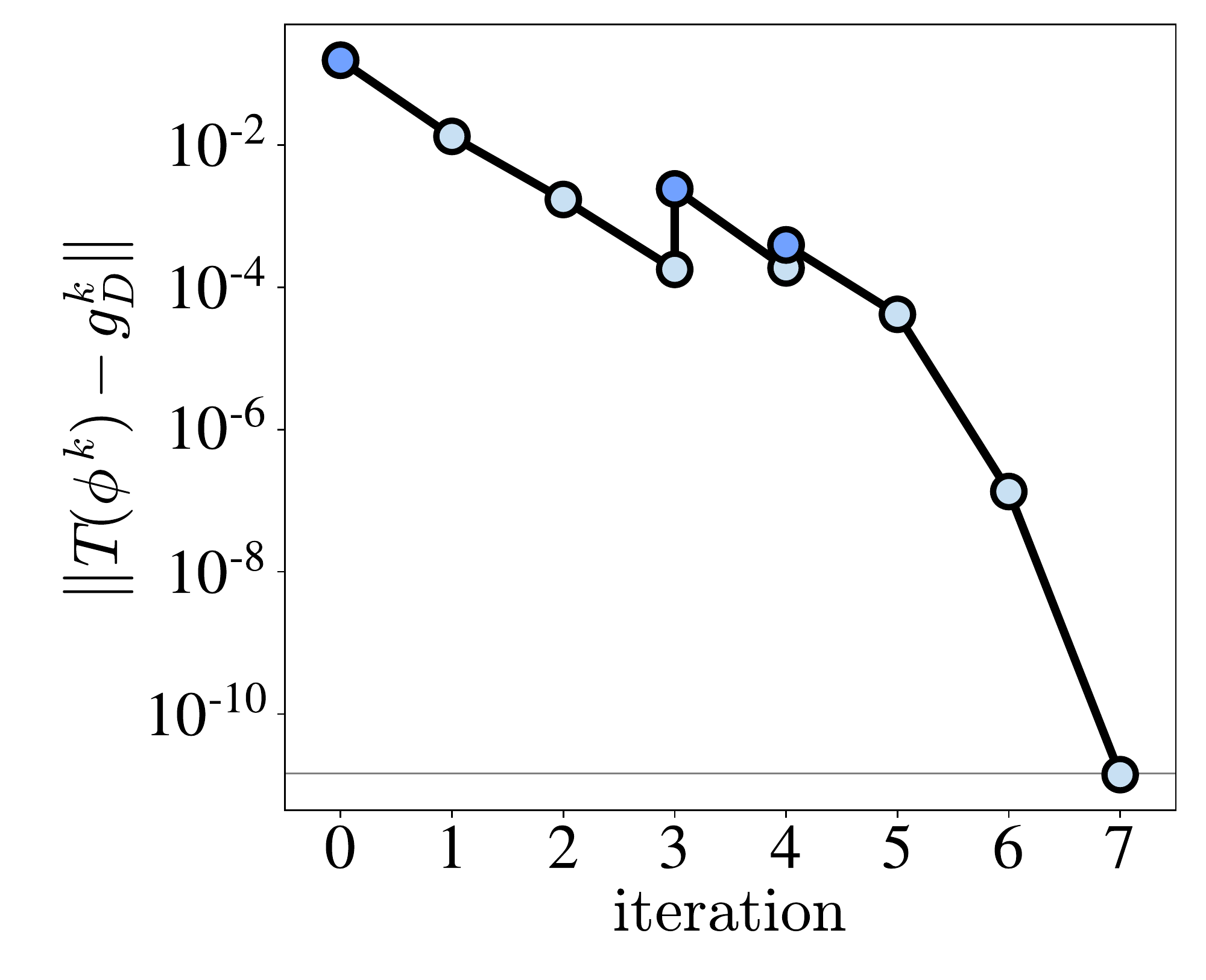}
		\caption{}
		\label{fig:spheresBL3_MF_AM_ATP4_Constraint}
	\end{subfigure}
	\hfill\hspace{0cm}
	\\
	\hfill
	\begin{subfigure}[b]{0.32\textwidth}
		\includegraphics[width=\textwidth]{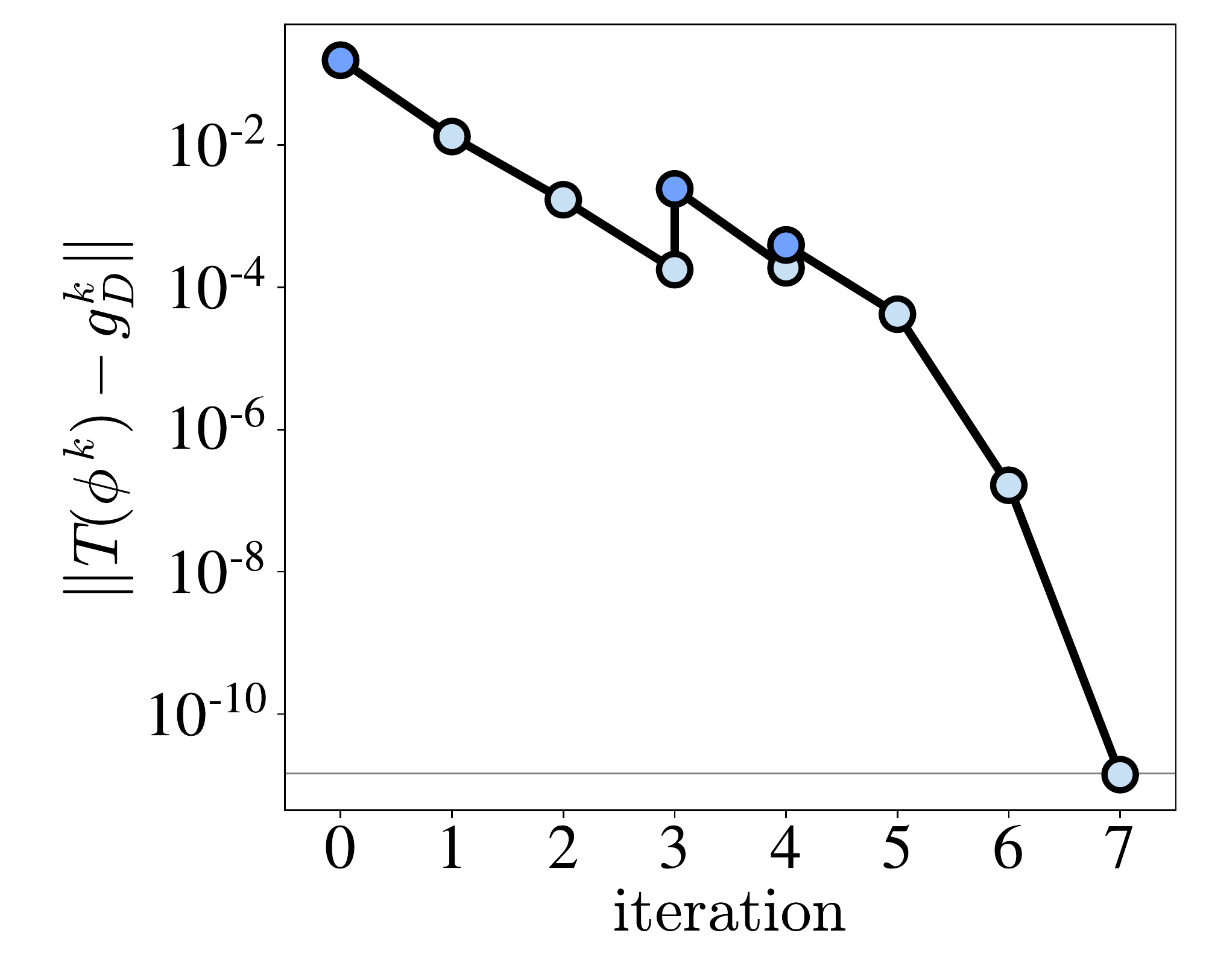}
		\caption{}
		\label{fig:spheresBL4_MF_AM_ATP4_Constraint}
	\end{subfigure}
	\hfill
	\begin{subfigure}[b]{0.32\textwidth}
		\includegraphics[width=\textwidth]{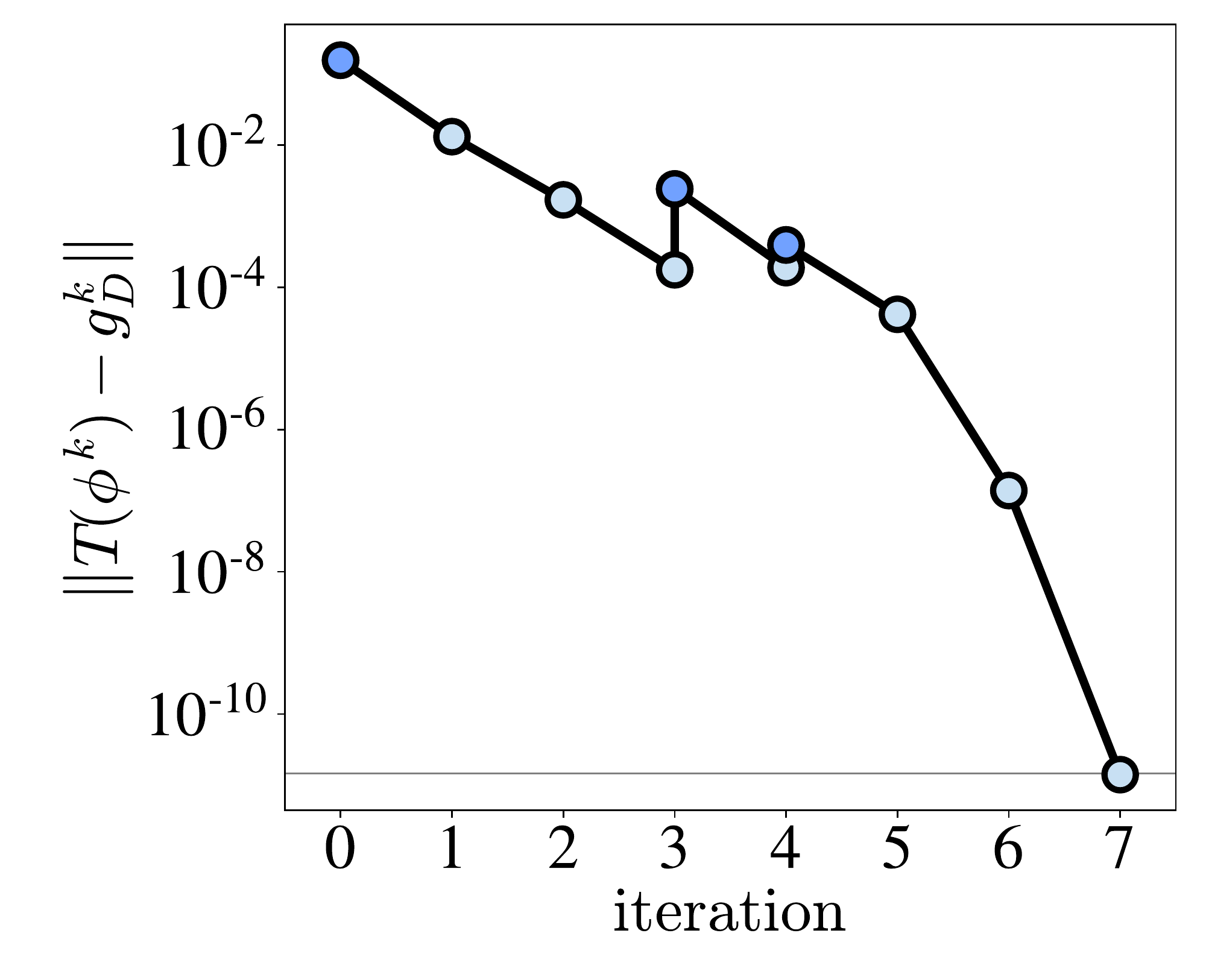}
		\caption{}
		\label{fig:spheresBL5_MF_AM_ATP4_Constraint}
	\end{subfigure}
	\hfill\hspace{0cm}
	\caption{Evolution of the norm of the constraint through the optimization process for the mesh optimized with boundary layer stretching of:
		(a) $1:1\cdot 10^1$;
		(b) $1:1\cdot 10^2$;
		(c) $1:1\cdot 10^3$;
		(d) $1:1\cdot 10^4$; and
		(e) $1:1\cdot 10^5$.}
	\label{fig:spheresBLConstraint}
\end{figure*}

\begin{figure*}[t!]
	\centering
	\begin{subfigure}[b]{0.66\textwidth}
		\includegraphics[width=\textwidth]{comparison_LabelsScaling}
	\end{subfigure}
	\\
	\hfill
	\begin{subfigure}[b]{0.32\textwidth}
		\includegraphics[width=\textwidth]{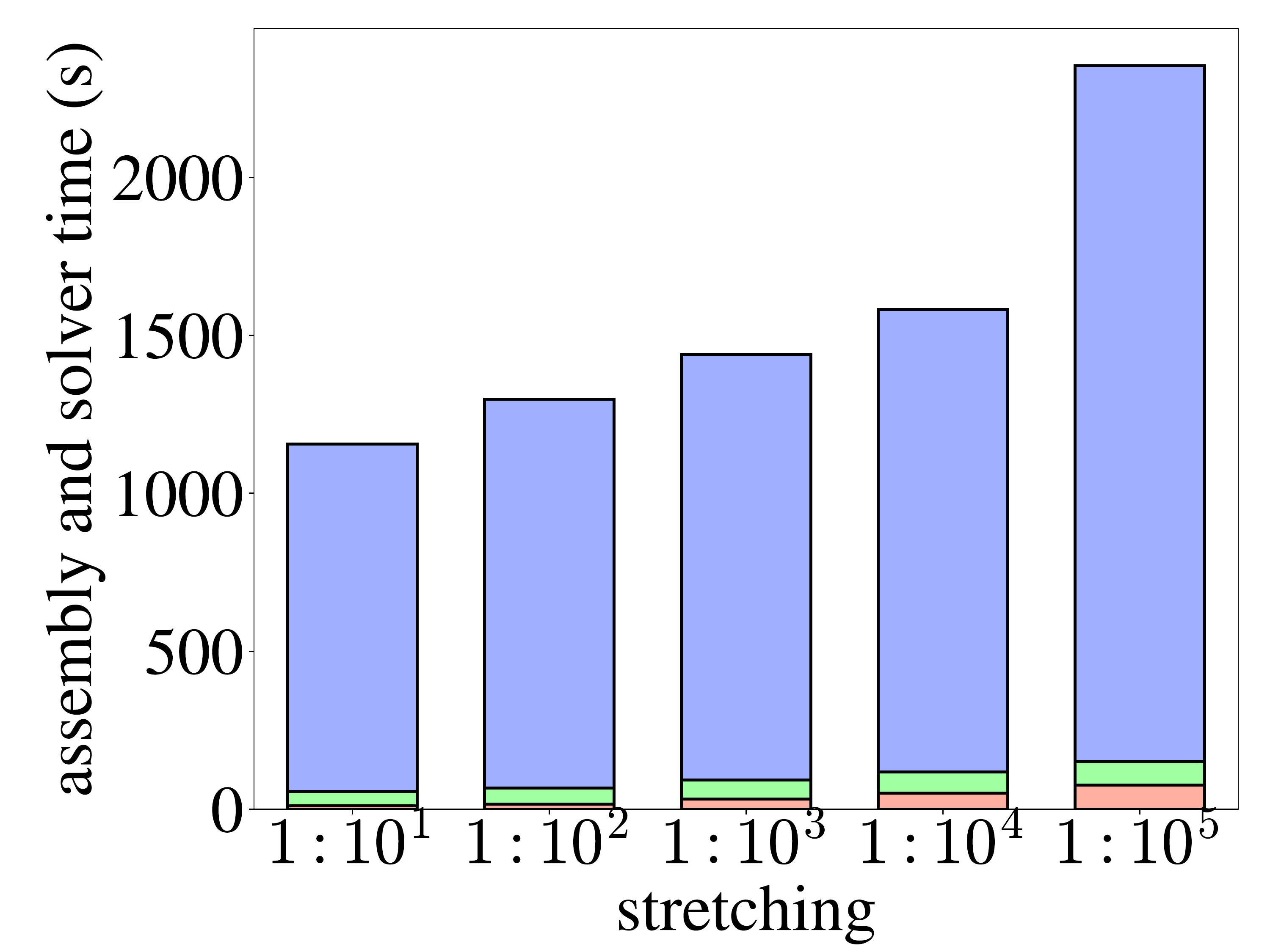}
		\caption{}
		\label{fig:scalingBL_NMF_TimeScaling}
	\end{subfigure}
	\hfill
	\begin{subfigure}[b]{0.32\textwidth}
		\includegraphics[width=\textwidth]{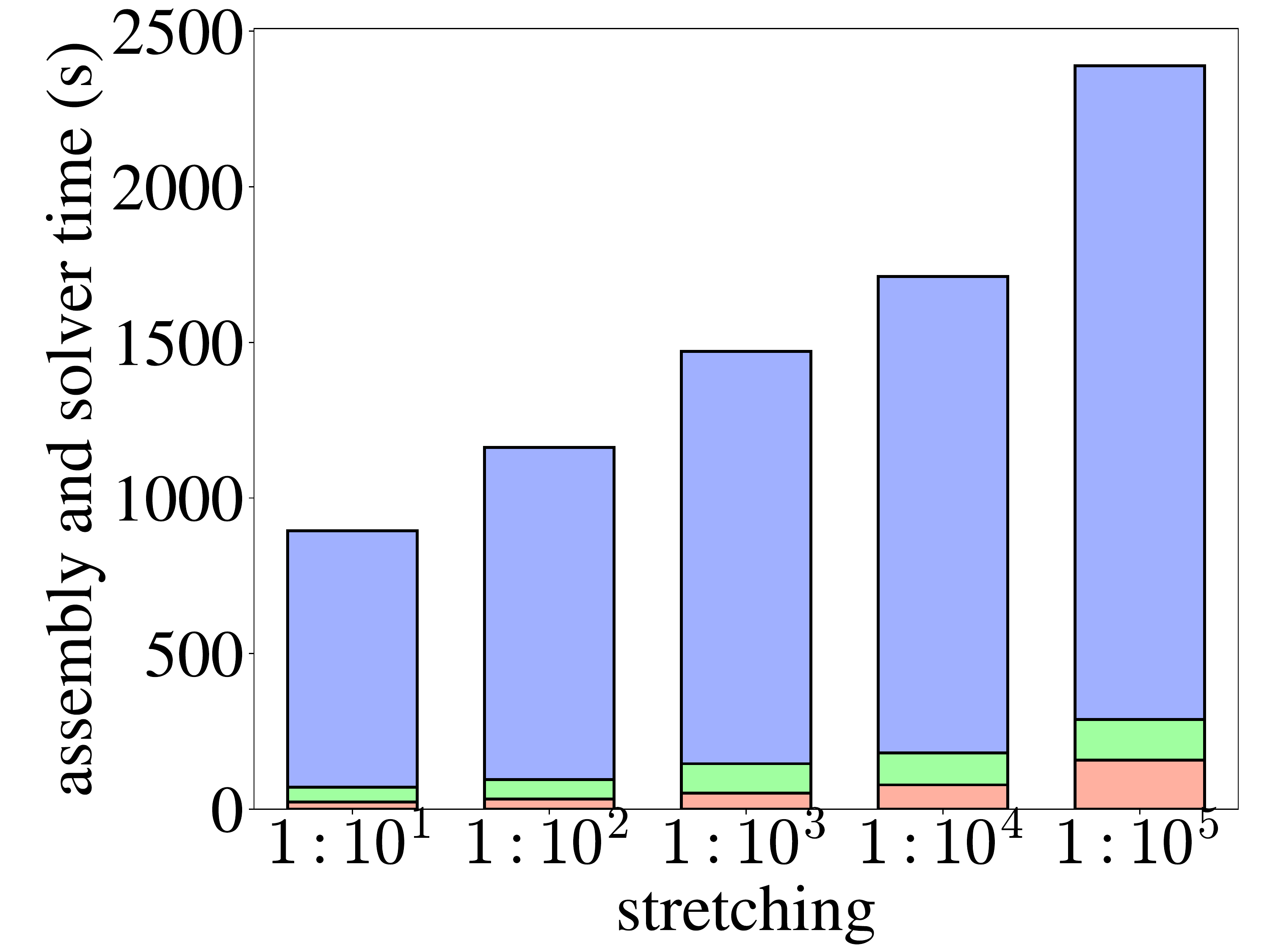}
		\caption{}
		\label{fig:scalingBL_MF_TimeScaling}
	\end{subfigure}
	\hfill\hspace{0cm}
	\\
	\hfill
	\begin{subfigure}[b]{0.32\textwidth}
		\includegraphics[width=\textwidth]{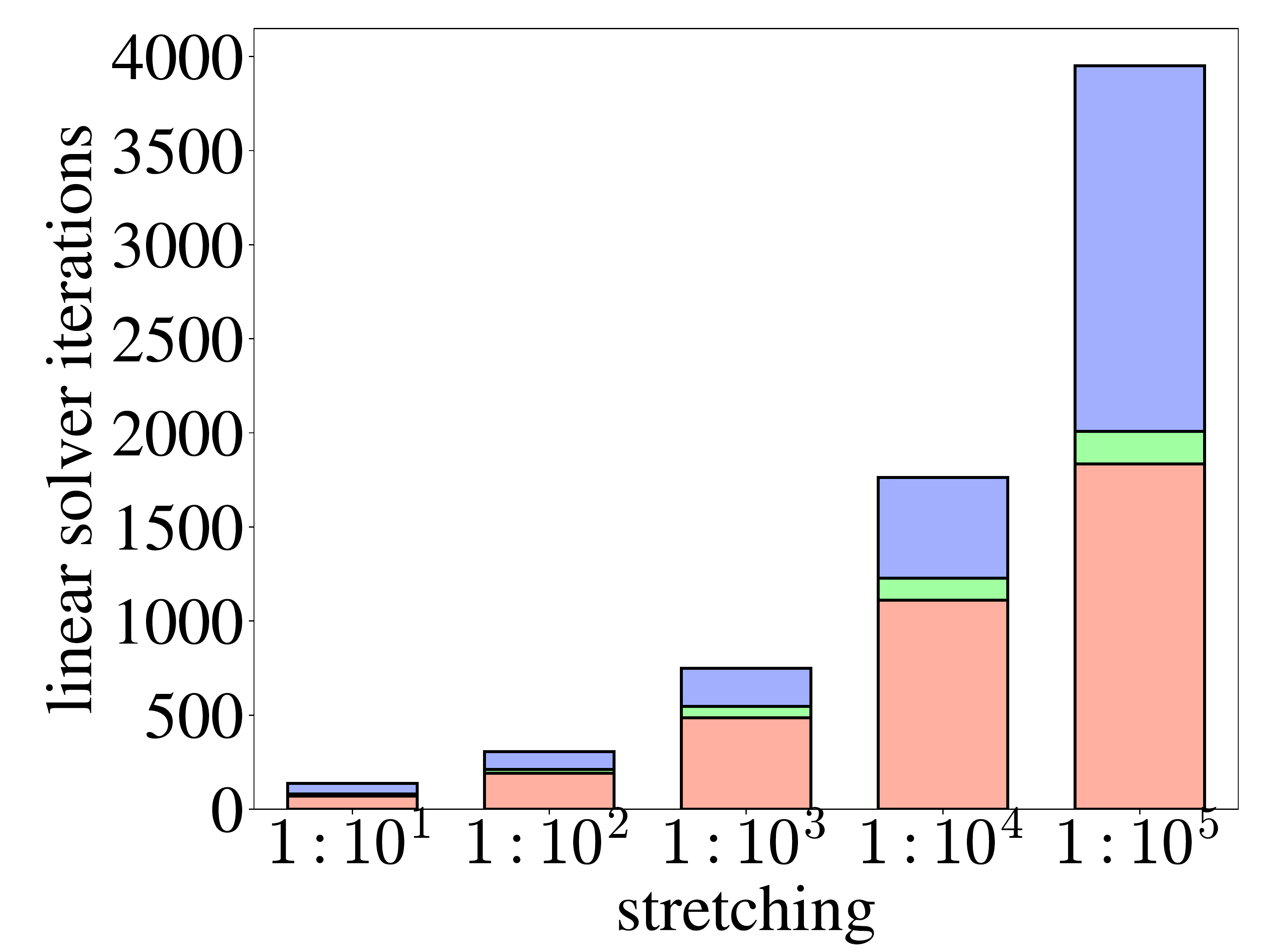}
		\caption{}
		\label{fig:scalingBL_NMF_IterationsScaling}
	\end{subfigure}
	\hfill
	\begin{subfigure}[b]{0.32\textwidth}
		\includegraphics[width=\textwidth]{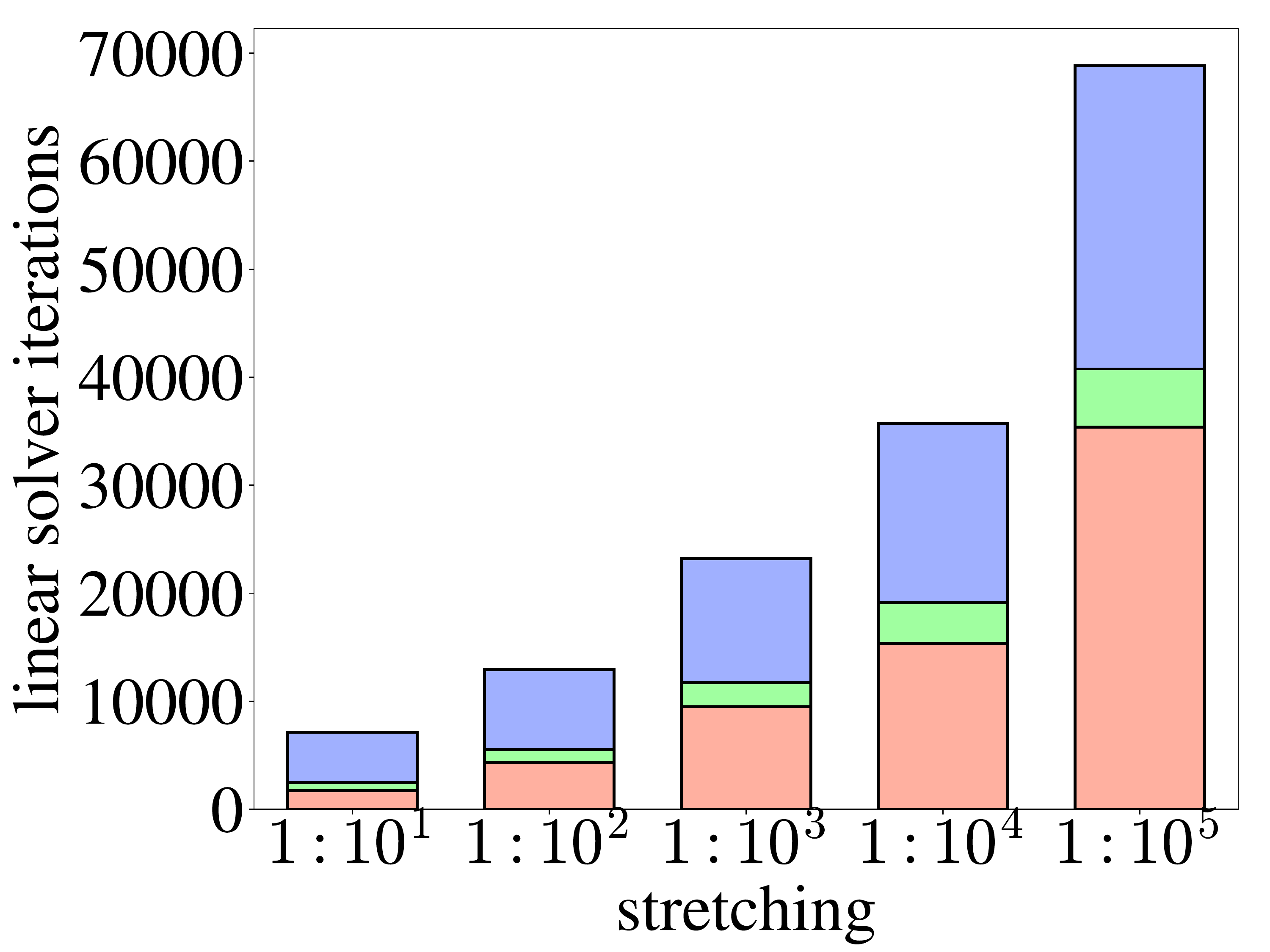}
		\caption{}
		\label{fig:scalingBL_MF_IterationsScaling}
	\end{subfigure}
	\hfill\hspace{0cm}
	\\
	\hfill
	\begin{subfigure}[b]{0.32\textwidth}
		\includegraphics[width=\textwidth]{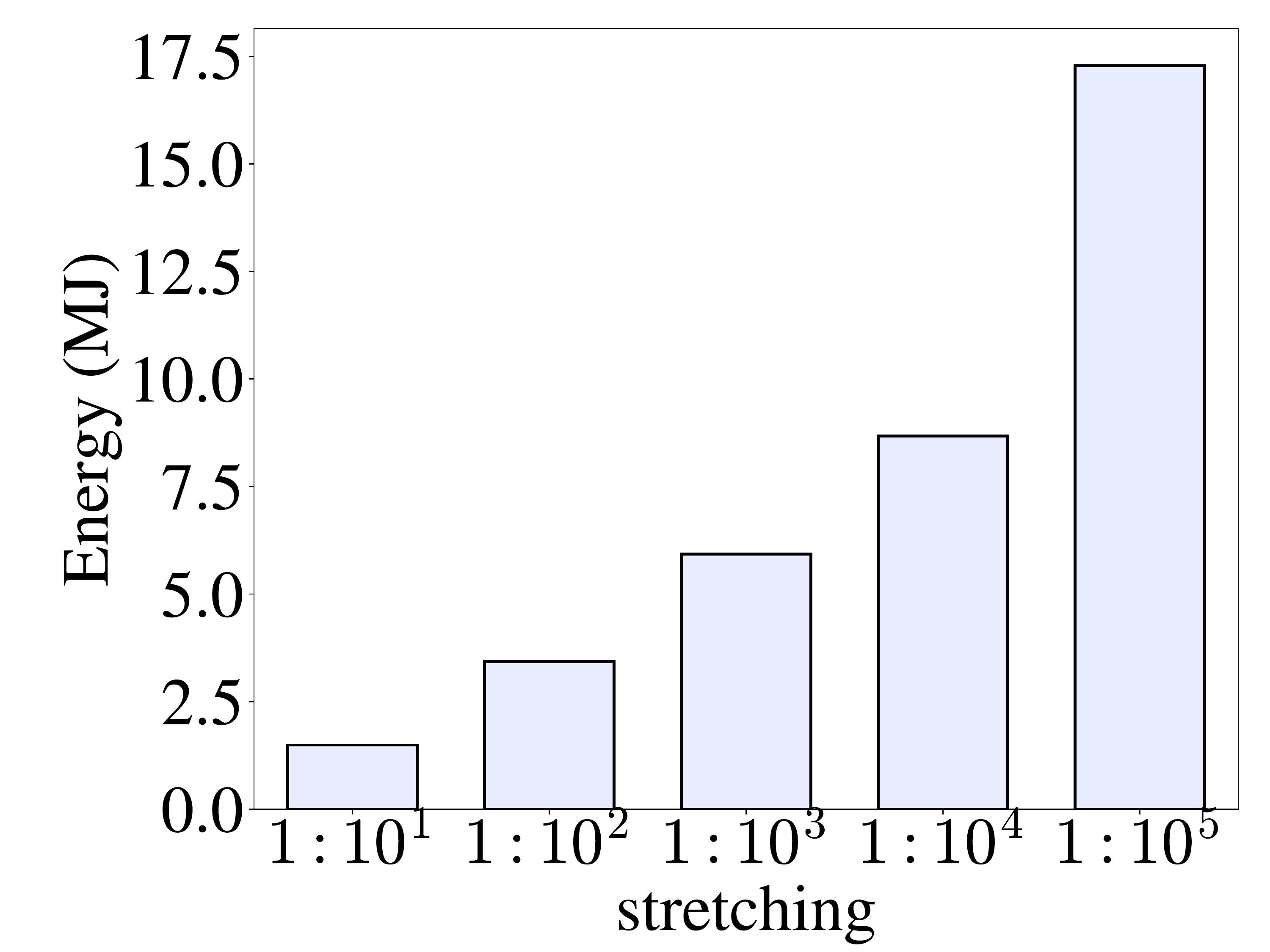}
		\caption{}
		\label{fig:scalingBL_NMF_Energy}
	\end{subfigure}
	\hfill
	\begin{subfigure}[b]{0.32\textwidth}
		\includegraphics[width=\textwidth]{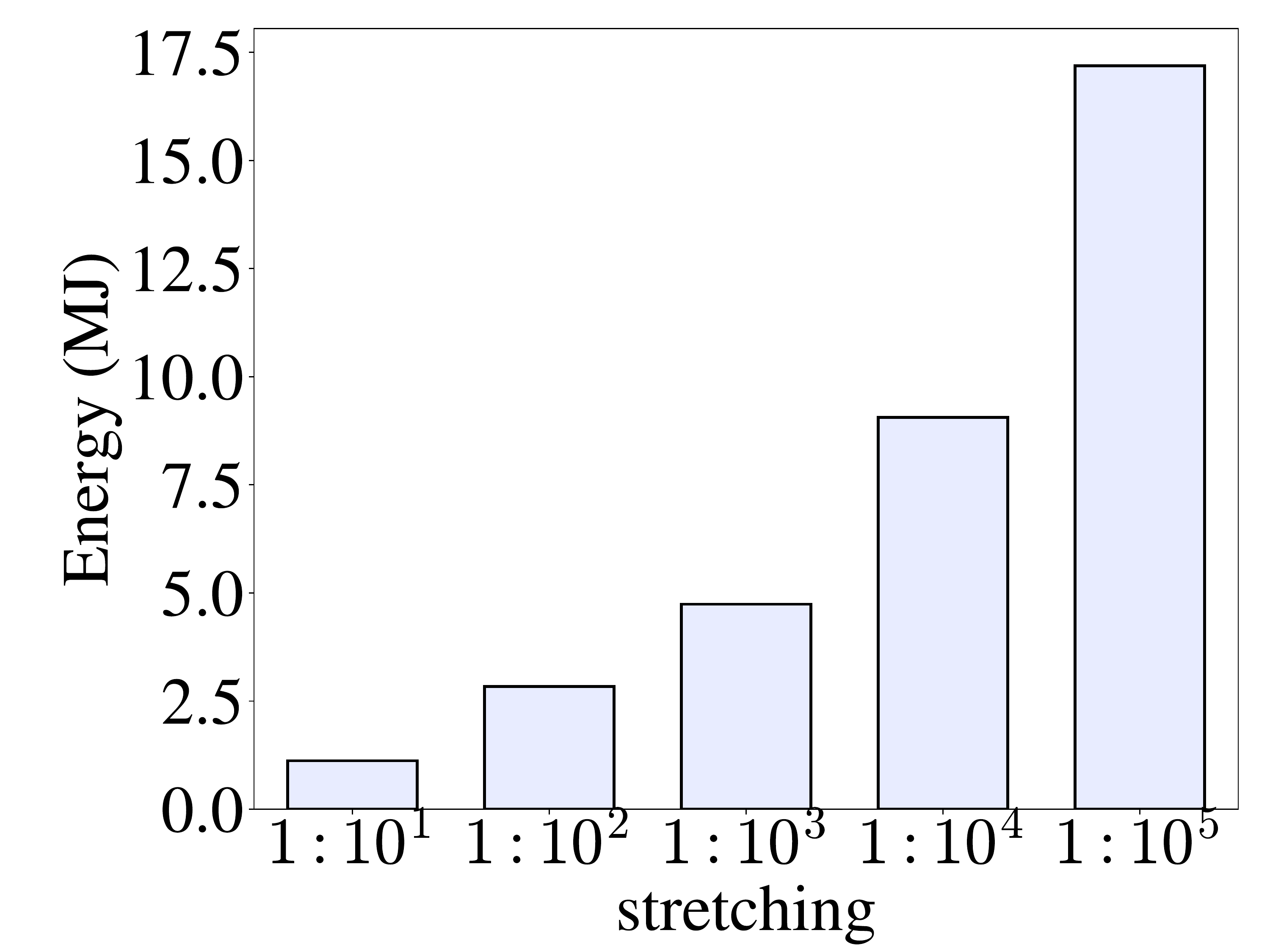}
		\caption{}
		\label{fig:scalingBL_MF_Energy}
	\end{subfigure}
	\hfill\hspace{0cm}
	\\
	\hfill
	\begin{subfigure}[b]{0.32\textwidth}
		\includegraphics[width=\textwidth]{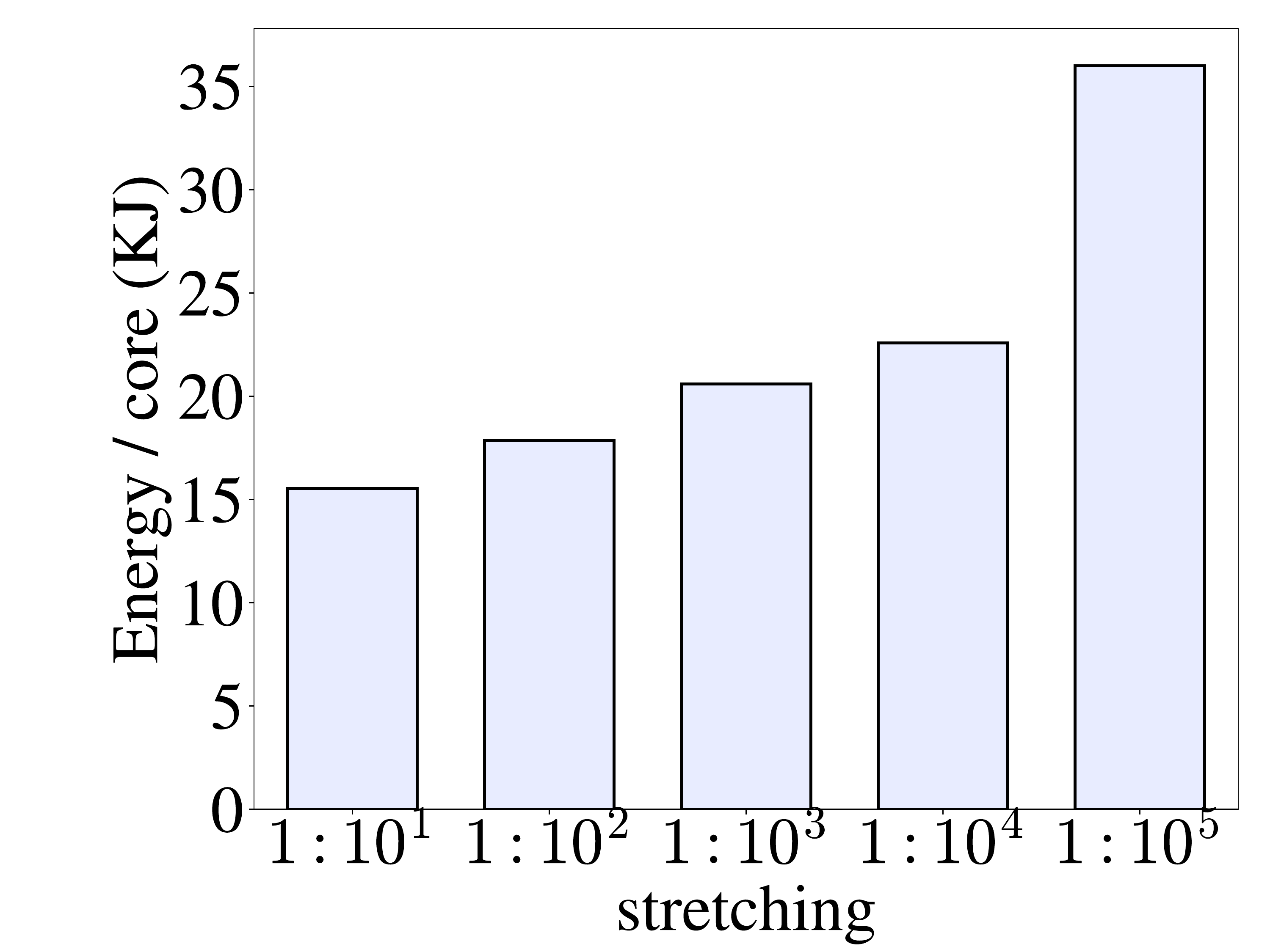}
		\caption{}
		\label{fig:scalingBL_NMF_EnergyPerCore}
	\end{subfigure}
	\hfill
	\begin{subfigure}[b]{0.32\textwidth}
		\includegraphics[width=\textwidth]{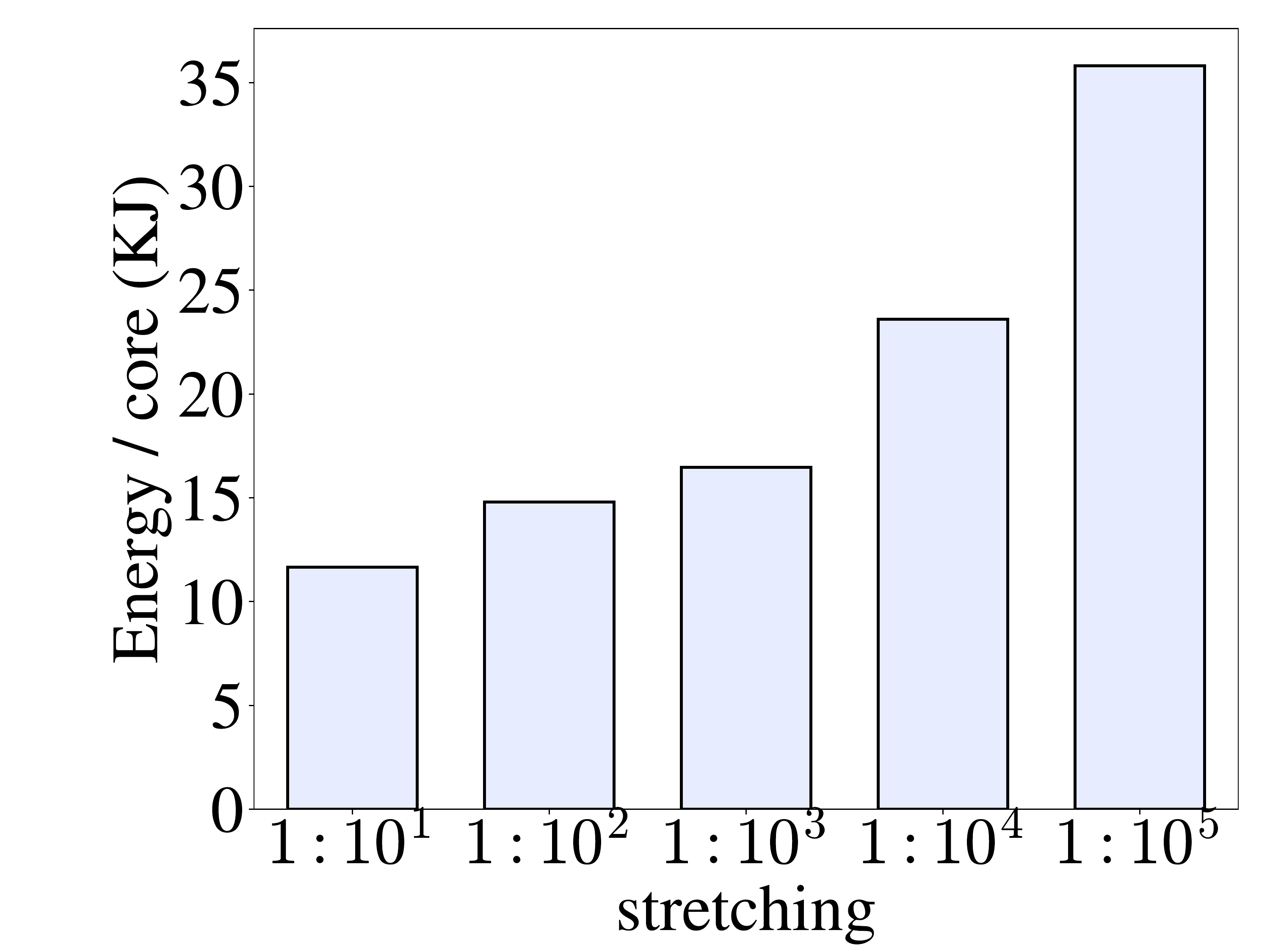}
		\caption{}
		\label{fig:scalingBL_MF_EnergyPerCore}
	\end{subfigure}
	\hfill\hspace{0cm}
	\caption{In columns, different pre-conditioners: 
		(a), (c), (e) and (g) ASDD(1)/SOR(2);
		(b), (d), (f) and (h) block-SOR(1)/ASDD(1)/SOR(2).
		In rows, (a) and (b) linear solver time;
		(c) and (d) linear solver iterations; 
		(e) and (f) total energy consumption; and
		(g) and (h) energy consumption per core.}
	\label{fig:spheresBL_Scalings}
\end{figure*}

\subsection{Influence of the boundary layer stretching}

In this example, we analyze the effect of the element stretching on the curving process. The domain is a sphere of radius four with a spherical hole in the center of radius one. We generate five meshes increasing the stretching and the number of layers of a boundary layer generated around the inner sphere. In all the meshes, the boundary mesh is the same. The growth factor of all the boundary layers is $10^{0.1} \simeq 1.259$, that ensures that every ten layers, the width of the layer is multiplied by ten. The maximum stretching of each mesh is $1:10^{1}$, $1:10^{2}$, $1:10^{3}$, $1:10^{4}$, $1:10^{5}$. The number of layers in the boundary layer has been chosen to obtain roughly 1500 elements per processor, and we have used 96, 192, 288, 384 and 480 cores. Specifically, the number of layers of each mesh is $10$, $23$, $37$, $47$ and $57$. This leads to meshes that are composed of $135\cdot10^3$, $291\cdot10^3$, $460\cdot10^3$, $581\cdot10^3$ and $702\cdot10^3$ elements. Figures \ref{fig:blScalingLevel1} to \ref{fig:blScalingLevel5} show the five curved high-order meshes of this example. The obtained meshes have converged with residual norms in the range of $1.43 \cdot 10^{-9}$ and $1.97 \cdot 10^{-9}$.

Figure \ref{fig:spheresBLConstraint} show the evolution of the constraint norm over the non-linear iterations for the five cases. In this example, the boundary mesh is the same for all the cases and therefore, the evolution for the five meshes is practically the same. That is, the proposed solver exhibits mesh independence at the non-linear level. When performing the mesh curving process for the polynomial degrees two and three, the constraint norm decreases geometrically with the non-linear iterations. This is especially important during the first iterations of each polynomial degree in the $p$-continuation technique. The main reason is that we predict a \emph{correct} value of the penalty parameter. Thus, the non-linear solver can perform the continuation of the solution when increasing the polynomial degree. When performing the curving of the polynomial degree four, we adapt the penalty parameter in order to accelerate the convergence of the non-linear penalty method. The norm of the constraint decreases more rapidly as the penalty method advances. This results in a decrease of the constraint norm of seven orders of magnitude in three iterations, and a reduction of the number of non-linear problems to be solved.

Figure \ref{fig:spheresBL_Scalings} shows the time to solve the linear problems, the total number of linear solver iterations, the energy consumption, and the energy consumption per core for the classical sparse matrix linear solver and the matrix-free linear solver. In both cases, the time to solve the linear systems becomes larger as the boundary layer stretching increases. The problem becomes more difficult to solve because the high-stretched elements increase the condition number of the linear systems. That is, the number of linear solver iterations increases with the boundary layer stretching. Similarly as in the previous example, the mesh curving process spends most of the linear solver iterations in the curving of the quadratic and cubic meshes.

Note that the number of linear iterations when using the proposed matrix-free linear solver is higher than when using the sparse matrix solver. Nevertheless, since the matrices involved in the proposed pre-conditioner are smaller than the matrices of the classical sparse matrix solver, the time to assemble and solve the linear systems and the energy consumption is similar in both cases. In addition, the proposed pre-conditioner for the matrix-free solver reduces the memory footprint by a factor of three.


This example shows that the proposed matrix-free linear solver combined with the block-SOR pre-conditioner are competitive in terms of wall-clock time and energy consumption for meshes with high-stretched elements, wile also reducing the memory footprint by a factor of three. The evolution of the constraint norm shows that the prediction of the penalty parameter allows reducing the number of non-linear problems to be solved and therefore, increase the efficiency of the proposed mesh curving algorithm.

\begin{figure*}[t!]
	\centering
	\begin{subfigure}[b]{0.66\textwidth}
		\includegraphics[width=\textwidth]{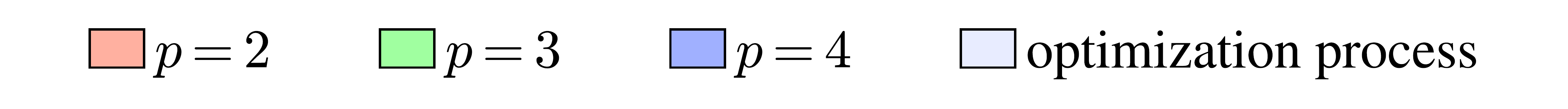}
	\end{subfigure}
	\\
	\hfill
	\begin{subfigure}[b]{0.32\textwidth}
		\includegraphics[width=\textwidth]{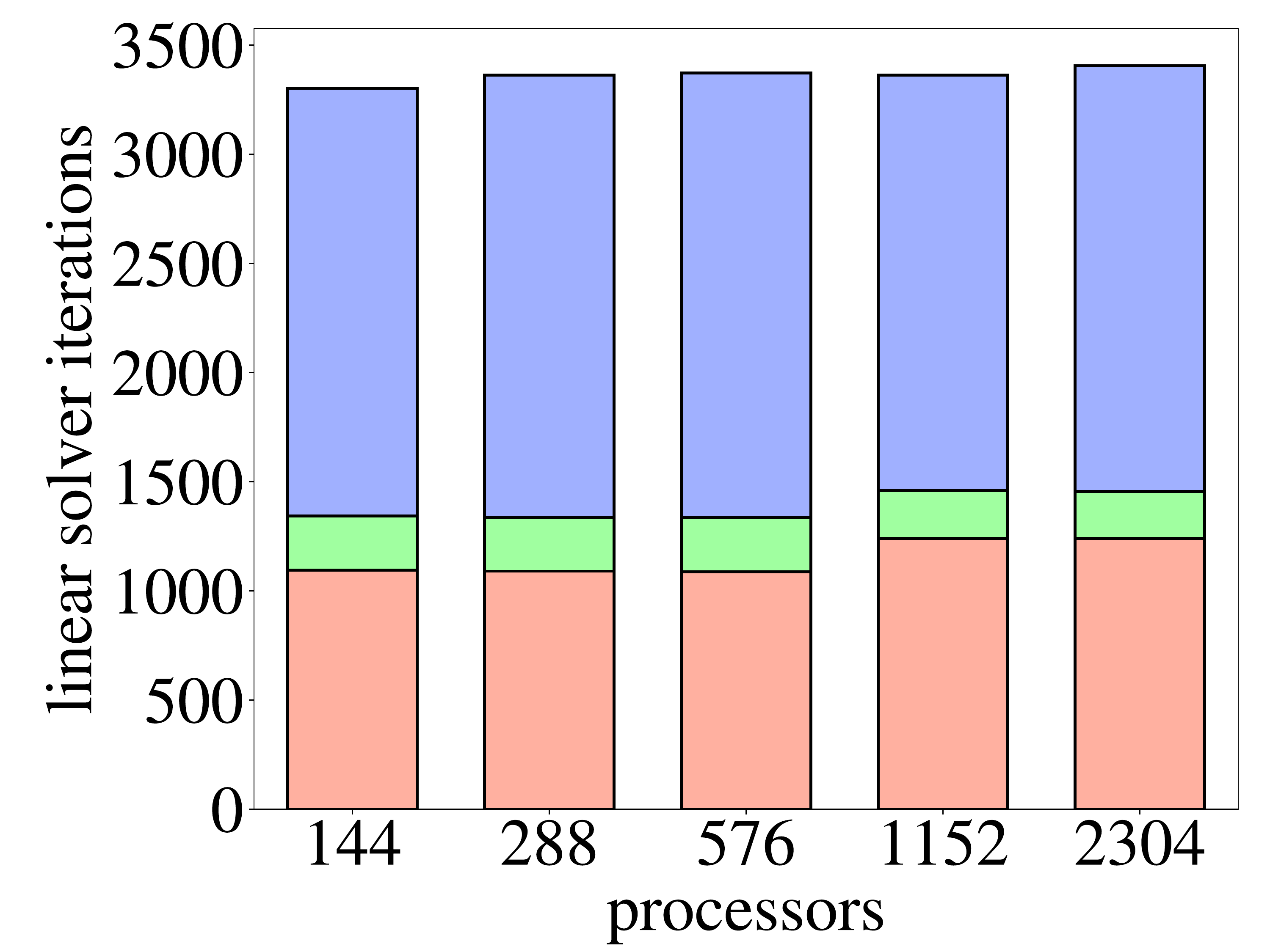}
		\caption{}
		\label{fig:StrongScaling_IterationsScaling}
	\end{subfigure}
	\hfill
	\begin{subfigure}[b]{0.32\textwidth}
		\includegraphics[width=\textwidth]{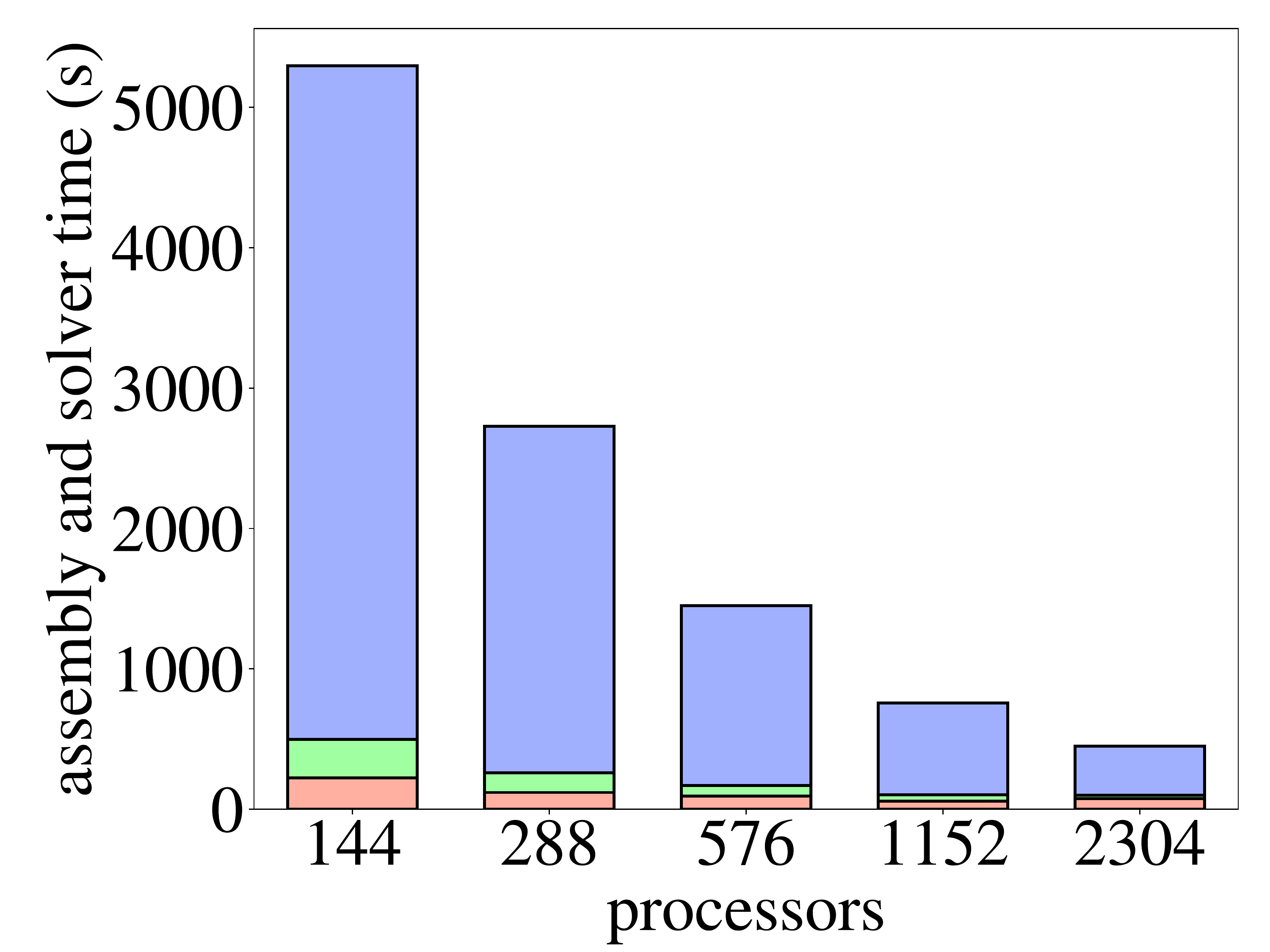}
		\caption{}
		\label{fig:StrongScaling_TimeScaling}
	\end{subfigure}
	\hfill\hspace{0cm}
	\\
	\hfill
	\begin{subfigure}[b]{0.32\textwidth}
		\includegraphics[width=\textwidth]{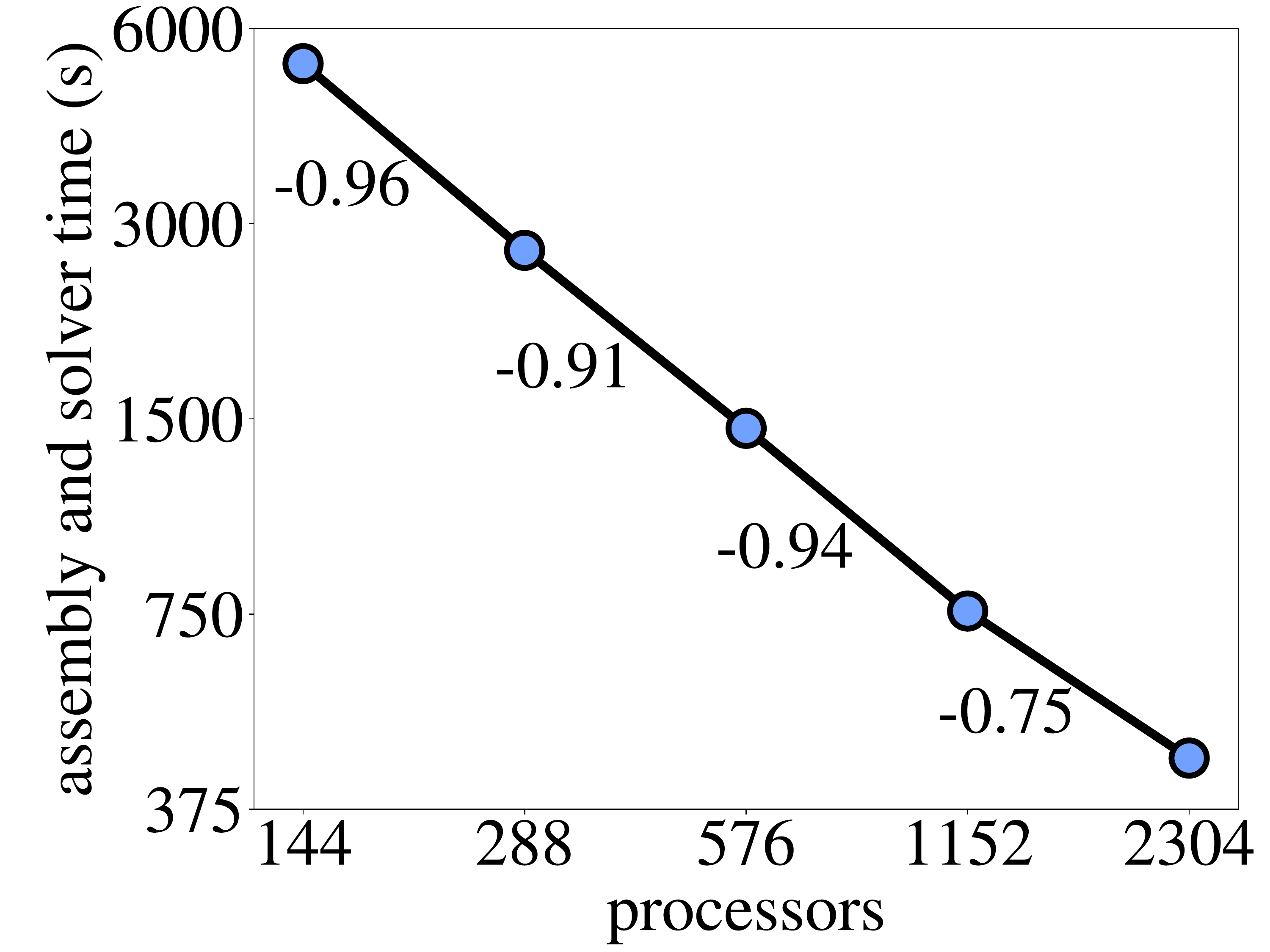}
		\caption{}
		\label{fig:StrongScaling_SpeedUp}
	\end{subfigure}
	\hfill
	\begin{subfigure}[b]{0.32\textwidth}
		\includegraphics[width=\textwidth]{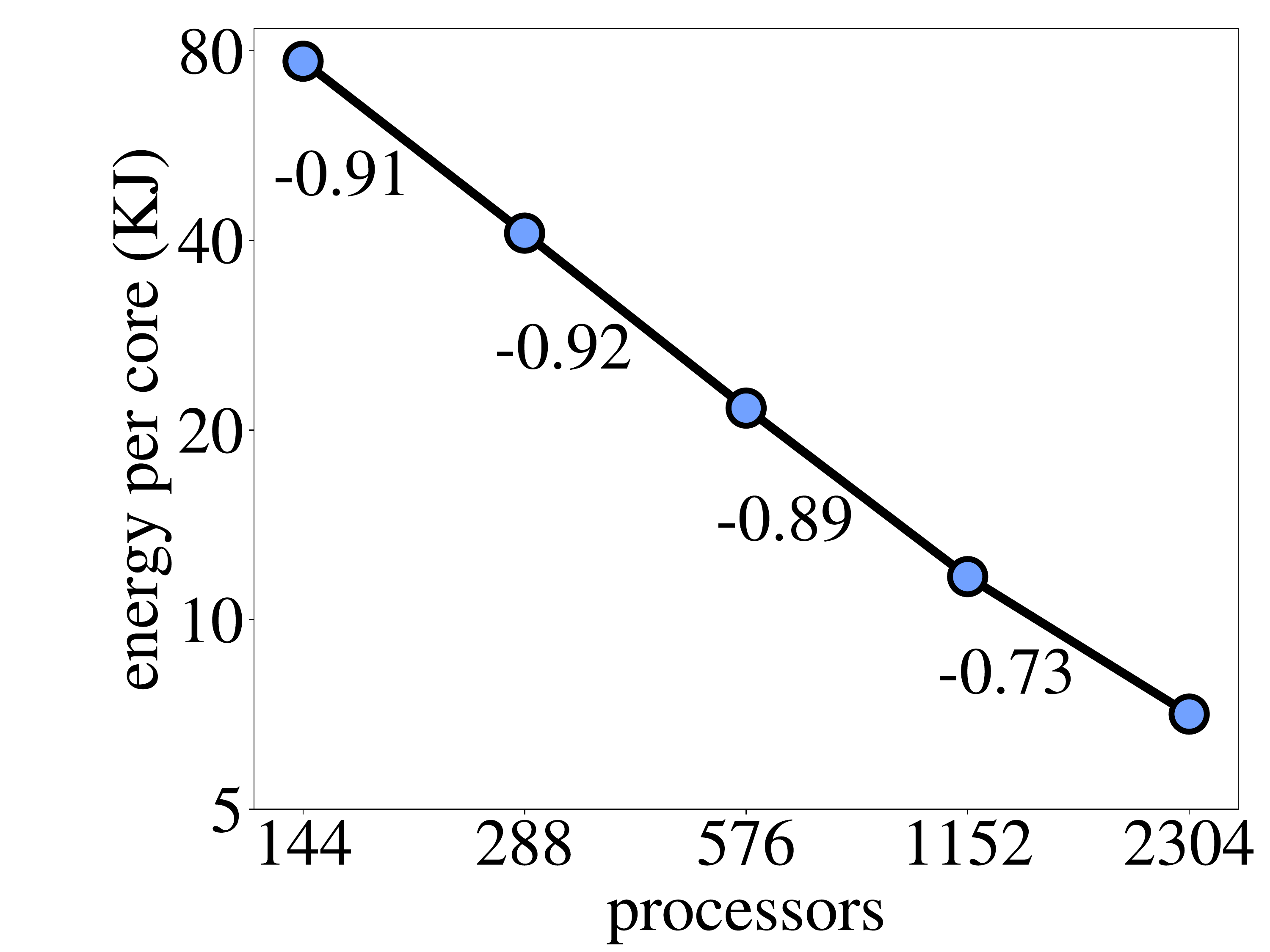}
		\caption{}
		\label{fig:StrongScaling_EnergyPerCore}
	\end{subfigure}
	\hfill
	\begin{subfigure}[b]{0.32\textwidth}
		\includegraphics[width=\textwidth]{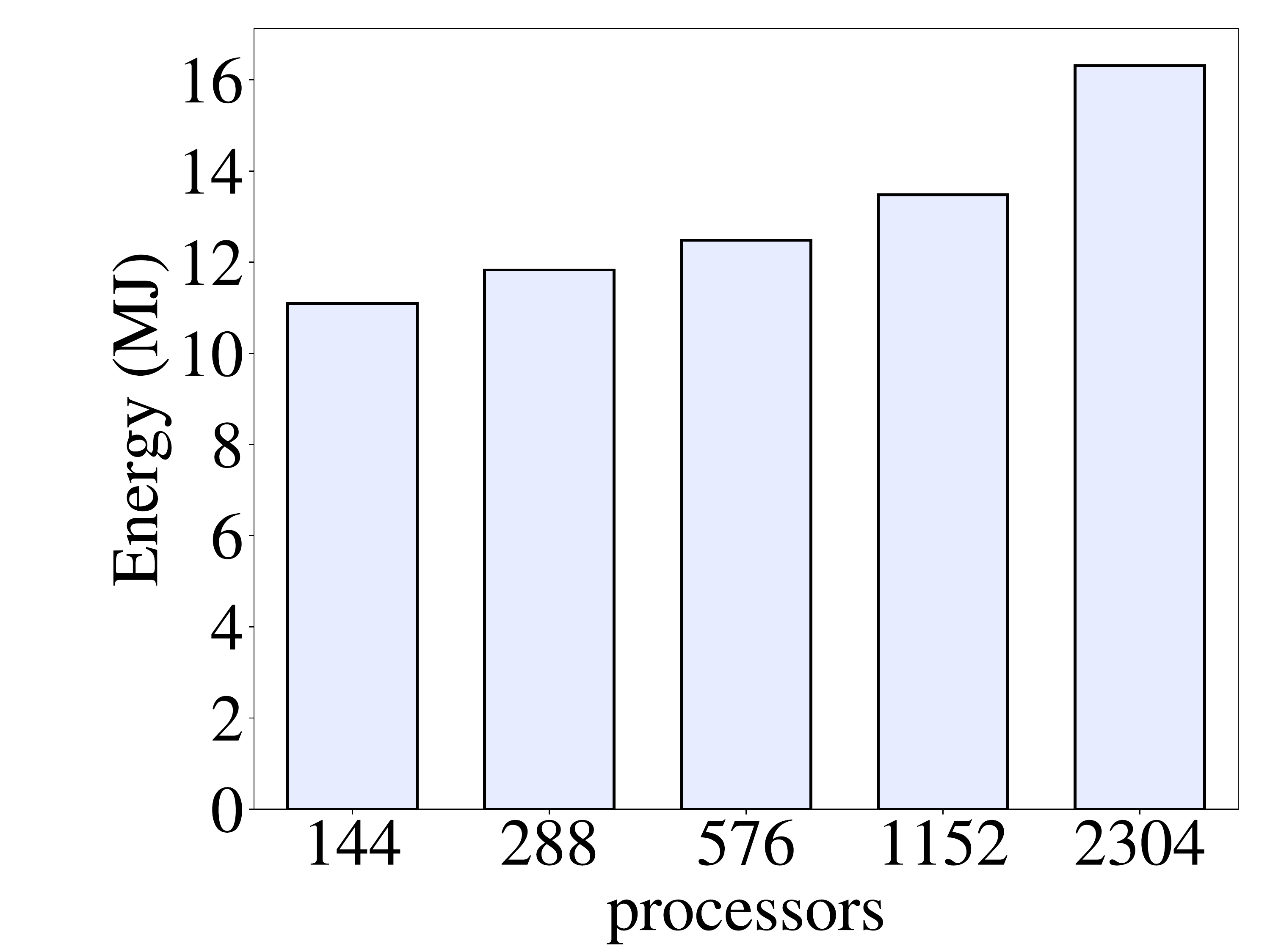}
		\caption{}
		\label{fig:StrongScaling_Energy}
	\end{subfigure}
	\hfill\hspace{0cm}
	\caption{Influence of the number of elements per processor:
		(a) linear solver iterations;
		(b) assembly and linear solver time;
		(c) assembly and linear solver time in logarithmic scale;
		(d) consumed energy per core in logarithmic scale; and
		(e) total consumed energy.}
	\label{fig:StrongScaling}
\end{figure*}

\subsection{Influence of the number of elements per processor}



In this example, we analyze the influence of the number of elements per processor to generate an isotropic mesh. To this end, we generate a mesh for a spherical domain of radius four with a spherical hole in the center of radius one. The mesh is composed of 1.44 million elements of polynomial degree four, and we perform the curving process with a different number of processors.
Specifically, we use 144, 288, 576, 1152, and 2304 processors, which leads to 10000, 5000, 2500, 1250, and 625 elements per processor, respectively.

The total number of linear iterations is similar in all the cases, see Figure \ref{fig:StrongScaling_IterationsScaling}. Moreover, the number of linear iterations to converge each polynomial degree is also similar. Thus, for the selected range of elements per processor, the increase in the communication load between processors does not hamper the efficiency of the proposed pre-conditioner.

As we increase the number of processors, the time to solve and assemble the linear systems is reduced, see Figure \ref{fig:StrongScaling_TimeScaling}. In Figure \ref{fig:StrongScaling_SpeedUp}, we plot the time to assemble and solve the linear systems against the number of processors on a logarithmic scale. The curve shows nearly an ideal strong scaling up until 1152 processors since the slopes of the linear segments are near $-1.0$. When performing the curving process with 2304 processors, the additional time spent communicating between processors is not negligible compared to the total curving time.

The consumed energy per core decreases since, as we increase the number of processors, each processor is working less time, see Figure \ref{fig:StrongScaling_EnergyPerCore}. The energy per core also presents a nearly ideal strong scaling up to 1152 processors. When using 2304 processors, the additional communication load significantly increases the energy consumption.

The total energy consumption to generate the curved meshes increases when we perform the curving process with more processors. The main reason is that with additional processors, there is an additional communication load between processors. This additional load leads to an increase in the consumed energy. Nevertheless, between 144 and 1152 processors, the energy consumption is similar, and the values are between 11MJ and 13.5MJ.

In the range of 10000 and 1250 elements per processor, the results show that we nearly obtain an ideal strong scaling. When using 2304 processors, we do not obtain an ideal strong scaling. Nevertheless, the time to assemble and solve the linear systems is also reduced.

\begin{figure*}[t!]
	\centering
	
	\begin{subfigure}[b]{0.75\textwidth}
		\includegraphics[width=\textwidth]{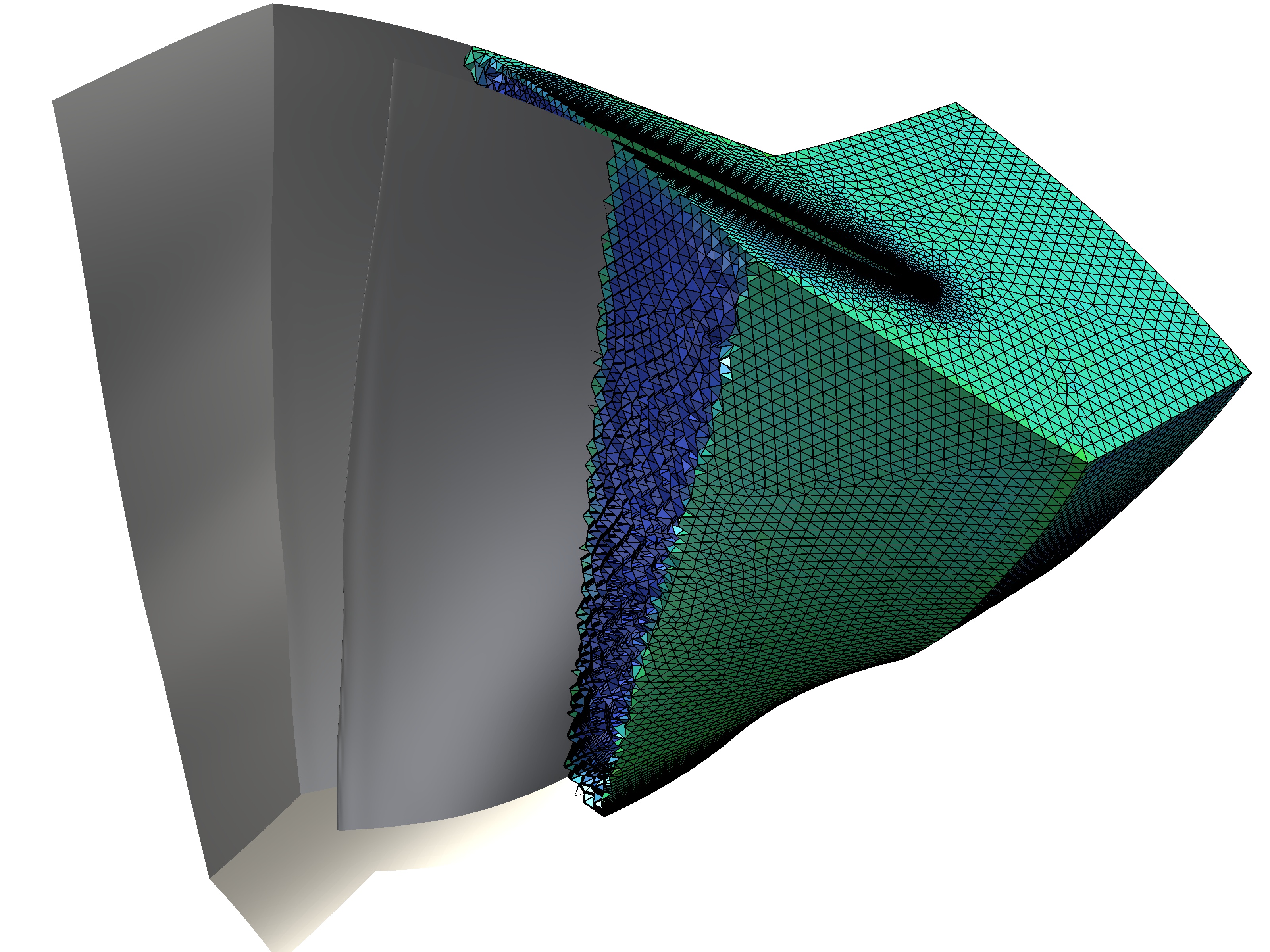}
		\caption{}
		\label{fig:rotor67_vCut}
	\end{subfigure}
	\\
	\begin{subfigure}[b]{0.75\textwidth}
		\includegraphics[width=\textwidth]{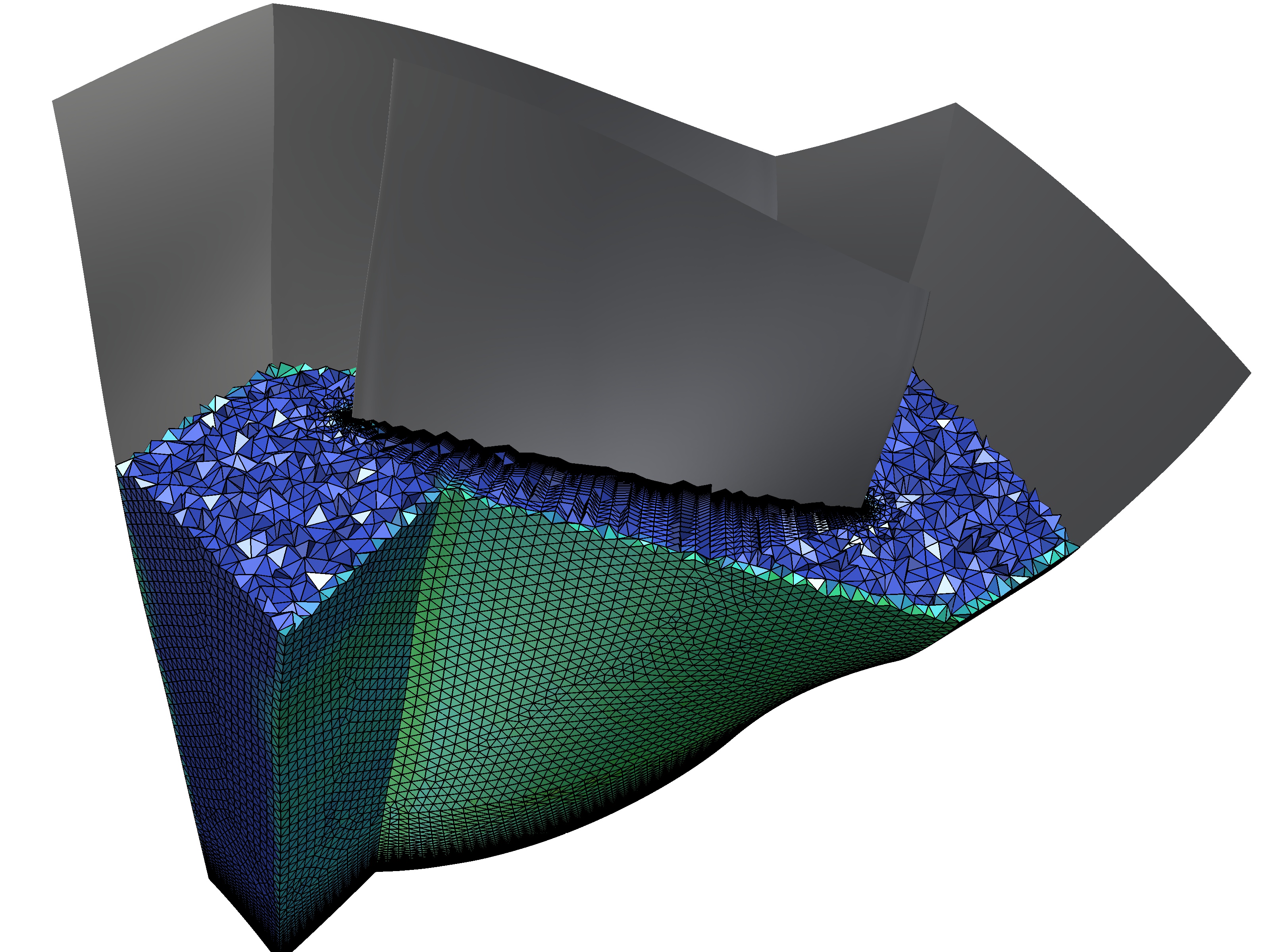}
		\caption{}
		\label{fig:rotor67_hCut}
	\end{subfigure}
	\begin{subfigure}[b]{0.75\textwidth}
		\includegraphics[width=\textwidth]{qualityLegend}
	\end{subfigure}
	\caption{Curved high order mesh of polynomial degree four for the NASA rotor 67:
		(a) vertical slice; and
		(b) horizontal slice.}
	\label{fig:rotor67}
\end{figure*}

\begin{figure*}[t!]
	\centering
	\hfill
	\begin{subfigure}[b]{0.45\textwidth}
		\includegraphics[width=\textwidth]{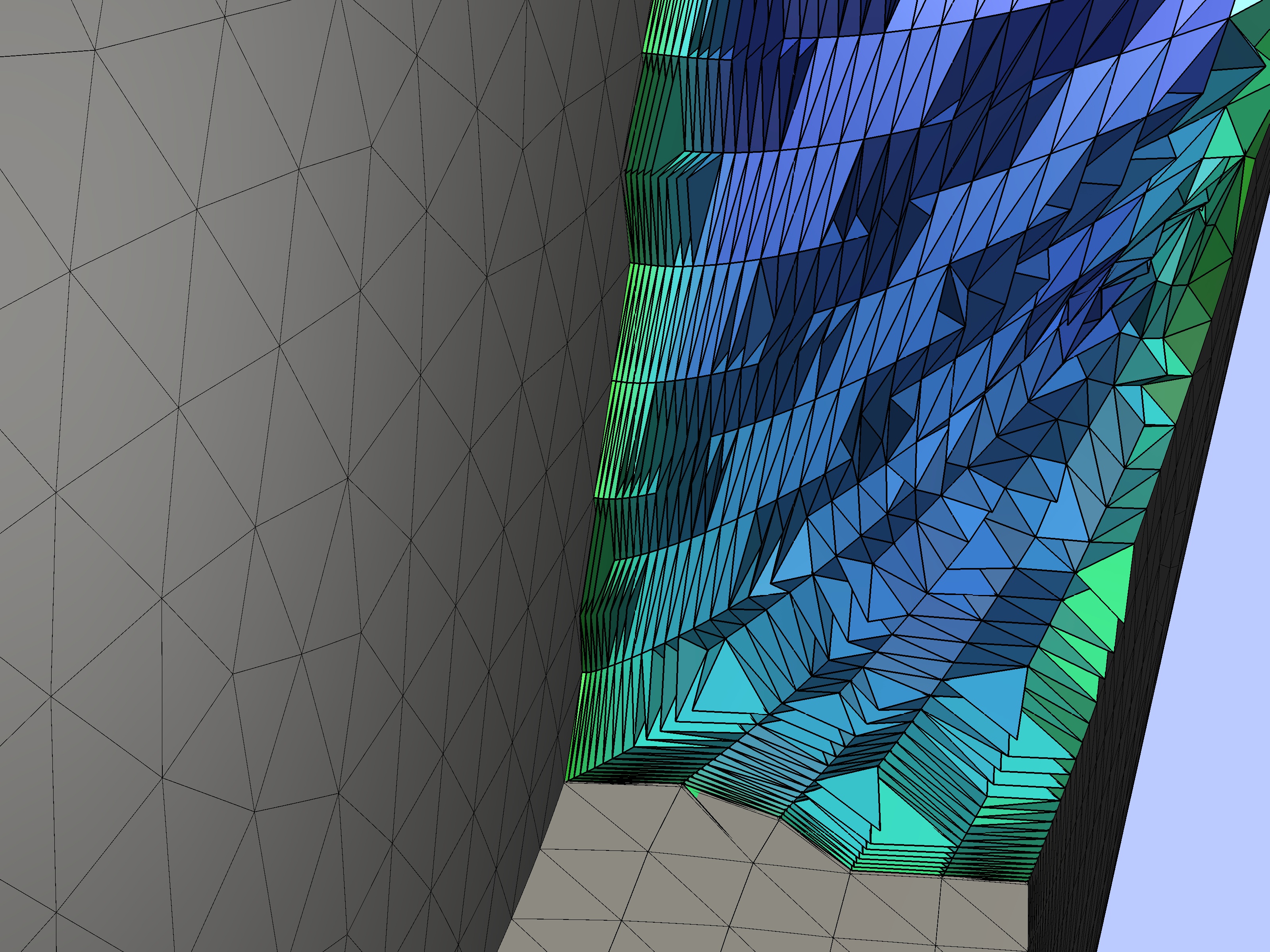}
		\caption{}
		\label{fig:rotor67_vCut_detail}
	\end{subfigure}
	\hfill
	\begin{subfigure}[b]{0.45\textwidth}
		\includegraphics[width=\textwidth]{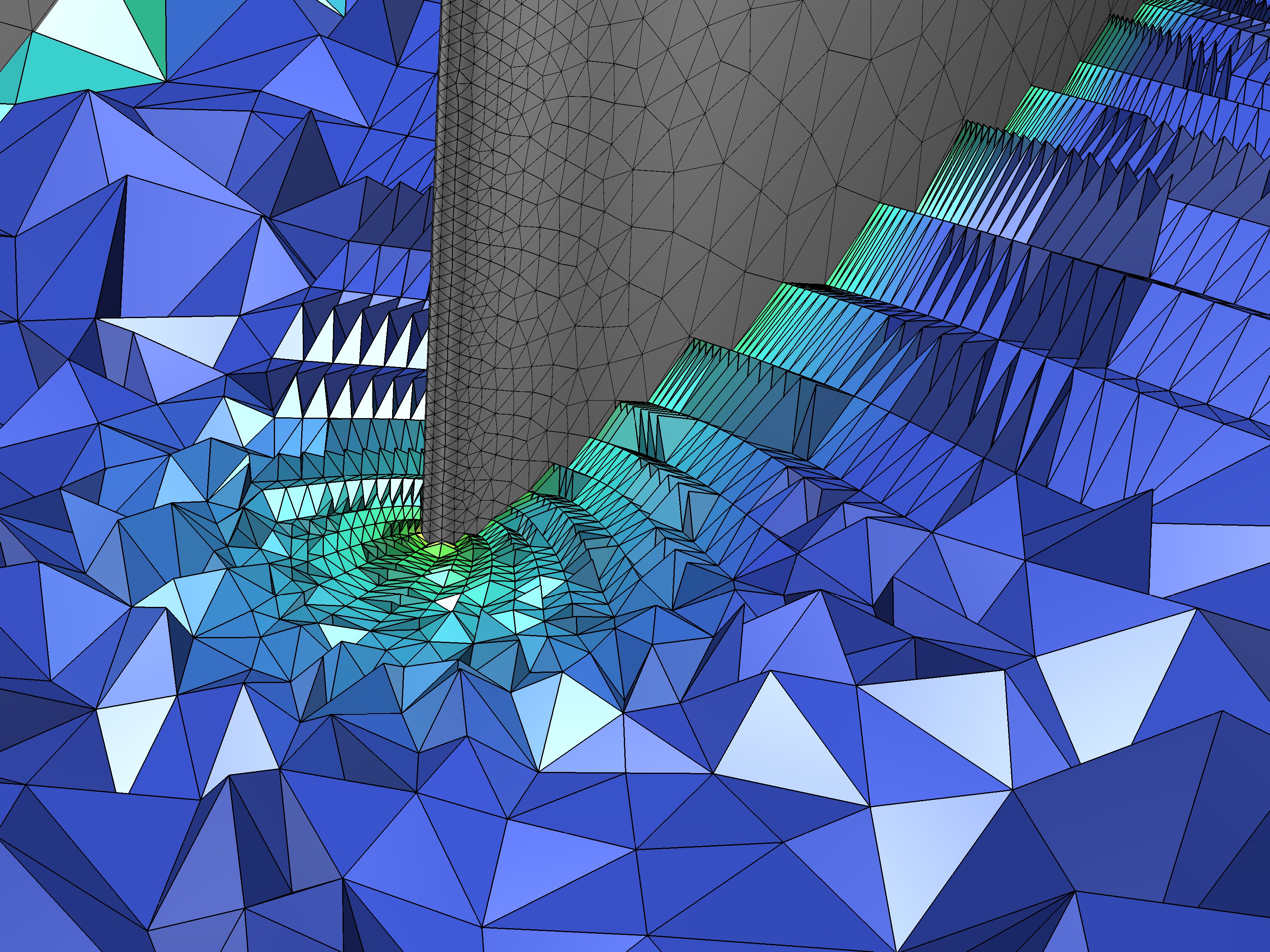}
		\caption{}
		\label{fig:rotor67_hCut_detail}
	\end{subfigure}
	\hfill\hspace{0cm}
	\\
	\begin{subfigure}[b]{0.75\textwidth}
		\includegraphics[width=\textwidth]{qualityLegend}
	\end{subfigure}
	\caption{Detail of the curved high order mesh of polynomial degree four for the NASA rotor 67:
		(a) at the blade-base joint; and
		(b) at the leading edge.}
	\label{fig:rotor67_detail}
\end{figure*}

\subsection{Periodic boundary condition: rotor 67}


This example shows an application of our mesh curving methodology to generate a periodic curved high-order mesh. To this end, we generate a mesh of polynomial degree four for the NASA rotor 67 model, a validation test case for turbomachinery CFD codes. The full rotor 67 is composed of 22 blades and, for efficiency purposes, we mesh a single blade. Therefore, we need to impose a periodic boundary condition that ensures that matching surfaces are meshed in the same manner. Thus, we need to generate an initial linear mesh in which the matching surfaces are meshed in the same manner.

To enforce the periodicity of the mesh, we impose that the non-linear boundary condition, $\vec g_D$ in Equation \eqref{eqn:constrained}, corresponds to a periodic condition. That is, when updating the boundary condition at each iteration of the fixed-point solver, we ensure that $\eval{\vec g_D}{\trace \mapB{}}$ results in a periodic configuration. When the curving process finishes, the boundary mesh, $\trace \mapB{}$, will be the same, up to the non-linear solver tolerance, as the boundary condition, $\eval{\vec g_D}{\trace \mapB{}}$, and therefore, the mesh will be periodic.

The main advantage of the proposed methodology is that we do not need to modify Newton's method to enforce a periodic update. Moreover, we do not need to modify the assembly of the linear system to impose the periodic boundary condition. That is, with the proposed approach to impose periodic boundary conditions, we only need to encode the mesh periodicity in the non-linear function $\vec g_D$.

To obtain matching surface meshes, we enforce that the nodes on source surfaces are mapped to nodes on the target surfaces. That is, to apply the periodic boundary condition, we obtain the nodes on target surfaces, $\vec x^t_i$, as mapped nodes on source surfaces $\vec x^s_i$ as:
\[
	\vec x^t_i = \eval{\varphi}{\vec x^s_i},
\]
being $\varphi$ the mapping that converts the source surfaces to the target surfaces. In the case of the rotor 67, the periodic mapping $\varphi$ corresponds to a rotation of $\pi/11$ radians along rotation axis of the rotor.

The final mesh for the rotor 67 is composed of 3.6 million elements of polynomial degree four, and contains a boundary layer of maximum stretching of $1:25$, see Figure \ref{fig:rotor67}. The minimum relative quality of the curved mesh is $0.987$. Figure \ref{fig:rotor67_vCut_detail} shows the boundary layer around the joint of the blade and the base, and Figure \ref{fig:rotor67_hCut_detail} shows a detail of the boundary layer around the leading edge. Although the convergence criterion in the residual norm is $10^{-8}$, the obtained mesh has converged with a residual norm of $4.63 \cdot 10^{-11}$.

To perform the optimization process, we have used 768 processors. Thus, each processors has around 4700 elements of polynomial degree four. The whole optimization process has taken 52 minutes to obtain the final mesh, and the process has used 31.5MJ of energy. We have used the matrix-free solver with the proposed block-SOR pre-conditioner, since the memory requirements of the sparse matrix linear solver do not allow to curve this mesh using 768 processors.

\subsection{Large meshes and complex geometry: common research model of the Third High-Lift Prediction Workshop}


\begin{figure*}[t!]
	\centering
	\begin{subfigure}[b]{0.95\textwidth}
		\includegraphics[width=\textwidth]{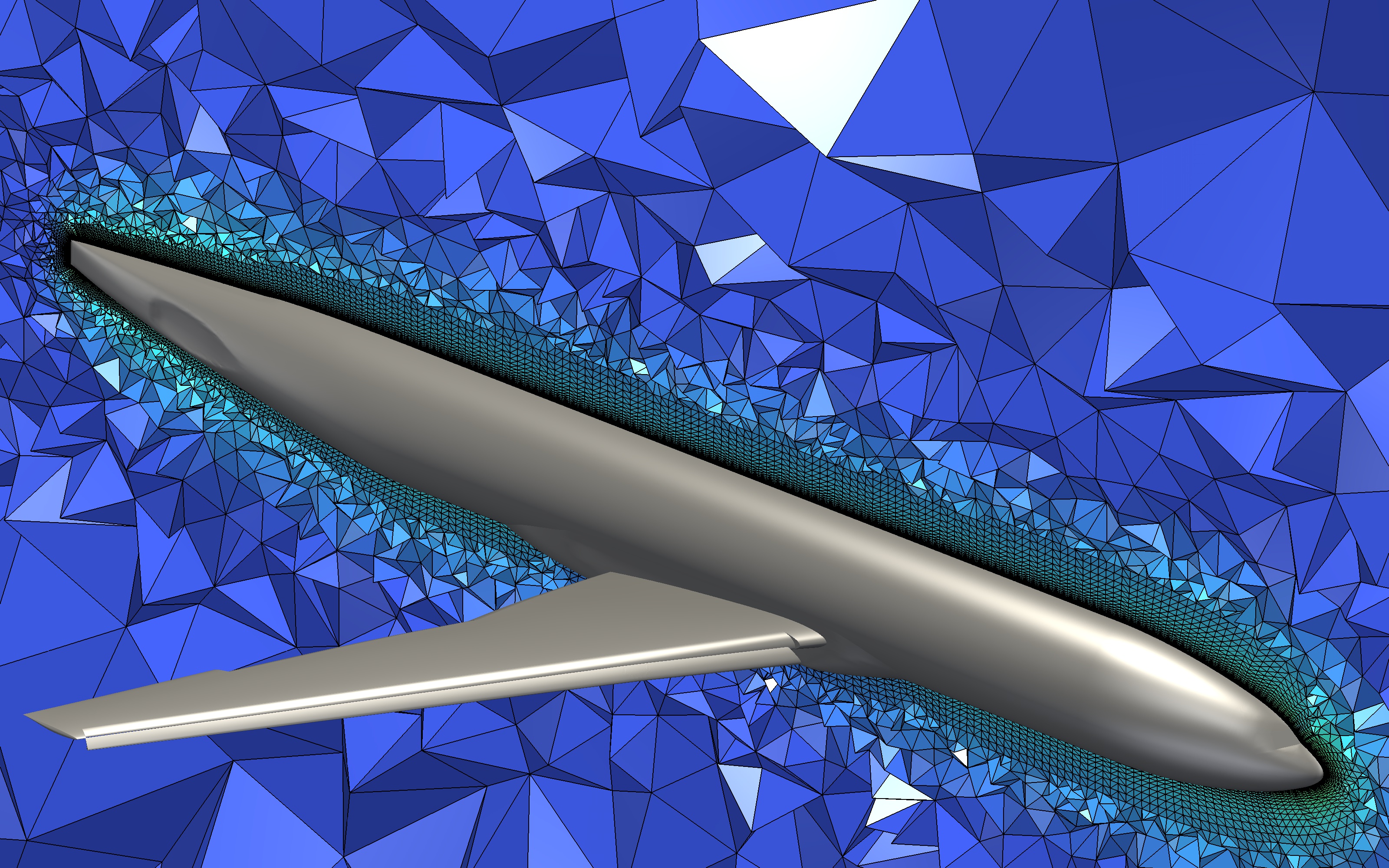}
		\caption{}
		\label{fig:HLPW3_sliceY}
	\end{subfigure}
	\\
	\begin{subfigure}[b]{0.95\textwidth}
		\includegraphics[width=\textwidth]{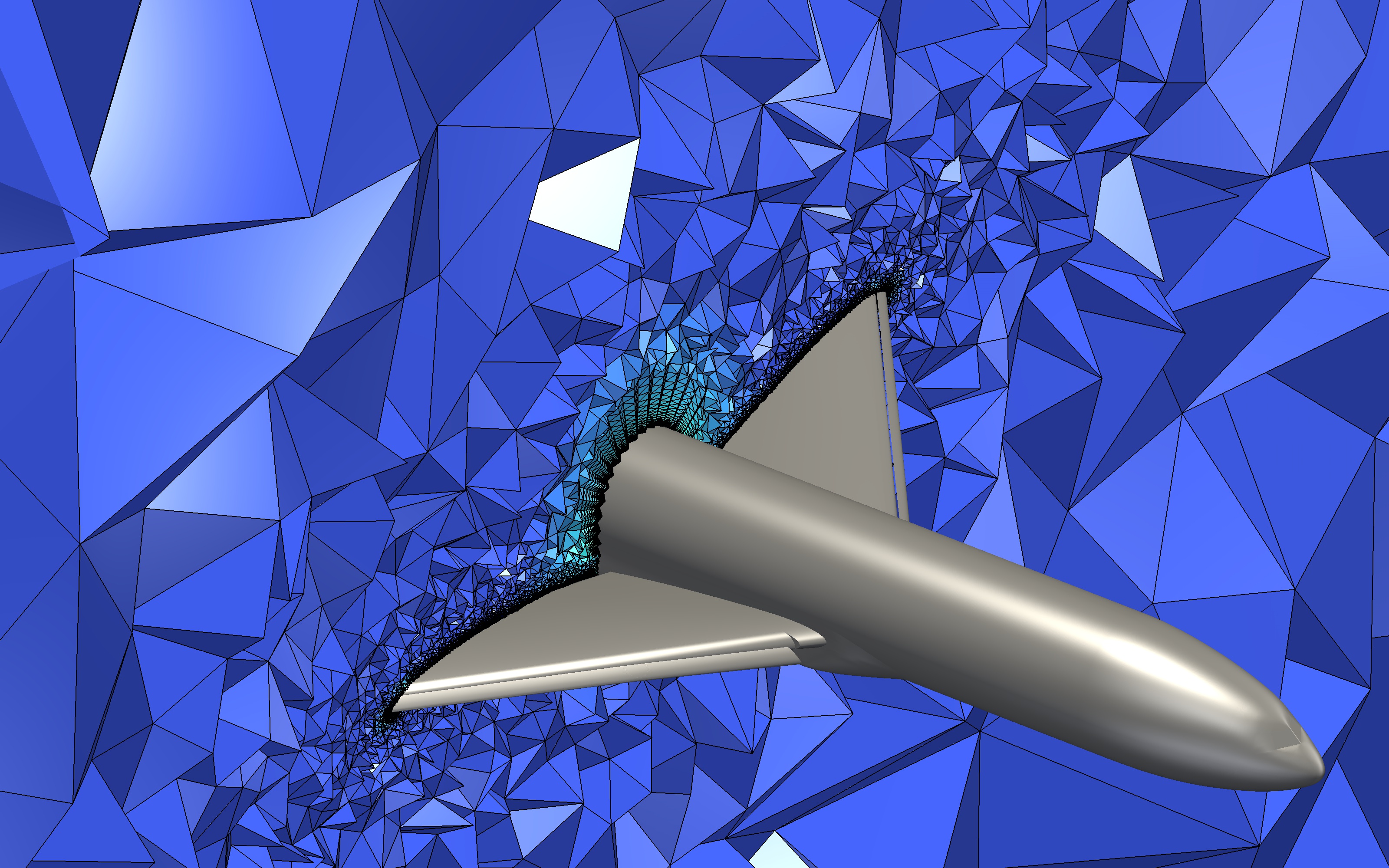}
		\caption{}
		\label{fig:HLPW3_sliceX}
	\end{subfigure}
	\begin{subfigure}[b]{0.75\textwidth}
		\includegraphics[width=\textwidth]{qualityLegend}
	\end{subfigure}
	\caption{Curved high order mesh of polynomial degree four for the common research model of the $3^{\text{rd}}$ High-Lift Prediction Workshop:
		(a) longitudinal cut; and
		(b) transversal cut.}
	\label{fig:HLPW3}
\end{figure*}

\begin{figure*}[t!]
	\centering
	\hfill
	\begin{subfigure}[b]{0.45\textwidth}
		\includegraphics[width=\textwidth]{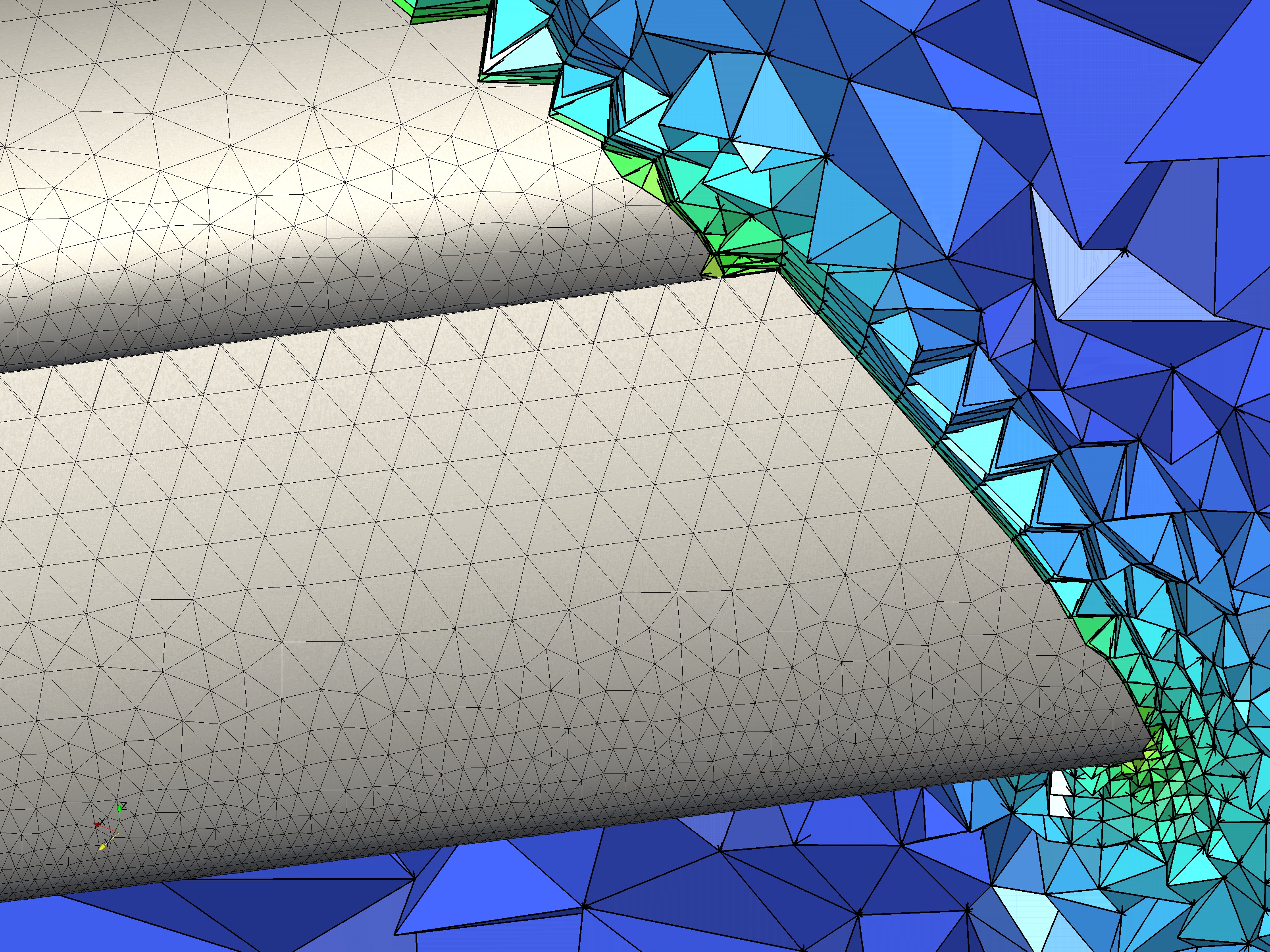}
		\caption{}
		\label{fig:HLPW3_slat}
	\end{subfigure}
	\hfill
	\begin{subfigure}[b]{0.45\textwidth}
		\includegraphics[width=\textwidth]{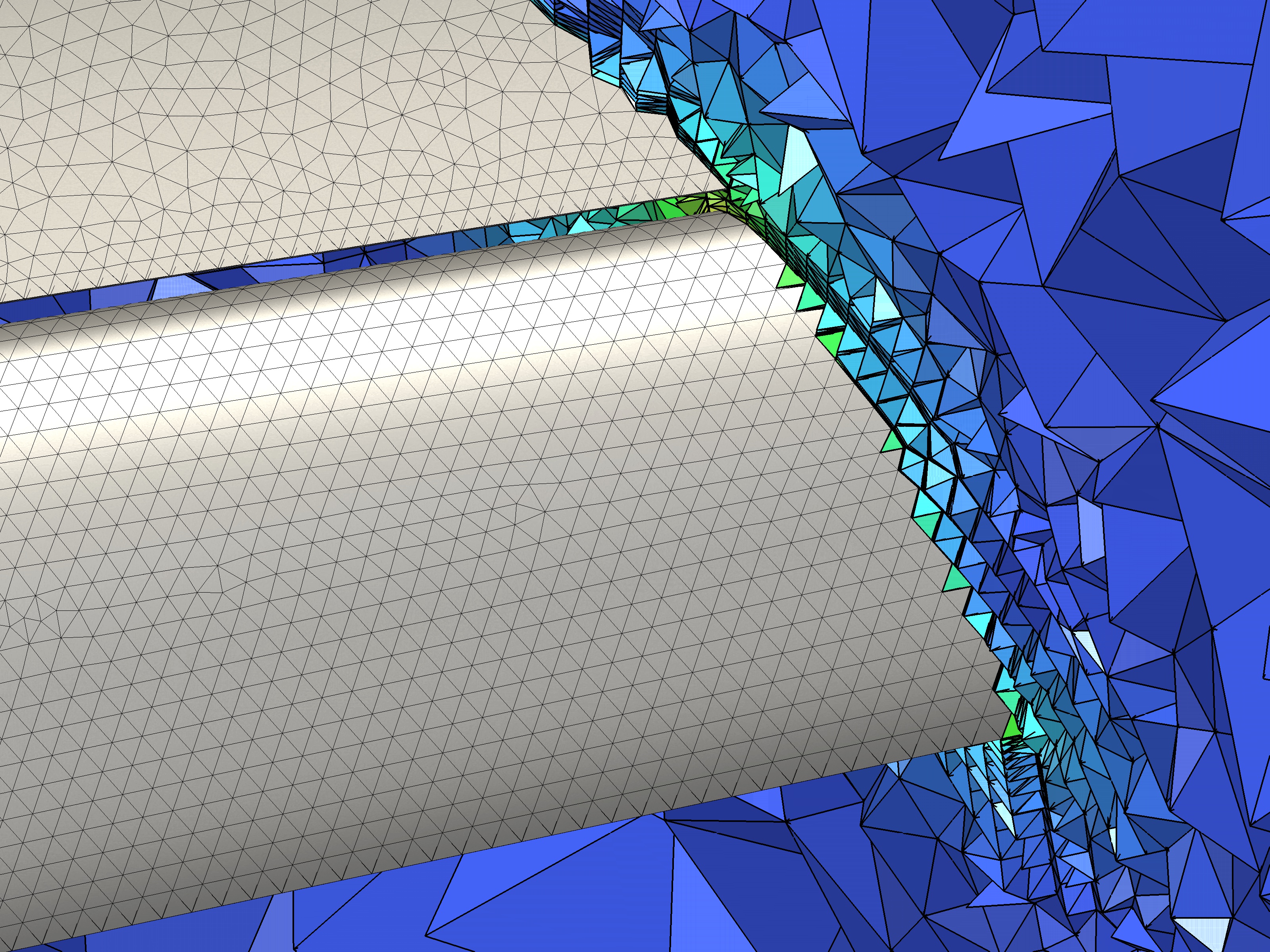}
		\caption{}
		\label{fig:HLPW3_flap}
	\end{subfigure}
	\hfill\hspace{0cm}
	\\
	\begin{subfigure}[b]{0.75\textwidth}
		\includegraphics[width=\textwidth]{qualityLegend}
	\end{subfigure}
	\caption{Detail of the aircraft wing:
	(a) at the slat; and
	(b) at the flap.}
	\label{fig:HLPW3_detail}
\end{figure*}

The objective of this example is to show the capabilities of the proposed formulation to deal with large-scale meshes generated for complex geometries. Thus, we generate a curved high-order mesh for the common research model presented in the $3^{\text{rd}}$ High-Lift Prediction Workshop \cite{rumsey2019:3HLPW}. We have generated a curved mesh composed of 12 million elements of polynomial degree four using 2400 processors. We have introduced a boundary layer of maximum stretching of $1:25$. The minimum relative quality is 0.885, and the optimization process has taken 135 minutes. The energy consumption to generate the curved high-order mesh is 177MJ. Although the convergence criterion in the residual norm is $10^{-8}$, the obtained mesh has converged with a residual norm at the last iteration of $5.25 \cdot 10^{-15}$. That is, all the entries of the residual vector in absolute value are less or equal than $5.25 \cdot 10^{-15}$, which means that we have obtained a fully converged mesh in the whole domain.

Figure \ref{fig:HLPW3} shows two global views of the curved high-order mesh, while Figures \ref{fig:HLPW3_slat} and \ref{fig:HLPW3_flap} show detailed views of the mesh of the aircraft slat and flap, respectively. Note that just the first layers close to the aircraft are curved and that the majority of the mesh contains almost straight-edged elements.

With the proposed matrix-free linear solver, we have been able to generate larger meshes without increasing the computational resources. The reduced memory footprint is the main advantage of the proposed block-SOR pre-conditioner. Specifically, with the proposed approach, each core has 5000 elements of polynomial degree four. Using the classical sparse matrix linear solver, it is not possible to store the sparse Hessian matrix using 2400 processors. Therefore, we would need to increase the memory per core, or the number of processors to be able to run this example using the sparse matrix linear solver.

%
%
%
%
%

\section{Discussion}
\label{sec:discussion}

After detailing the methods and results to generate curved meshes, we present a discussion in many aspects related to the nodes re-projection, mesh independence, the $p$-continuation, the penalty parameter, the forcing term, the mesh optimization, the distributed parallel solver, and the CAD-based mesh generation. We finalize this section with the future work.

\subsection{Re-projection of the nodes}

\newtext{At each iteration of the penalty method, we compute a new target boundary configuration. To compute this configuration, we project the boundary nodes of the current mesh onto the target CAD. This re-projection allows avoiding the computation of the projection derivatives needed in Newton’s method. To this end, we fix this boundary condition for the whole penalty iteration, and then we proceed to solve the corresponding minimization problem. }

\newtext{Although the target boundary configuration may be invalid, our method enforces a valid volume mesh. During the iterations of the penalty method, the nodal projections may lead to a target boundary configuration that is invalid. In our applications, we have seen that this issue is more probable in the first iterations when the mesh boundary is far from the target geometry. On the contrary, we have seen that as the curving process evolves, the mesh boundary is closer to the target geometry, and thus, the nodal projections lead to valid target boundary configurations. Nevertheless, our method is devised to enforce the boundary condition as much as possible without inverting the volume elements.}

\subsection{Mesh independence}

The second and third examples show that the presented formulation shows mesh independence at the non-linear level. In each example, the evolution of the curving process is almost the same. Specifically, for sufficiently fine meshes, the number of iterations of the penalty method is the same, and the value of the constraint norm at each iteration is practically the same. Moreover, in the example of the boundary layer stretching, we show that the evolution of the curving process is not dependent on the element stretching, up to an element stretching of $1:10^5$.

Nevertheless, this is not the case when solving the obtained linear systems. As the number of elements increase and the elements become more stretched, the linear systems are harder to solve because the condition number of the matrices increases. It is especially important to highlight the difficulty increase when the elements become more stretched. Thus, the main difficulty of the mesh curving process is solving the linear systems, especially when stretched elements appear in the mesh.

\subsection{$p$-continuation}

We have proposed a $p$-continuation technique to reduce the computational resources of the mesh curving solver while increasing its robustness. The main idea is to use the solution of a polynomial degree as a starting position for the next polynomial degree. Thus, the $p$-continuation technique enriches the solution starting from the function space of lower polynomial degree, up to the function space of highest polynomial degree. Since the function spaces are nested, we only need to compute a \emph{correction} term to obtain the optimal mesh in each space.

Alternatively, the $p$-continuation technique can be interpreted as a methodology to compute an initial configuration of the mesh for the last polynomial degree. Since the initial position of the nodes for the last polynomial degree are close to the optimal position, the computational time to optimize the mesh is reduced and the robustness of the optimization process is increased.

\subsection{Penalty Parameter}

The penalty method can be interpreted as a continuation method that continuously depends on the penalty parameter to enforce the boundary condition. As the penalty parameter tends to infinity, the boundary condition is fully enforced and we obtain the final mesh. In practical applications, the penalty parameter has to be large enough to enforce the boundary condition up to a tolerance threshold.

The penalty parameter plays an important role in the robustness and efficiency of the mesh curving process. In the proposed penalty method, we prefer small increments of the penalty parameter in the initial iterations because these iterations are the hardest to solve. Specifically, when the forcing term is not used, the first linear systems that are solved are the ones that need more iterations to converge. Even when the number of non-linear problems to be solved would be reduced, increasing faster the penalty parameter would make the initial linear systems harder to solve. In our applications, we have observed that faster increases of the penalty parameter can lead to non-converging linear and non-linear problems. This is especially true when approximating complex geometries with highly stretched elements.

The importance of the proposed penalty parameter adaption is that we ensure when it is possible to increase more rapidly the penalty parameter to reduce the number of non-linear problems to be solved, without hampering the robustness of the mesh curving process.

\subsection{Forcing term}


When all the linear systems are solved with the same accuracy, the initial linear systems are the ones that require more iterations to be solved. Thus, the forcing term allows reducing the number of linear iterations to solve such linear systems. Since it is not necessary to obtain an accurate solution of the initial linear systems, the forcing term avoids obtaining an over-converged solution in the initial steps of the penalty method.

In the first example, we show that the forcing term reduces the number of linear iterations to curve the quadratic and cubic mesh more than two times. Thus, the forcing term is an important feature when curving quadratic meshes. This is so because all the linear solver iterations are performed in quadratic meshes and the total number of linear solver iterations is reduced.

The generation of a quadratic mesh is an important step when generating high-order meshes. The majority of problems related to the generation of the linear mesh can be detected and solved when generating a quadratic mesh. Thus, the forcing term allows obtaining quadratic meshes more efficiently, and therefore, it allows to reduce the time of the whole process of high-order mesh generation.


\subsection{Mesh Optimization}

In this work, we are performing a full optimization process of the whole mesh. Although this approach may seem expensive, we ensure that the final curved high-order mesh is optimal in all the elements. This is an important point because non-optimized elements may introduce spurious oscillations in the solution of a simulation process. Nevertheless, some implementations only optimize the worst quality elements. Specifically, they optimize the elements with quality lower than a given threshold, and some additional elements around to increase the feasible locations of the high-order nodes. Therefore, these methods require to select the threshold value in such a manner that the optimized mesh is good enough for the simulation, and how many elements to additionally optimize to obtain a feasible solution. By optimizing the whole mesh, we ensure the optimality of all the elements without additional parameters.

The optimization of the whole mesh is especially important in cases with a highly-stretched boundary layer.  In these cases, it is of major importance to obtain optimal elements in the boundary layer and therefore, the whole boundary layer should be optimized. Since the number of elements in the boundary layer can be a significant fraction of the total elements, it is not clear the reduction in computational resources of optimizing only a fraction of the elements.

Although the existent literature comparing local and global mesh optimization does not deal with piece-wise polynomial curved meshes and highly stretched elements, it suggests that for mesh curving a specific-purpose global optimization method might be preferred. Existent literature in local and global optimization methods for linear meshes shares a common conclusion. When highly optimized and accurate meshes are required, especially in isotropic meshes featuring high gradations of the element size, a specific-purpose global feasible Newton method outperforms local optimization methods. This setting also corresponds to the general mesh curving problem where we need to exploit the quadratic convergence of Newton's method since we want high-precision to both approximate the curved geometries and deal with highly stretched elements. Furthermore, we want to apply our mesh curving method to meshes featuring smaller sizes close to the objects immersed in the fluid stream and bigger sizes in the far-field, and thus, we need to deal with high gradations of the element size. Nevertheless, in the near future, it could be interesting to compare our global parallel method featuring a global quadratic convergence rate with a local parallel method featuring a local quadratic convergence rate.

\subsection{Distributed Parallel}

To generate curved meshes, we use all the memory available in each computing node and thus, use a small number of processors for coarse meshes and a larger number of processors for fine meshes. That is, instead of fixing the problem size and use more computing resources, we increase both the problem size and the number of processors. This allows the generation of large meshes as long as we have the required processors.

In our formulation, we need to store the Hessian matrix or the pre-conditioner matrices in a distributed fashion. Thus, these memory requirements determine the number of computing nodes required for the curving process. This need is so since each computing node has a fixed amount of main memory to deal with a maximum number of non-zero entries of the matrices, determined by the polynomial degree and the number of elements per processor.

\subsection{CAD-based Mesh Generation}

The process of generating a curved high-order mesh given a CAD model has three stages. The first stage is the geometry healing and defeaturing of a given CAD model. The second stage is the generation of an initial linear mesh with elements of the desired shape and size. Finally, the third stage is the actual high-order mesh curving by optimizing the mesh distortion. In our particular case, we spend most of the time in the first two stages. Specifically, the first two stages require the work of trained personnel and specialized software.

Usually, if the initial linear mesh is \emph{good} enough, the optimization process is performed without incident. In our experience, the high-order mesh needs enough resolution to represent the underlying CAD geometry. In addition, to avoid invalid elements, there should not be any element with two faces on the same simulation surface (CAD surfaces in the virtual geometry group). Since the intent is to group surfaces with (almost) normal continuity, the curving process will lead to (almost) null Jacobians in the shared edge.

\subsection{Future work}

\newtext{In the near future, we will investigate reducing the memory requirements further using a mixed-order approach. In this approach, the elements with negligible curving are represented with degree one, and the significantly curved ones with the target degree. During the curving process, a significant number of elements could remain linear. Accordingly, this would lead to fewer degrees of freedom and thus to smaller matrices. At the end of the mesh curving process, the linear elements can be represented with the target polynomial degree.}

\section{Concluding Remarks}
\label{sec:concludingRemarks}

We have presented a $p$-continuation method on a penalty-based second-order optimizer. To obtain a fully converged mesh in all the domain, we have used a global and tight tolerance that ensures a converged residual.  We have combined three ingredients that reduce the memory footprint, the waiting time, and the energy consumption of generating curved meshes. The matrix-free GMRES with the block-SOR pre-conditioner reduces three times the memory footprint. Moreover, The penalty parameter adaption decreases the number of non-linear problems. Finally, the indicator of the required tolerance to solve the linear systems reduces the number of linear iterations.

Our method generates large-scale meshes composed of millions of quartic elements with high stretching for complex virtual geometries using thousands of processors. This capability is critical to perform high-fidelity simulations on complex domains using unstructured high-order methods.

\section{Acknowledgements}
\label{sec:acknowledgements}

This project has received funding from the European Research Council (ERC) under the European Union's Horizon 2020 research and innovation programme under grant agreement No 715546. This work has also received funding from the Generalitat de Catalunya under grant number 2017 SGR 1731. The work of Xevi Roca has been partially supported by the Spanish Ministerio de Econom\'ia y Competitividad under the personal grant agreement RYC-2015-01633. We acknowledge PRACE for awarding us access to MareNostrum at Barcelona Supercomputing Center (BSC), Spain.

\biboptions{sort&compress}
\bibliography{bibliography}

\listOfTodo

\end{document}